\documentclass[11pt]{article}
\pdfoutput=1
\usepackage{jheppub}
\usepackage{amsthm}

\usepackage{hyperref}

\usepackage{xcolor}
\usepackage{mathtools}

\usepackage{tikz}
\usetikzlibrary{arrows.meta}
\usetikzlibrary{decorations.markings}
\tikzset{->-/.style={decoration={
			markings,
			mark=at position #1 with {\arrow{>}}},postaction={decorate}}}

\usepackage{float}

\usepackage{graphicx}
\usepackage{caption}
\usepackage{subfigure}
\usetikzlibrary{calc,fadings,decorations.pathreplacing}
\usepackage{verbatim}
\usepackage[normalem]{ulem}

\preprint{LITP-26-22}

\title{Dressed AdS$_4$ black holes in M-theory from the 3d superconformal index}

\author[a]{Evan Deddo,}

\author[b]{Seppe Geukens,}

\author[a]{Leopoldo A.~Pando Zayas,}

\author[b]{Christoph F.~Uhlemann,}

\author[a,c]{and Wenjie Zhou}

\emailAdd{evdedd@umich.edu, seppe.geukens@vub.be, lpandoz@umich.edu, christoph.uhlemann@vub.be, wzhou51@jhu.edu}

\affiliation[a]{Leinweber Institute for Theoretical Physics, 
University of Michigan, Ann Arbor, MI 48109, USA}

\affiliation[b]{Theoretische Natuurkunde, Vrije Universiteit Brussel and The International Solvay Institutes,\\ Pleinlaan 2, B-1050 Brussels, Belgium}

\affiliation[c]{Department of Physics and Astronomy, Johns Hopkins University, Baltimore, MD 21218, USA}

\abstract{The superconformal index of 4d $\mathcal N\,{=}\,4$ SYM has been shown to exhibit a rich phase structure, whose holographic interpretation in AdS$_5$ involves (dual-)dressed black holes, gray galaxies, and graviton gas contributions. Here we extend the analysis to 3d CFTs, focusing on ABJM theory and the $\mathcal N=4$ ADHM theory. We study the microcanonical superconformal index with two independent charges at finite $N$ and obtain exact signed state degeneracies up to large charges. The results exhibit the black hole physics expected to emerge at large-$N$, including isolated black holes and black holes dressed by additional matter.}

\begin{document}

\maketitle

\clearpage

\parskip 1mm

\section{Introduction}

The interplay between quantum field theory, string theory, and holography in elucidating the microscopic description of black holes has a fruitful history. Accounts for the entropy of supersymmetric black holes through state counting date back to~\cite{Strominger:1996sh} in the D1/D5 system, and an additional angle emerged from derivations for the entropy of static AdS$_4$ black holes through the topologically twisted index of 3d SCFTs in \cite{Benini:2015eyy}. A microscopic accounting of the entropy of rotating AdS$_5$ black holes initially faced obstacles \cite{Kinney:2005ej} which were overcome in \cite{Cabo-Bizet:2018ehj,Choi:2018vbz,Benini:2018ywd}, where the entropy was reproduced from the superconformal index of 4d $\mathcal N=4$ SYM.

The scope of these investigations was expanded recently in \cite{Choi:2025lck,Deddo:2025jrg}: The superconformal index (SCI) as a tool to count supersymmetric states is not restricted to black hole states. Indeed, it was shown in \cite{Choi:2025lck} that the SCI of 4d $\mathcal N=4$ SYM exhibits a rich phase structure through which it can recover and support the existence of supersymmetric gravitational configurations where the total charge and spin are carried by a black hole with supporting structures. These gravitational configurations include gray galaxies and dual-dressed black holes, and were constructed previously in \cite{Kim:2023sig,Choi:2024xnv,Bajaj:2024utv}. The original field-theoretic evidence for such configurations comes from the structure of supercharge cohomologies in 4d  $\mathcal N=4$ SYM with gauge group $\mathrm{SU}(N)$ at small values of $N$ \cite{Choi:2023znd}. These results establish the SCI as a powerful diagnostic for more general gravitational configurations beyond isolated black holes.

Here we take the above investigations to the realm of 3d SCFTs and gravitational saddles in AdS$_4$. Compared to previous investigations of 4d $\mathcal N=4$ SYM,  this provides a complementary angle: On the gravitational side the holographic duals admit truncations to 4d supergravities, which tend to be more tractable than the 5d theories relevant for 4d $\mathcal N=4$ SYM, e.g.\ with fewer independent curvature invariants. On the field theory side, on the other hand, the supersymmetric indices involve non-perturbative states and the partition functions include additional sums over flux configurations. This adds a complication compared to 4d $\mathcal N=4$ SYM where the full SCI arises as plethystic exponentiation of the single-letter index. A further difference is that the accounting for the degrees of freedom becomes more involved; while conformal anomaly coefficients give an easily accessible way to count degrees of freedom in 4d $\mathcal N=4$ SYM, the 3d free energy contains dynamical information and is harder to access.

The concrete 3d theories we analyze are both realized in M-theory. The first is ABJM theory \cite{Aharony:2008ug}, the $\mathrm{U}(N)_k\times \mathrm{U}(N)_{-k}$ Chern-Simons-matter theory realized by $N$ M2 branes probing a $\mathbb{C}^4/\mathbb{Z}_k$ singularity. The second is ADHM theory \cite{deBoer:1996mp,deBoer:1996ck}, which is 3d $\mathcal N=4$ $\mathrm{U}(N)$ SYM with an adjoint hypermultiplet and $N_f$ fundamental hypermultiplets. The IR fixed point of the latter describes $N$ M2 branes at an $A_{N_f-1}$ singularity $\mathbb{C}^2\times (\mathbb{C}^2/\mathbb{Z}_{N_f})$. For $N_f=1$ it coincides with $k=1$ ABJM theory. The holographic duals in M-theory are AdS$_4\times S^7/\mathbb{Z}_k$ and AdS$_4\times S^7/\mathbb{Z}_{N_f}$, distinguished by different orbifold actions on the $S^7$ for $(k,N_f)\neq (1,1)$. The orbifold action is free for ABJM and with a fixed $S^3$ for ADHM (see \cite{Hatsuda:2021oxa,Chester:2023qwo} for reviews).

To explore the emergence of black hole physics from the CFTs, we study the microcanonical superconformal index at finite $N$ and with two independent charges for the two aforementioned 3d SCFTs. Compared to the recent analysis of $N_f=1$ ADHM in \cite{Hwang:2025hfs} we incorporate two independent charges rather than restricting to a single charge combination, and push to larger charges. This allows for detailed comparisons to large-$N$ expectations based on the dual gravitational descriptions, including both isolated black holes and multi-component configurations where part of the total charge is carried outside the black hole.

The relevant AdS$_4$ black hole solutions were constructed in \cite{Cvetic:2005zi,Hristov:2019mqp} (and used e.g.\ in \cite{Choi:2018fdc,Cassani:2019mms}), and can be uplifted to the M-theory duals for ABJM and ADHM. 
The more general gravitational (i.e.\ large $N$) black hole phase diagram in $AdS_4$ was discussed in \cite{Jones:2025gno}, with minimal assumptions on the nature of additional components dressing a central black hole. Natural candidates include graviton gas configurations \cite{Kim:2023sig} and dual giant gravitons, studied in empty $AdS_4$ as probes \cite{Grisaru:2000zn,Janssen:2002cf,Bhattacharyya:2007sa,Nishioka:2008ib} and as backreacted solutions \cite{Lin:2004nb}. However, to our knowledge no backreacted configurations combining black holes with giant gravitons have been constructed. On the gravitational side we will also work with mild assumptions and at the level of entropy functions. We will show that a simple extension of the discussion in \cite{Jones:2025gno}, which incorporates some anticipated effects of the backreaction of additional dressing components, aligns very well with the finite $N$ index results, which remain our main focus.

\bigskip

\textbf{Outline:} In section~\ref{sec:3dCFTs} we introduce the relevant aspects of the 3d field theories and their superconformal indices. In section~\ref{sec:entropy-functions} we review the dual black holes and set up entropy functions motivated by the gravitational description of large-$N$ theories. In section~\ref{sec:results} we present the superconformal index results in the form of indicial entropies. In section~\ref{sec:comparison} we compare the rigorous finite-$N$ results to expectations based on gravitational intuition.

\section{ABJM and ADHM superconformal indices}\label{sec:3dCFTs}

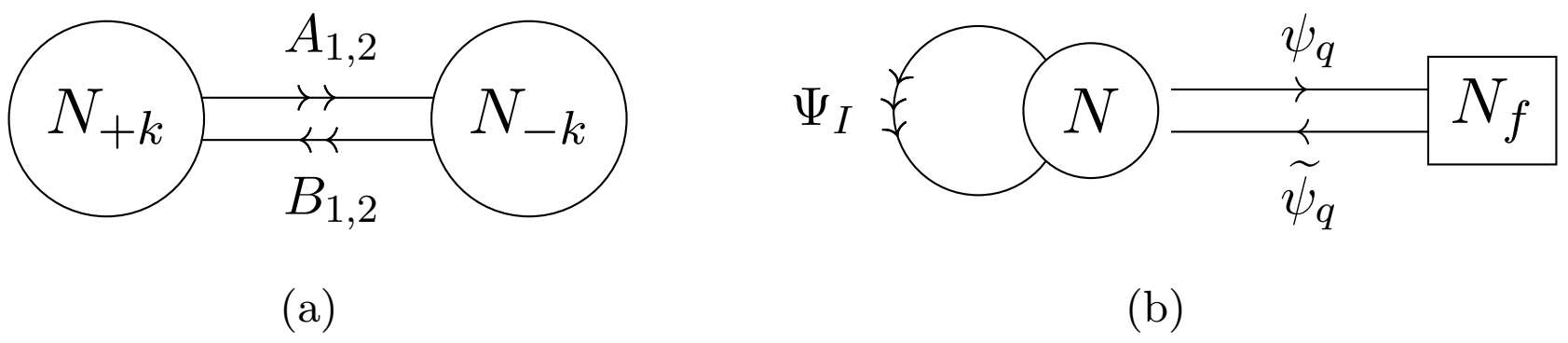
\begin{figure}
	\centering
	\subfigure[][]{\label{fig:ABJMquiver}
		\begin{tikzpicture}
			\draw[->-=0.47,->-=0.57] (0.57,0.15) -- (2.43,0.15);
			\draw[->-=0.47,->-=0.57] (2.43,-0.15) -- (0.57,-0.15);
			\node at (0,0) [circle,draw,scale=1.2,fill=white] {$N_{+k}$};
			\node at (3,0) [circle,draw,scale=1.2,fill=white] {$N_{-k}$};
			\node at (1.6,0.55) {$A_{1,2}$};
			\node at (1.6,-0.6) {$B_{1,2}$};
		\end{tikzpicture}
	}\qquad
	\subfigure[][]{
		\begin{tikzpicture}\label{fig:ADHMquiver}
			\draw[->-=0.52] (0.57,0.15) -- (2.43,0.15);
			\draw[->-=0.52] (2.43,-0.15) -- (0.57,-0.15);
			\draw[->-=0.45,->-=0.5,->-=0.55] (-0.2,0) arc (0:360:0.6);
			\node at (0,0) [circle,draw,scale=1.2,fill=white] {$N$};
			\node at (2.85,0) [rectangle,draw,scale=1.2,fill=white] {$N_f$};
			\node at (-1.9,0) {$\Psi_I$};
			\node at (1.55,0.5) {$\psi_q$};
			\node at (1.55,-0.6) {$\widetilde{\psi}_q$};
		\end{tikzpicture}
	}
	\caption{Quiver diagrams in 3d $\mathcal N=2$ notation for ABJM theory (left) and ADHM theory (right).\label{fig:quivers}}
\end{figure}

We study two prominent families of 3d SCFTs which can both be engineered by M2-branes in M-theory. The first is the ABJM Chern-Simons-matter theory of \cite{Aharony:2008ug}, and the second theory is the 3d Yang-Mills gauge theory known as ADHM \cite{deBoer:1996mp,deBoer:1996ck}, where the SCFT arises as an IR fixed point. Concretely, the theories are defined as follows:
\begin{itemize}
	\item[--] ABJM theory is a $\mathrm{U}(N)_k\times \mathrm{U}(N)_{-k}$ Chern--Simons theory coupled in $\mathcal N=2$ language to four bifundamental chiral multiplets $A_{1,2}$ and $B_{1,2}$, two in each orientation, with superpotential
	\begin{align}
		W_{\mathrm{ABJM}}
		&=
		\operatorname{Tr}
		\left[
		A_1B_1 A_2B_2
		-
		A_1B_2 A_2B_1
		\right] \, .
	\end{align}
	It has $\mathcal N=6$ supersymmetry for general $k$, which enhances to $\mathcal N=8$ for $k=1,2$ \cite{Bashkirov:2010kz}.
	
	\item[--]  ADHM theory is an $\mathcal N=4$ $\mathrm{U}(N)$ Yang-Mills theory with an adjoint hypermultiplet and $N_f$ fundamental hypermultiplets. In $\mathcal N=2$ terms it has three adjoint chiral multiplets $\Psi_{1,2,3}$, with $\Psi_{1,2}$ forming an $\mathcal N=4$ hypermultiplet and $\Psi_3$ belonging to the $\mathcal N=4$ vector multiplet, and $N_f$ fundamental/anti-fundamental pairs $(\psi_q,\widetilde\psi_q)$. 
	The superpotential is 
	\begin{align}
		W_{\mathrm{ADHM}}
		&=
		\operatorname{Tr}
		\left[
		\sum_{q=1}^{N_f}
		\widetilde\psi_q\Psi_3\psi_q
		+
		\Psi_3[\Psi_1,\Psi_2]
		\right] \, .
	\end{align}
\end{itemize}
The particle content is summarized in the quiver diagrams in Figure~\ref{fig:quivers}.

In M-theory, ABJM and ADHM theory are both realized by $N$ M2-branes probing different singularities. Namely,
\begin{align}\label{eq:M2-sing}
	\text{ABJM: \ } &\mathbb{C}^4/\mathbb{Z}_k~, & \text{ADHM: \ } & \mathbb{C}^2\times \mathbb{C}^2/\mathbb{Z}_{N_f}~.
\end{align}
The two setups are equivalent for $N_f=k=1$, where the supersymmetry enhances to $\mathcal N=8$ with $\mathrm{SO}(8)$ R-symmetry. For $(N_f,k)\neq (1,1)$ they are different.
The holographic descriptions of the (IR) SCFTs in M-theory involve AdS$_4\times S^7$ with a freely acting $\mathbb{Z}_k$ orbifold acting on $S^7$ for ABJM and a $\mathbb{Z}_{N_f}$ orbifold with a fixed locus for ADHM \cite{Aharony:2008ug,Mezei:2013gqa}.

\paragraph{Free energies:} The exact finite-$N$ $S^3$ partition functions for ABJM and ADHM admit closely related matrix-model representations. Namely, for ABJM \cite{Marino:2011eh} and ADHM \cite{Mezei:2013gqa}, respectively, as
\begin{align}
	Z^{\mathrm{ABJM}}_{S^3}[N,k]
	&=
	\frac{1}{(2\pi)^N N!}
	\int\left[\prod_{a=1}^N d\lambda_a\right]
	\det\left[\frac{1}{\cosh(\lambda_a-\lambda_b)}\right]
	\prod_{a=1}^N\frac{1}{2\cosh(k\lambda_a)} \, ,
	\label{eq:ZS3_ABJM_exact}
\\
	Z^{\mathrm{ADHM}}_{S^3}[N,N_f]
	&=
	\frac{1}{(2\pi)^N N!}
	\int\left[\prod_{a=1}^N d\lambda_a\right]
	\det\left[\frac{1}{\cosh(\lambda_a-\lambda_b)}\right]
	\prod_{a=1}^N
	\left(\frac{1}{2\cosh\lambda_a}\right)^{N_f} \, .
	\label{eq:ZS3_ADHM_exact}
\end{align}
Writing $F=-\log Z_{S^3}$, their leading large-$N$ free energies take the form
\begin{align}\label{eq:large-N F}
	F^{(0)}_{\mathrm{ABJM}}
	&=
	\frac{\sqrt{2}\pi}{3}\sqrt{k}\,N^{3/2} \, ,
	&
	F^{(0)}_{\mathrm{ADHM}}
	&=
	\frac{\sqrt{2}\pi}{3}\sqrt{N_f}\,N^{3/2} \, .
\end{align}
The matrix models (\ref{eq:ZS3_ABJM_exact}), (\ref{eq:ZS3_ADHM_exact}) agree for $N_f=k=1$, as expected. Moreover, upon identifying $N_f=k$, the free energies agree in their leading large-$N$ behavior, though they differ starting at the first subleading order unless $N_f=k=1$ \cite{Mezei:2013gqa}. For ABJM, exact finite-$N$ values were computed in \cite{Hatsuda:2012dt} for $(k,N)=(1,44), (2,20), (3,18), (4,16)$ and $(6,14)$. 

The perturbative large-$N$ expansions of these integrals may also be resummed to all orders into an Airy function, providing two examples of the Airy conjecture for holographic M2-brane SCFTs \cite{Marino:2011eh,Geukens:2024zmt,Bobev:2025ltz}. The Airy function approximations take the form
\begin{equation}\label{eq:Airy_Z}
    Z_{S^3}^\text{(Airy)}= C^{-1 / 3} e^{A} \text { Ai}\left[C^{-1 / 3}\left(N-B\right)\right],
\end{equation}
where $B$ and $C$ for ABJM \cite{Hatsuda:2012dt} and ADHM \cite{Hatsuda:2014vsa} are
\begin{align}
    B_\text{ABJM}(k)&=\frac{k}{24}+\frac{1}{3 k}, &\quad C_\text{ABJM}(k)&=\frac{2}{\pi^2 k},\nonumber\\
    B_\text{ADHM}(N_f)&=\frac{1}{2 N_f}-\frac{N_f}{8}, &\quad C_\text{ADHM}(N_f)&=\frac{2}{\pi^2 N_f}.
\end{align}
For both theories, $A$ is provided in \cite{Hatsuda:2014vsa}, with $A_\text{ADHM}(N_f)$ expressible in terms of the ABJM function:
\begin{align}
    A_\text{ABJM}(k)&=\frac{2 \zeta(3)}{\pi^2 k}\left(1-\frac{k^3}{16}\right)+\frac{k^2}{\pi^2} \int_0^{\infty} d x \frac{x}{e^{k x}-1} \log \left(1-e^{-2 x}\right), \nonumber\\
    A_\text{ADHM}\left(N_f\right)&=\frac{1}{2}\left[A_\text {ABJM}\left(N_f\right)+A_\text {ABJM}(1) N_f^2\right].
\end{align}

\paragraph{Free sectors:} A feature that will play a role below is the presence of decoupled sectors. Their existence can be inferred from the brane picture (cf.\ \eqref{eq:M2-sing}). For $N_f=k=1$ the M2-branes probe $\mathbb C^4$, and there is a $\mathrm{U}(1)$ theory as decoupled sector corresponding to overall motion of the M2-branes. ADHM with $N_f>1$  retains a free hypermultiplet associated with the unquotiented $\mathbb C^2$, while ABJM with $k>1$ has no analogous decoupled sector. 

The free sector in $\mathrm{U}(N)$ ADHM arises as follows: The $\mathrm{U}(1)$ part of the $\mathcal N=4$ adjoint hypermultiplet has no gauge interactions and decouples for all $N$, $N_f$. For $N_f=1$, the theory in addition has scalar monopole operators with $\Delta=\frac{1}{2}$ for all $N$, as shown in \cite[section 3.4]{Bashkirov:2010kz}. They saturate the unitarity bound and are consequently free, and their supersymmetry completion provides a decoupled twisted $\mathcal N=4$ hypermultiplet. Since a twisted multiplet also is the IR limit of a $\mathrm{U}(1)$ gauge theory with one fundamental hypermultiplet, this provides the remaining fields for a full $\mathrm{U}(1)$ ADHM decoupled sector arising in the IR.

The partition functions for the decoupled sectors are
\begin{equation}\label{eq:Z_decoupled}
    Z_{S^3\;\text{(dec)}}^\text{ABJM}[k]=
    \begin{cases}
        1/4, \quad &k=1\\
        1, \quad &k>1
    \end{cases} \, ,
    \qquad\qquad
    Z_{S^3\;\text{(dec)}}^\text{ADHM}[N_f]=
    \begin{cases}
        1/4, \quad &N_f=1\\
        1/2, \quad &N_f>1
    \end{cases} \, ,
\end{equation}
following from \eqref{eq:ZS3_ADHM_exact} with $N=1$ for $k=1$ or $N_f=1$ and from the hypermultiplet contribution derived in \cite{Gulotta:2012yd} for $N_f>1$.

\subsection{The ABJM index}\label{sec:ABJM_index}

ABJM theory has bosonic global symmetry $\mathrm{SO}(6)_R \times \mathrm{U}(1)_T \simeq \mathrm{SU}(4)_R\times \mathrm{U}(1)_T$
which enhances to $\mathrm{SO}(8)_R$ for $k=1,2$ \cite{Aharony:2008ug,Hayashi:2022ldo}. 
Only a subgroup is manifest in the index, whose definition singles out an $\mathcal N=2$ supercharge. For $k>2$, the subgroup commuting with this supercharge is
\begin{equation}
\mathrm{SU}(2)_{A}\times \mathrm{SU}(2)_B\times \mathrm{U}(1)_T \, ,
\end{equation}
where $\mathrm{SU}(2)_A$ and $\mathrm{SU}(2)_B$ rotate the two pairs of chiral multiplets $A_{1,2}$ and $B_{1,2}$, respectively. The remaining $\mathrm{U}(1)_R\subset \mathrm{SU}(4)_R$ is the superconformal R-symmetry. For $k=1,2$, monopole operators enhance the symmetry commuting with the index supercharge to an $\mathrm{SU}(4)$ subgroup of $\mathrm{SO}(8)_R$. This $\mathrm{SU}(4)$ is distinct from the $\mathrm{SU}(4)_R$ which is part of the $\mathcal N=6$ superconformal symmetry present for $k>2$.

With $F_A$ and $F_B$ denoting the Cartans of $\mathrm{SU}(2)_A$ and $\mathrm{SU}(2)_B$, and $T$ denoting the topological charge, and using the conventions in \cite{Bobev:2022wem}, the index takes the form
\begin{align}\label{eq:ABJM_index_tr}
    \mathcal I_{\mathrm{ABJM}}
(x,\xi_A,\xi_B,\xi_T)
=
\operatorname{Tr}_{\mathrm{BPS}}\left[
(-1)^{\mathsf F}
x^{4J+2R}
\xi_A^{F_A}
\xi_B^{F_B}
\xi_T^T
\right] \, .
\end{align}
It is convenient to overparametrize the fugacities by introducing
\begin{align}
\nu_1&=\xi_T^{-1/k}\xi_A \, , 
&
\nu_2&=\xi_T^{-1/k}\xi_A^{-1} \,  ,
&
\nu_3&=\xi_T^{1/k}\xi_B \, ,
&
\nu_4&=\xi_T^{1/k}\xi_B^{-1} \, , 
\end{align}
which by definition obey
\begin{equation}
	\nu_1\nu_2\nu_3\nu_4=1 \,.
\end{equation}
The corresponding charges $Q_I$ are defined by
\begin{align} \label{eq:geo_charges}
    F_A&=Q_1-Q_2 \, ,
    &
    F_B&=Q_3-Q_4 \, ,
    &
    kT&=Q_3+Q_4-Q_1-Q_2 \, , 
    &
    2R&=\sum_{I=1}^4Q_I \, .
\end{align}
In this charge basis, the BPS equation reads
\begin{equation}\label{eq:BPS condition}
	E=J+R=J+\frac12 \sum_{I=1}^4Q_I \, ,
\end{equation}
and the index becomes
\begin{align}\label{eq:ABJM_index_nu}
	\mathcal I_{\mathrm{ABJM}}(x,\nu_I)
	&=
	\operatorname{Tr}_{\mathrm{BPS}}\left[
	(-1)^{\mathsf F}
	x^Q
	\prod_{I=1}^4\nu_I^{Q_I}
	\right] \, ,
	&
	Q&=4J+\sum_{I=1}^4Q_I \, .
\end{align}
In the M-theory realization of ABJM theory the charges $Q_I$ generate individual phase rotations of the four complex coordinates of $\mathbb{C}^4$ probed by the M2-branes. These symmetries are preserved by the $\mathbb{Z}_k$ orbifold which acts freely and by a common phase identification on the complex coordinates. In the holographic dual the $S^7$ captures the angular directions in $\mathbb{C}^4$ and the $Q_I$ are the charges carried by the black holes to be discussed in section \ref{sec:entropy-functions}.

The localized matrix model is \cite{Kim:2009wb} 
\begin{align}\label{eq:ABJM_index}
    \mathcal I_{\mathrm{ABJM}}(x,\nu_I)
    =&\frac{1}{(N!)^2}
    \sum_{\mathfrak m,\widetilde{\mathfrak m}\in\mathbb Z^N}
    x^{2\epsilon_0}\oint \prod_{i=1}^N
    \frac{dz_i}{2\pi\mathrm i z_i}
    \frac{d\widetilde z_i}{2\pi\mathrm i\widetilde z_i}
    z_i^{k\mathfrak m_i}\widetilde z_i^{-k\widetilde{\mathfrak m}_i}
    \nonumber\\
    &\times\prod_{i\neq j}
    \left(1-\frac{z_i}{z_j}x^{2|\mathfrak m_i-\mathfrak m_j|}\right)
    \left(1-\frac{\widetilde z_i}{\widetilde z_j}
    x^{2|\widetilde{\mathfrak m}_i-\widetilde{\mathfrak m}_j|}\right)
    \nonumber\\
    &\times\prod_{i,j=1}^N\operatorname{PE}\left[
    x^{2|\mathfrak m_i-\widetilde{\mathfrak m}_j|}
    \left(
    \frac{\widetilde z_j}{z_i}f^+(x,\nu_I)
    +\frac{z_i}{\widetilde z_j}f^-(x,\nu_I)
    \right)\right] \, .
\end{align}
Here $\mathfrak m_i,\widetilde{\mathfrak m}_i$ are the GNO magnetic fluxes and $z_i,\widetilde z_i$ the corresponding gauge holonomies, and 
\begin{align}
    f^+(x,\nu_I)
    =&\frac{x(\nu_1+\nu_2)-x^3(\nu_3^{-1}+\nu_4^{-1})}
    {1-x^4} \, ,
    \quad
    f^-(x,\nu_I)
    =\frac{x(\nu_3+\nu_4)-x^3(\nu_1^{-1}+\nu_2^{-1})}
    {1-x^4} \,,
    \nonumber\\
    \epsilon_0
    =&\sum_{i,j=1}^N|\mathfrak m_i-\widetilde{\mathfrak m}_j|
    -\sum_{i<j}|\mathfrak m_i-\mathfrak m_j|
    -\sum_{i<j}|\widetilde{\mathfrak m}_i-
    \widetilde{\mathfrak m}_j| \, .
\end{align}
In terms of the original fugacities, the single-letter indices take the form
\begin{equation}\label{eq:phys_fug}
\begin{aligned}
    f^+(x,\xi_A,\xi_B,\xi_T)
    &=\xi_T^{-1/k}
    \frac{x(\xi_A+\xi_A^{-1})-x^3(\xi_B+\xi_B^{-1})}
    {1-x^4} \, ,
    \\
    f^-(x,\xi_A,\xi_B,\xi_T)
    &=\xi_T^{1/k}
    \frac{x(\xi_B+\xi_B^{-1})-x^3(\xi_A+\xi_A^{-1})}
    {1-x^4} \, .
\end{aligned}
\end{equation}
This form makes the $\mathrm{SU}(2)_A\times \mathrm{SU}(2)_B$ manifest through the appearance of the corresponding characters. The separation into two pairs reflects the $\mathcal N=2$ description used for localization.

\subsubsection{(2,2) and (1,3) two-fugacity specializations} \label{sec:ABJM-index-slices}

While the index with all fugacities carries the most detailed information, it is computationally challenging. A much simpler object is the unrefined index with all flavor fugacities set to one, 
\begin{equation} \label{eq:single_fugacity_index}
	\mathcal I_{\mathrm{ABJM}}(x) = 
	\operatorname{Tr}_{\mathrm{BPS}}
	\left[(-1)^{\mathsf F}x^Q\right]= 	\sum_{Q\geq0}d(Q)x^Q \, ,
\end{equation}
where $d(Q)$ are the signed degeneracies.
It retains only coarse information.
Our focus will instead be on two intermediate specializations which retain two independent fugacities. They contain more fine-grained information than the unrefined index but less than the full index. 

\paragraph{(2,2) specialization:}
A natural one-parameter restriction is to pairwise identify fugacities as follows,
\begin{equation}
    (\nu_1,\nu_2,\nu_3,\nu_4)
    =\left(y,y^{-1},y,y^{-1}\right) \, .
\end{equation}
In terms of the original fugacities this implies
\begin{align}
    \xi_A&=y \, ,
    &
    \xi_B&=y \, ,
    &
    \xi_T&=1 \, ,
\end{align}
and the index takes the form
\begin{align}\label{eq:(2,2) index}
    \mathcal I_{\mathrm{ABJM}}(x,y)
    &=\operatorname{Tr}_{\mathrm{BPS}}\left[
    (-1)^{\mathsf F}x^Q
    y^{(Q_1-Q_2+Q_3-Q_4)}\right] \, .
\end{align}
This restriction retains the diagonal Cartan of $\mathrm{SU}(2)_A\times \mathrm{SU}(2)_B$, and the resulting index is invariant under $y \rightarrow y^{-1}$.

\paragraph{(1,3) specialization:}
A second natural restriction singles out one fugacity and assigns a
common value to the other three,
\begin{equation}\label{eq:nu_to_y}
    (\nu_1,\nu_2,\nu_3,\nu_4)
    =\left(y^{2},y^{-2/3},y^{-2/3},y^{-2/3}\right) \, .
\end{equation}
In terms of the original fugacities this implies
\begin{align}
    \xi_A&=y^{4/3} \, ,
    &
    \xi_B&=1 \, ,
    &
    \xi_T&=y^{-2k/3} \, ,
\end{align}
and the index takes the form\footnote{The overall scaling of $y$ in \eqref{eq:nu_to_y} is a matter of convention. Our choice of exponents will prove convenient for comparison with the entropy of the corresponding AdS$_4$ configurations. In the bulk we will impose constraints that result in the charges associated with $y$ in \eqref{eq:(2,2) index} and \eqref{eq:(1,3) index} becoming identical.}
\begin{align}\label{eq:(1,3) index}
    \mathcal I_{\mathrm{ABJM}}(x,y)
    &=\operatorname{Tr}_{\mathrm{BPS}}\left[
    (-1)^{\mathsf F}x^Q
    y^{2Q_1-\frac23(Q_2+Q_3+Q_4)}\right] \, .
\end{align}
For $k=1,2$, all choices of 3 fugacities among the $\nu_I$ to identify are equivalent, since the charges can be rotated by $\mathrm{SO}(8)$. For $k>2$, the index \eqref{eq:ABJM_index} is still invariant under $\nu_1\leftrightarrow\nu_2$ and $\nu_3\leftrightarrow\nu_4$, while  exchanging the pairs $(\nu_1,\nu_2)\leftrightarrow(\nu_3,\nu_4)$ exchanges $f^+\leftrightarrow f^-$ and can be compensated by exchanging the two gauge nodes and reversing their magnetic fluxes.

\subsection{The ADHM index}
\label{sec:ADHM_index}

The ADHM theory possesses $\mathrm{SU}(2)_{\ell} \times \mathrm{SU}(2)_{R_1} \times \mathrm{SU}(2)_{R_2} \times \mathrm{U}(1)_T \times \mathrm{U}(N_f)$ global symmetry, where $\mathrm{SU}(2)_{R_1} \times \mathrm{SU}(2)_{R_2}$ is the R-symmetry, $\mathrm{SU}(2)_{\ell}$ is a flavor symmetry rotating $\Phi_{1,2}$, $\mathrm{U}(1)_T$ is the topological symmetry, and $\mathrm{U}(N_f)$ is the flavor symmetry acting on the $N_f$ fundamental hypermultiplets.
In the M2-brane description, the $\mathrm{SU}(2)_{\ell} \times \mathrm{SU}(2)_{R_1}\sim\mathrm{SO}(4)$ symmetry acts on one $\mathbb{R}^4$ in the transverse $\mathbb{R}^8$, and the $\mathrm{SU}(2)_{R_2} \times \mathrm{U}(1)_T$ acts on the other $\mathbb{R}^4$. For $N_f=1$ the global symmetry enhances to $\mathrm{SO}(8)$ at the IR fixed point. For $N_f=2$ there is no supersymmetry enhancement but $\mathrm{U}(1)_T$ enhances to $\mathrm{SU}(2)$.

In $\mathcal{N}=2$ language, the R-charge is $R\equiv R_1+R_2$ while $F\equiv R_1-R_2$ becomes another flavor charge, and the trace formula for the ADHM superconformal index is
\begin{equation}\label{eq:ADHM_index_tr}
   \mathcal{I}_{\mathrm{ADHM}}\left(x, \xi_{\ell}, \xi_F, \xi_T\right)=\operatorname{Tr}_{\mathrm{BPS}}\left[(-1)^{\mathsf F} x^{4 J+2R}\left(\xi_{\ell}\right)^{\ell}\left(\xi_F\right)^F\left(\xi_T\right)^T\right], 
\end{equation}
where $J$ is the third component of angular momentum, and fugacities $x$, $\xi_\ell$, $\xi_F$, and $\xi_T$ have been introduced for the charges defined above. We will consider two equivalent expressions for the ADHM index: the matrix integral and a factorized form. 

\medskip

\paragraph{ADHM index as matrix integral:} The matrix model for the ADHM index was given in \cite{Gang:2011xp} and we follow the conventions of \cite{Bobev:2022wem}, where
\begin{align}\label{eq:ADHM_index}
\mathcal{I}_\mathrm{ADHM}\left(x, \xi_{\ell}, \xi_F, \xi_T\right) & =\frac{1}{N!} \sum_{\mathfrak{m} \in \mathbb{Z}^N} \oint\left(\prod_{i=1}^N \frac{d z_i}{2 \pi \mathrm{i} z_i} \xi_T^{\mathfrak{m}_i}\right) \prod_{i \neq j}^N\left(1-\frac{z_i}{z_j} x^{2\left|\mathfrak{m}_i-\mathfrak{m}_j\right|}\right)\nonumber\\
&\times\prod_{i,j=1}^N\text{PE}\left[x^{2|\mathfrak{m}_i-\mathfrak{m}_j|}\frac{z_i}{z_j}\left(\frac{x \left( \sqrt{\xi_\ell \xi_F} + \sqrt{\frac{\xi_F}{\xi_\ell}} \right) - x^3 \left( \frac{1}{\sqrt{\xi_\ell \xi_F}} + \sqrt{\frac{\xi_\ell}{\xi_F}} \right)}{1 - x^4}\right)\right]\nonumber\\
&\times\prod_{i,j=1}^N\text{PE}\left[x^{2|\mathfrak{m}_i-\mathfrak{m}_j|}\frac{z_i}{z_j}\left(\frac1{\xi_F}-\xi_F\right)\frac{x^2}{1-x^4}\right]\nonumber\\
&\times\left[\prod_{i=1}^N\left(\frac{x}{\sqrt\xi_F}\right)^{|\mathfrak{m}_i|}\text{PE}\left[x^{2|\mathfrak{m}_i|}\left(z_i+\frac1{z_i}\right)\frac{x\sqrt{\xi_F}-x^3\frac1{\sqrt{\xi_F}}}{1-x^4}\right]\right]^{N_f} \, .
\end{align}
The second, third, and fourth lines correspond to the adjoint hypermultiplet, vector multiplet, and fundamental hypermultiplet, respectively. Following the conventions of \cite{Hwang:2025hfs}, we have rescaled the fugacity $x$ such that the associated charge $Q\equiv4J+2R$ is integer quantized.

\paragraph{Factorized form of ADHM index:}
The ADHM index can also be expressed in a factorized form, following \cite{Hwang:2012jh}. We restrict the  discussion here to $N_f=1$. To derive the factorized form, the residues of the matrix integral are reorganized into a sum of terms corresponding to Higgs vacua of the theory. Each Higgs vacuum is labeled by a Young tableau of size $N$, and the sum over tableaux $\mathcal{Y}$ takes the form \cite{Choi:2019zpz}
\begin{equation}\label{eq:factorized_schematic}
    \mathcal{I}_\text{ADHM}(x, \xi_{\ell}, \xi_F, \xi_T)=\sum_{|\mathcal{Y}|=N} Z_{\text {pert}}^{\mathcal{Y}}(x, \xi_{\ell}, \xi_F) Z_{\text {vort}}^{\mathcal{Y}}(x, \xi_{\ell}, \xi_F, \xi_T) Z_{\text {vort}}^{\mathcal{Y}}\left(x^{-1}, \xi_{\ell}^{-1}, \xi_F^{-1}, \xi_T^{-1}\right) \, .
\end{equation}
Each term in the sum is a product of three terms corresponding to perturbative, vortex, and anti-vortex partition functions. The explicit expressions for $Z_{\text {pert}}^{\mathcal{Y}}$ and $Z_{\text {vort}}^{\mathcal{Y}}$ are\footnote{Our fugacity conventions are related by  $q=x^2$, $z=\sqrt{\xi_\ell}$, $t=\xi_F$, $w=\xi_T$ to those of \cite{Hwang:2025hfs}, where the factorized form was used to evaluate the single-fugacity index and compare to black holes.}
\begin{align} \label{eq:Z_pert}
Z_{\mathrm{pert}}^{\mathcal{Y}}(x, \xi_{\ell}, \xi_F)= & \prod_{\mathrm{a} \neq \mathrm{b} \in \mathcal{Y}}\left(1-v_{\mathrm{a}}^{-1} v_{\mathrm{b}}\right)\times \prod_{\mathrm{a} \in \mathcal{Y}} \frac{\left(v_{\mathrm{a}} x^4 ; x^4\right)}{\left(v_{\mathrm{a}}^{-1} ; x^4\right)^\prime} \frac{\left(v_{\mathrm{a}}^{-1}\xi_F^{-1} x^2 ; x^4\right)}{\left(v_{\mathrm{a}} \xi_F x^2 ; x^4\right)}\times
 \nonumber\\
& \prod_{\mathrm{a}, \mathrm{b} \in \mathcal{Y}} \frac{\left(v_{\mathrm{a}} v_{\mathrm{b}}^{-1} \xi_F x^2 ; x^4\right)^\prime\Big(v_{\mathrm{a}} v_{\mathrm{b}}^{-1}\xi_\ell^{-\frac{1}{2}} \xi_F^{-\frac{1}{2}} x^3 ; x^4\Big)\Big(v_{\mathrm{a}} v_{\mathrm{b}}^{-1}\xi_\ell^{\frac{1}{2}} \xi_F^{-\frac{1}{2}} x^3 ; x^4\Big)}{\left(v_{\mathrm{a}}^{-1} v_{\mathrm{b}}\xi_F^{-1} x^2 ; x^4\right)\Big(v_{\mathrm{a}}^{-1} v_{\mathrm{b}} \xi_\ell^{\frac{1}{2}} \xi_F^{\frac{1}{2}} x ; x^4\Big)^\prime\Big(v_{\mathrm{a}}^{-1} v_{\mathrm{b}} \xi_\ell^{-\frac{1}{2}} \xi_F^{\frac{1}{2}} x ; x^4\Big)^\prime} \, ,
\end{align}
and
\begin{align}
\label{eq:Z_vort}
\nonumber\\
Z_{\text {vort }}^{\mathcal{Y}}(x, \xi_{\ell}, \xi_F, \xi_T)= & \sum_{k_{\mathrm{a}}}\left(\xi_T \xi_F^{\frac{1}{2}} x^{-1}\right)^{\sum_{\mathrm{a} \in \mathcal{Y}} k_{\mathrm{a}}}\prod_{\mathrm{a} \in \mathcal{Y}} \frac{\left(v_{\mathrm{a}}^{-1} ; x^4\right)_{-k_{\mathrm{a}}}}{\left(v_{\mathrm{a}}^{-1} \xi_F^{-1} x^2 ; x^4\right)_{-k_{\mathrm{a}}}} \nonumber\\
& \times\prod_{\mathrm{a} \neq \mathrm{b} \in \mathcal{Y}} \frac{\Big(v_{\mathrm{a}}^{-1} v_{\mathrm{b}} \xi_\ell^{\frac{1}{2}} \xi_F^{\frac{1}{2}} x ; x^4\Big)_{-k_{\mathrm{a}}+k_{\mathrm{b}}}\left(v_{\mathrm{a}}^{-1} v_{\mathrm{b}} \xi_F^{-1} x^2 ; x^4\right)_{-k_{\mathrm{a}}+k_{\mathrm{b}}}}{\left(v_{\mathrm{a}}^{-1} v_{\mathrm{b}} ; x^4\right)_{-k_{\mathrm{a}}+k_{\mathrm{b}}}\Big(v_{\mathrm{a}}^{-1} v_{\mathrm{b}} \xi_\ell^{\frac{1}{2}} \xi_F^{-\frac{1}{2}} x^3 ; x^4\Big)_{-k_{\mathrm{a}}+k_{\mathrm{b}}}} \, ,
\end{align}
where for each box $\mathrm{a}\in\mathcal Y$ with coordinates $\text {(i(a), j(a))}$ we define
\begin{equation}
    v_{\mathrm{a}}\equiv\xi_\ell^{\frac12(\mathrm{i}(\mathrm{a})-\mathrm{j}(\mathrm{a}))}(\xi_F x^2)^{\frac{1}{2}(\mathrm{i}(\mathrm{a})+\mathrm{j}(\mathrm{a})-2)} \, .
\end{equation}
The first sum in $Z_{\text {vort}}^{\mathcal{Y}}$ is taken over all $N$-tuples $k_{\mathrm{a}}$ that assign each box a non-negative integer in a non-decreasing fashion across all columns and rows. The primed Pochhammer symbols in $Z_{\text {pert}}^{\mathcal{Y}}$ denote the removal of vanishing factors. 

\subsubsection{(2,2) and (1,3) two-fugacity specializations}

For the discussion of specializations of the ADHM index we introduce a charge basis which matches the ABJM description for $N_f=1$. We keep the notation $Q_I$ with $I=1,\ldots 4$, and parametrize the $\mathrm{SU}(2)_{R_1}\times \mathrm{SU}(2)_{R_2}\times \mathrm{SU}(2)_\ell\times \mathrm{U}(1)_T$ charges as follows,
\begin{align}\label{eq:geo_charges_ADHM}
	R_1&=(Q_1+Q_3)/2\,,
	& \ell&=(Q_1-Q_3)/2\,,
	\nonumber\\
	R_2&=(Q_2+Q_4)/2\,,&
	T&=Q_2-Q_4\,.
\end{align}
These expressions associate $Q_1$, $Q_3$ with $\mathrm{SU}(2)_{R_1}\times \mathrm{SU}(2)_\ell$, and $Q_2$, $Q_4$ with $\mathrm{SU}(2)_{R_2}\times \mathrm{U}(1)_T$.
The fugacities are correspondingly redefined as
\begin{align}
    \nu_1&\equiv\sqrt{\xi_F\xi_\ell}\,, &
    \nu_2&\equiv\frac{\xi_T}{\sqrt{\xi_F}}\,,& 
    \nu_3&\equiv\sqrt\frac{\xi_F}{\xi_\ell}\,,&
    \nu_4&\equiv\frac 1{\xi_T\sqrt{\xi_F}}\,.
\end{align}
This brings the $N_f=1$ ADHM index into the same form as \eqref{eq:ABJM_index_nu}. Two-fugacity specializations can now be defined in full analogy to the ABJM discussion as
\begin{align}
	(2,2):\quad\nu_1&=\nu_3=y\,,&\nu_2&=\nu_4=y^{-1}\,,&&
	\nonumber\\
	(1,3):\quad\nu_1&=y^2\,&\nu_2&=\nu_3=\nu_4=y^{-2/3}\,.
\end{align}
The (2,2) identification implies $\xi_\ell=\xi_T=1$, while the (1,3) identification fixes $\xi_T=1$, $\xi_\ell=\xi_F^2$. For $N_f=1$, when the global symmetry is enhanced to $\mathrm{SO}(8)$, the index is symmetric under permutations of the $\nu_I$, and different choices of which fugacities are to be identified pairwise for (2,2) or as a triplet for the (1,3) specialization yield equivalent information. For $N_f\neq 1$, different fugacity identifications capture different information about the spectrum. The (2,2) specialization with $\nu_1=\nu_3$ and $\nu_2=\nu_4$ retains precisely the R-charge fugacities, and is in that sense universal. We will restrict the discussion for $N_f>1$ to this specialization.

For $N_f=1$, the (2,2) ADHM index is invariant under $y \rightarrow y^{-1}$, consistent with the matching to ABJM theory, while for $N_f>1$ this symmetry is no longer expected since the Higgs-branch and Coulomb branch flavor symmetries differ.

\subsection{Mapping the charge space}\label{sec:tetrahedron}
The ABJM \eqref{eq:ABJM_index_tr} and ADHM \eqref{eq:ADHM_index_tr} indices are sensitive to four independent charges. When the fugacity relations defining the (2,2) and (1,3) index specializations in Sec.~\ref{sec:ABJM-index-slices} are imposed, classes of BPS states which lie in linear subspaces defined by charge relations are identified in the index. To discuss this in more detail we first recall the distribution of BPS states in $(J,Q_I)$ charge space. For simplicity we focus on ADHM with $N_f=1$ which is equivalent to $k=1$ ABJM theory.

The space of supersymmetric states for $k=1$ ABJM and $N_f=1$ ADHM is constrained by the 3d $\mathcal{N}=8$ superconformal algebra. As reviewed in Appendix \ref{app:3d_N=8_SCA}, each of the 16 different supercharges $\mathcal{Q}$ yields a commutation relation of the form
\begin{equation}\label{eq:{Q,Qd}}
    \left\{\mathcal{Q}, \mathcal{Q}^{\dagger}\right\}=E \pm J+\frac{1}{2}\left( \pm Q_1 \pm Q_2 \pm Q_3 \pm Q_4\right) \, . 
\end{equation}
Each choice of $\mathcal{Q}$ realizes a unique sign combination with an even number of plus signs preceding the $Q_I$. The defining supercharge for the index corresponds to the combination with all minus signs. All supercharges satisfy $\{\mathcal Q,\mathcal Q^\dagger\}\geq0$ on all states.
For states satisfying \eqref{eq:BPS condition}, $\{\mathcal Q,\mathcal Q^\dagger\}\geq0$ implies $J\geq0$ and $Q_I+Q_K\geq0$ for $I\neq K$. 

It will be convenient to introduce a charge basis $h_I$, $I=1,2,3,4$, similar to \cite{Kim:2009wb}, where the schematic relation \eqref{eq:{Q,Qd}} takes the form $\left\{\mathcal{Q}, \mathcal{Q}^{\dagger}\right\}=E\pm J\pm h_I$ and
\begin{align}
    h_1&=\frac12\left(Q_1-Q_2-Q_3+Q_4\right) \, ,
    &h_2&=\frac12\left(-Q_1+Q_2-Q_3+Q_4\right)\nonumber \, , \\
    h_3&=\frac12\left(-Q_1-Q_2+Q_3+Q_4\right) \, ,
    &h_4&=\frac12\left(Q_1+Q_2+Q_3+Q_4\right) \, .
\end{align}
Defining fugacities $y_{1,2,3}$ corresponding to $h_{1,2,3}$, the index may be expressed as
\begin{equation}\label{eq:ADHM_index_tr_y}
   \mathcal{I}_{\mathrm{ADHM}}\left(x, y_1,y_2,y_3\right)=\operatorname{Tr}_{\mathrm{BPS}}\left[(-1)^F x^{4 J+2h_4}y_1^{h_1}y_2^{h_2}y_3^{h_3}\right] \, ,
\end{equation}
where
\begin{equation}
    y_1=\frac{\sqrt{\xi_\ell}}{\xi_T}\,,\qquad y_2=\frac1{\xi_F}\,,\qquad y_3=\frac1{\xi_T\sqrt{\xi_\ell}}\,.
\end{equation}
In the $h$ basis, $\{\mathcal Q,\mathcal Q^\dagger\}\geq0$ yields the 16 constraints
\begin{equation} \label{eq:BPS_bounds}
   \qquad E\geq\pm J\pm h_I\,,\quad I=1,2,3,4\,.
\end{equation}
The BPS condition $E=J+h_4$ in tandem with the other $I=4$ inequalities constrains
\begin{equation}
    J\geq0 \, ,\quad h_4\geq0 \quad \implies \quad Q\equiv4J+2h_4\geq 0\,,
\end{equation}
and the $i=1,2,3$ inequalities are equivalent to
\begin{equation}
    |h_i|\leq h_4\leq Q/2\,.
\end{equation}
For fixed $Q$, states are therefore restricted to lie inside a cube of side length $Q$ in $(h_1,h_2,h_3)$ space.
States in the interior of the cube are in general $1/16$-BPS, while states on the boundaries possess enhanced supersymmetry. The six faces $h_i=\pm Q/2$ for $i=1,2,3$ each saturate an additional BPS equality of \eqref{eq:BPS_bounds} in addition to the original BPS condition $E=J+h_4$:
\begin{align}\label{eq:cube_BPS_constr}
    h_i=\pm(2J+h_4)\quad\implies\quad E=-J\pm h_i\,.
\end{align}

\begin{figure}
	\centering
	\includegraphics[width=0.5\linewidth]{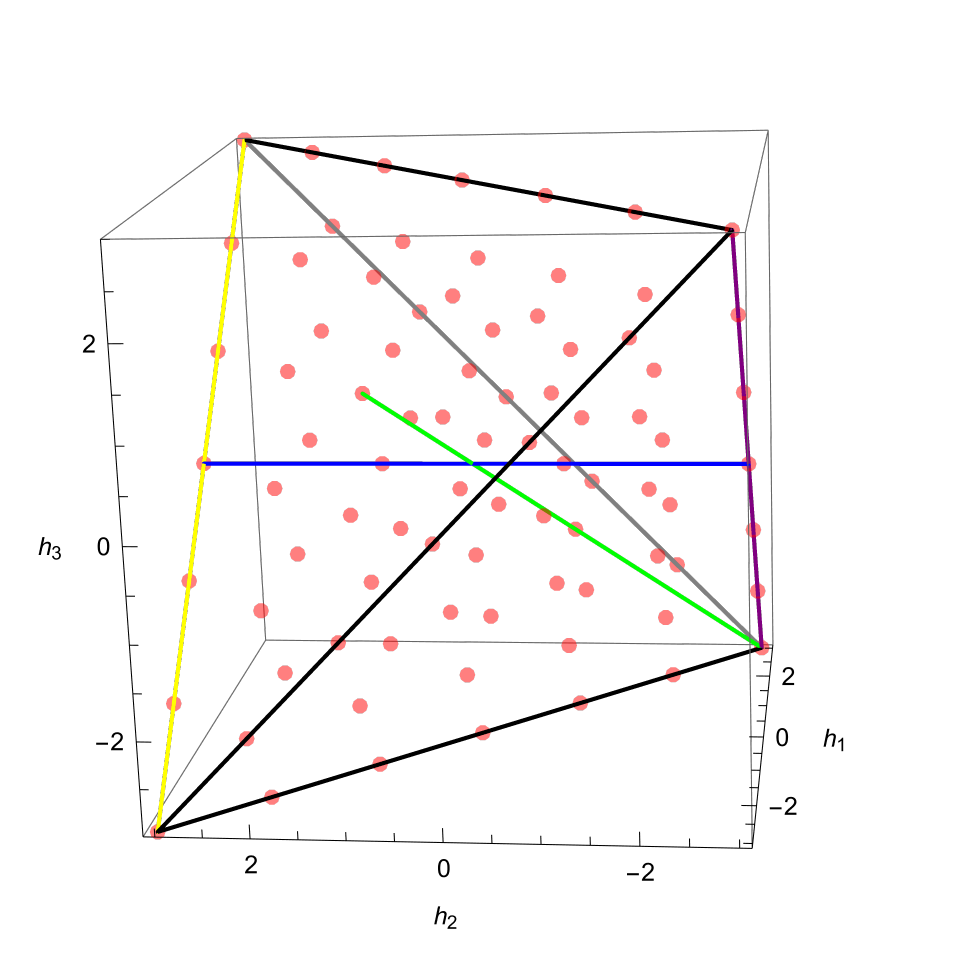}
	\includegraphics[width=0.4\linewidth]{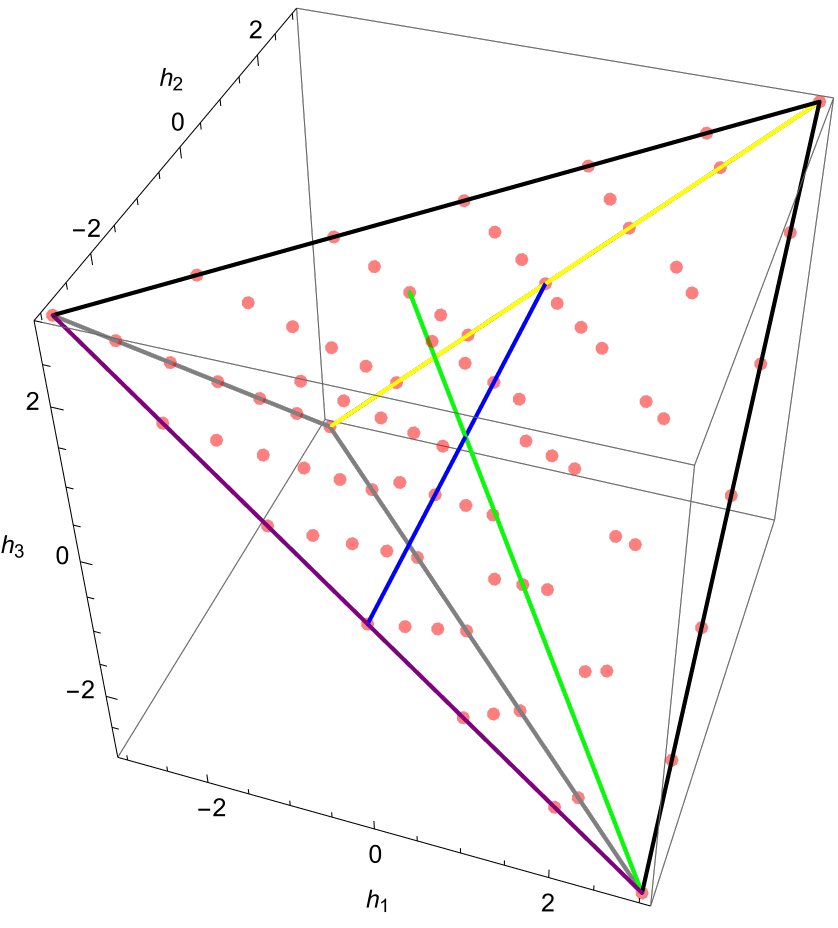}
	\caption{Points in $(h_1,h_2,h_3)$ space where the index is nonzero (red) for $N=2$ and $Q=6$. The occupied states form a tetrahedral region. The (2,2) specialization corresponds to projection onto the blue line, and the (1,3) specialization to projection onto the green line. The Coulomb limit corresponds to the yellow line, and the Higgs limit to the purple line.}
	\label{fig:charge_space}
\end{figure}

Within the charge space allowed by the superconformal algebra, further restrictions appear depending on the theory under consideration. To illustrate this for $N_f=1$ ADHM theory, Fig.~\ref{fig:charge_space} shows the lattice points in charge space where the index is non-zero for $(N,Q)=(2,6)$. Rather than filling the entire $|h_i|\leq Q/2$ cube, they only populate the tetrahedral region
\begin{align}\label{eq:tetrahedron_faces}
    h_1+h_2-h_3\leq Q/2\,,\qquad &-h_1+h_2+h_3\leq Q/2\,,\nonumber\\
    h_1-h_2+h_3\leq Q/2\,,\qquad
    &-h_1-h_2-h_3\leq Q/2\,.
\end{align}
Using the BPS condition, the four inequalities \eqref{eq:tetrahedron_faces} equivalently take the simple form
\begin{align}\label{eq:tetrahedron_J}
	J+Q_I\geq 0
\end{align} 
in the original charge basis. Compared to the superconformal algebra bounds $J\geq0$ and $Q_I+Q_K\geq0$, negative charges $Q_I$ are only allowed when compensated by angular momentum, and states with $J=0$ have $Q_I\geq0$.
The restrictions \eqref{eq:tetrahedron_faces} and \eqref{eq:tetrahedron_J} are tighter than those implied by the superconformal algebra. They are a result of the specific building blocks appearing in the index \eqref{eq:ADHM_index}. This includes letters in the plethystic exponential and terms outside the plethystic exponential for which the overall sum over $\mathfrak{m}$ and the product over $i,j=1,\ldots,N$ effectively constructs powers and products.

The charges of the letters define vectors $(Q,h_1,h_2,h_3)$, and states built from these letters correspond to linear combinations of the basic vectors.
Consider the terms inside the first plethystic exponential in \eqref{eq:ADHM_index}. Dropping the geometric series denominator $(1-x^4)$ corresponding to descendant states, the letters associated with the adjoint hypermultiplet scalars are
\begin{align}
    x^{1+2|\mathfrak{m}_i-\mathfrak{m}_j|} \left( \sqrt{\xi_\ell \xi_F} + \sqrt{\frac{\xi_F}{\xi_l}} \right)&=x^{1+2|\mathfrak{m}_i-\mathfrak{m}_j|}\left(\sqrt{\frac{y_1}{y_2y_3}}+\sqrt{\frac{y_3}{y_1y_2}}\right) \, .
\end{align}
Due to the products over $i,j$, each integer pair $(i,j)$ labels a distinct letter. Reading off the charges from the exponents of the $(x,y_1,y_2,y_3)$ fugacities yields two vectors
\begin{equation}\label{eq:Higgs_vert}
    (Q,h_1,h_2,h_3)=
    \begin{cases}
        \left(1+2|\mathfrak{m}_i-\mathfrak{m}_j|,\frac12,-\frac12,-\frac12\right)\\
        \left(1+2|\mathfrak{m}_i-\mathfrak{m}_j|,-\frac12,-\frac12,\frac12\right)
    \end{cases} \, .
\end{equation}
When $\mathfrak{m}_i=\mathfrak{m}_j$, both vectors saturate three of the inequalities \eqref{eq:tetrahedron_faces}, and thus point toward the vertices bounding the purple edge of the tetrahedron. Similarly, the scalar contribution from the fundamental hypermultiplet and the fermionic component of the vector in \eqref{eq:ADHM_index} may saturate two conditions in \eqref{eq:tetrahedron_faces} and also point to this edge. No descendant states may saturate any of the inequalities. Altogether, states on the purple edge of the tetrahedral region must be built from linear combinations of these hypermultiplet scalar letters and fermionic letter, all with $\mathfrak{m}_i=\mathfrak{m}_j$. This edge corresponds to the Higgs limit of the index.

We now turn to the yellow edge of the tetrahedron in Fig.~\ref{fig:charge_space}. The vertices bounding this edge arise from terms outside the plethystic exponential. For each tuple $\mathfrak{m}$:
\begin{equation}
\prod_{i=1}^N\xi_T^{\mathfrak{m}_i}\left(\frac{x}{\sqrt{\xi_F}}\right)^{|\mathfrak{m}_i|}
\implies
(Q,h_1,h_2,h_3)=\sum_{i=1}^N\begin{cases}
    |\mathfrak{m}_i|\left(1,-\frac12,\frac12,-\frac12\right)\, ,\qquad &\mathfrak{m}_i\geq 0\\
    |\mathfrak{m}_i|\left(1,\frac12,\frac12,\frac12\right)\, ,\qquad &\mathfrak{m}_i< 0
\end{cases} \, .
\end{equation}
The two basic vectors $(1,-\frac12,\frac12,-\frac12)$ and $(1,\frac12,\frac12,\frac12)$ each satisfy three of the constraints \eqref{eq:tetrahedron_faces}, yielding the mentioned vertices. The linear combinations produced by arbitrary $\mathfrak{m}$ populate a full diagonal in the $(h_1,h_3)$ plane. Including the additional bosonic contribution from the vector multiplet in \eqref{eq:ADHM_index}, the resulting edge corresponds to the Coulomb limit of the index (Fig.~\ref{fig:charge_space}, yellow).

\paragraph{(2,2) and (1,3) indices:} The (2,2) specialization was defined in \eqref{eq:(2,2) index} and the charges associated with the retained fugacities are $(Q,-2h_2)$. This index does not track $h_1$ or $h_3$, and counts states projected onto the $h_2$ axis, shown as blue line in Fig.~\ref{fig:charge_space}. The (1,3) specialization was defined in \eqref{eq:(1,3) index} and the charges are $(Q,\frac43(h_1-h_2-h_3))$. The corresponding projection is onto the green line in Fig.~\ref{fig:charge_space}. We introduce a common notation for the degeneracies as 
\begin{align}\label{eq:Qh-def}
	&&&d(Q,h) \, , & h&=\begin{cases} -2h_2 \, , & \text{(2,2) specialization}\\ \frac{4}{3}(h_1-h_2-h_3) \, , & \text{(1,3) specialization.}\end{cases}
\end{align}
For the (2,2) index $d(Q,h)$ is nonzero for $|h|\leq Q$, and given the symmetries of the tetrahedron it is symmetric under $h\to-h$. For the (1,3) index $d(Q,h)$ has support on $-2Q/3\leq h\leq 2Q$ and is not symmetric. These features will be exhibited explicitly in Sec.~\ref{sec:results}.

\paragraph{The $(Q,h)$-index locus}: The charge lattice in Fig.~\ref{fig:charge_space} captures the $(h_1,h_2,h_3)$ charges, with $h_4$ suppressed. Each point in the figure therefore represents a line in the full charge space. We briefly discuss this aspect.
The (2,2)-coefficient $d(Q,h)$ counts short multiplets whose charges satisfy
\begin{align}
    (Q_1+Q_3)-(Q_2+Q_4)&=h \, ,
    &
    Q_1+Q_3,\ Q_2+Q_4&\geq0 \, ,
    &
    \sum_{I=1}^4 Q_I&\leq Q \, .
    \label{eq:pairwise_index_line}
\end{align}
We refer to this as the $(Q,h)$-index locus. Since $Q$ and $h$ are fixed, it is represented by a single point of the blue line in Fig.~\ref{fig:charge_space}, onto which the (2,2) index projects, namely $h_2=-h/2$. It extends in the direction of $J$ or equivalently $h_4$, which is not displayed in the figure. The differences $Q_1-Q_3$ and $Q_2-Q_4$, corresponding to $h_1$ and $h_3$, are not probed by the index. While at the endpoints $|h|=Q$ the inequalities in \eqref{eq:pairwise_index_line} leave only a single value of $(Q_1+Q_3,\,Q_2+Q_4)$, for $|h|<Q$ they allow a range and the index locus can be populated by states with different $(Q_1+Q_3,\,Q_2+Q_4)$.

\subsection{Coulomb and Higgs limits}\label{subsec:Coulomb-Higgs}

We now return to general $N_f$ and $k$, and discuss the Coulomb and Higgs limits for the (2,2) specialization. To do so, we first introduce some notation. Setting $y_1=y_3=1$ in \eqref{eq:ADHM_index_tr_y} gives
\begin{align}
	\mathcal{I}(x,y) &= \sum_{Q,h}  d(Q,h) x^Q y^h = \sum_{Q,h} d(Q,h) g^{(Q+h)/2} c^{(Q-h)/2} \, ,
\end{align}
where we defined
\begin{align}
	y^2&\equiv y_2^{-1}=\xi_F \, , & g&\equiv xy \, , & c&\equiv xy^{-1} \, .
\end{align}
The degeneracies are related to those in the single-fugacity index in \eqref{eq:single_fugacity_index} by $d(Q)= \sum_h d(Q,h)$. 
The Higgs and Coulomb limits now amount to taking
\begin{align}
	\text{Higgs Limit:}\qquad&  x \rightarrow 0\,,\quad y \rightarrow \infty\ \  \text{ with $g$ fixed,} \nonumber\\ 
	\text{Coulomb Limit:}\qquad&  x \rightarrow 0\, ,\quad y\rightarrow 0\ \;\ \   \text{ with $c$ fixed.}
\end{align}
This isolates the Coulomb and Higgs branch operators and gives their Hilbert series.
In view of the support $|h|\leq Q$ implied by \eqref{eq:tetrahedron_faces}, the Higgs and Coulomb indices reduce to
\begin{align} \label{eq:endpointlimit}
	\mathcal{I}^{\text{H}}(g)
	&\equiv
	\sum_{Q} d_{\text{H}}(Q) g^{Q}
	\qquad \text{with} \qquad
	J=Q_2=Q_4=0 \, , \nonumber\\
	\mathcal{I}^{\text{C}}(c)
	&\equiv
	\sum_{Q} d_{\text{C}}(Q) c^{Q}
	\qquad \text{with} \qquad
	J=Q_1=Q_3=0 \, .
\end{align}
The limits retain the coefficients with $h= \pm Q$ at the boundary of the populated region of the (2,2) index, and $d_{\text{H}}(Q)$ and $d_{\text{C}}(Q)$ count the protected operators at the two endpoints. In particular, the endpoints also enforce $J=0$. 
In the ADHM matrix model \eqref{eq:ADHM_index}, the bare-monopole factor in a non-zero flux sector $\mathfrak m$ contains
\begin{align}
	\left(\frac{x}{y}\right)^{N_f\sum_i|\mathfrak m_i|}=c^{N_f\sum_i|\mathfrak m_i|} \, .
	\label{eq:ADHM_bare_monopole_pairwise}
\end{align}
Non-zero flux sectors, i.e.\ contributions from monopole operators, are thus retained in the Coulomb limit and projected out in the Higgs limit.
The limits have a transparent realization in Figure \ref{fig:charge_space}. Higgs-branch operators obey
\begin{align}
	J=Q_2=Q_4=0 \, ,
	\qquad
	Q_1+Q_3=Q \, ,
\end{align}
and populate the purple edge at $h_2=-Q/2$. Coulomb-branch operators instead obey
\begin{align}
	J=Q_1=Q_3=0 \, ,
	\qquad
	Q_2+Q_4=Q \, ,
\end{align}
and populate the yellow edge at $h_2=+Q/2$. Since the (2,2) index only retains $h=-2h_2$ and not $h_1$ and $h_3$, the purple and yellow lines project to the endpoints of the blue line at $h=+Q$ and $h=-Q$, respectively.

We close with the large-charge behavior of the endpoint contributions, given by $d_{\text{C}}(Q)$ and $d_{\text{H}}(Q)$, defined in \eqref{eq:endpointlimit}. The Coulomb and Higgs limit reduce the ABJM and ADHM indices to Hilbert series \cite{Hayashi:2022ldo,Hwang:2025hfs}, whose pole orders determine the polynomial growth of these degeneracies at fixed $N$. A pole of order $D$ yields polynomial growth of degree $D-1$ in the charge. The Coulomb branch of both theories and the ABJM Higgs branch have complex dimension $2N$, while the ADHM Higgs branch has dimension $2N N_f$. Consequently,
\begin{align} \label{eq:large_hilbert_series}
\text{ABJM:}&&	d_{\text{C}}(Q) &\sim  Q^{2N-1}\,,  & d_{\text{H}}(Q)&\sim Q^{2N-1}\,,&&
\nonumber\\
\text{ADHM:}&&	d_{\text{C}}(Q) &\sim  Q^{2N-1}\,,  & d_{\text{H}}(Q)&\sim Q^{2N N_f-1}\,.&&
\end{align}
These expressions hold at fixed $N$ and fixed $N_f$ or $k$, for charges with nonzero degeneracy. In ABJM, $k$ affects the normalization and allowed charges, but not the leading order behavior.

\subsection{Flux and vorticity sectors} \label{sec:flux-vortex}

In this section we discuss flux and vorticity sectors. We start with ADHM theory and turn to ABJM at the end. The index forms in \eqref{eq:ADHM_index} and \eqref{eq:factorized_schematic} involve infinite sums over flux sectors and vorticities, respectively. For practical computations these sums have to be truncated. 

We start the discussion with the form in \eqref{eq:ADHM_index}, which can be organized as follows,
\begin{align}
	\mathcal I(x,\xi_\ell,\xi_F,\xi_T)
	&=
	\sum_{\mathfrak m\in\mathbb Z^N}
	\xi_T^{T(\mathfrak m)}
	\mathcal I_{\mathfrak m}(x,\xi_\ell,\xi_F) \, , 
	&
	T(\mathfrak m)&=\sum_{i=1}^N\mathfrak m_i\,,
	\label{eq:ADHM_flux_decomposition}
\end{align}
so that $\mathcal I_{\mathfrak m}$ denotes the contribution of the flux sector $\mathfrak m$, which is independent of $\xi_T$. We now define the truncated index
\begin{align}
	\mathcal I_m^{(p)}(x,\xi_\ell,\xi_F,\xi_T)
	&=
	\sum_{\mathfrak m\in\mathbb Z^N, |\mathfrak{m}|\leq p }
	\xi_T^{T(\mathfrak m)}
	\mathcal I_{\mathfrak m}(x,\xi_\ell,\xi_F) \, , 
	&
	|\mathfrak m|&=\sum_{i=1}^N|\mathfrak m_i| \, ,
	\label{eq:ADHM_flux_decomposition_trunc}
\end{align}
where the subscript $m$ indicates that this truncation is tied to the matrix model form \eqref{eq:ADHM_index}.
To capture terms up to order $x^{Q_{\rm max}}$ accurately, it is sufficient to take
\begin{align}\label{eq:p-Nf-rel}
	p\geq \frac{Q_{\rm max}}{N_f} \, .
\end{align}
This follows directly from \eqref{eq:ADHM_index}, where the bare monopole factor in the last line contributes $x^{N_f\sum_{i=1}^N|\mathfrak m_i|}$ while all letters and other terms contribute non-negative powers of $x$. So contributions with $|\mathfrak m|>Q_{\rm max}/N_f$ involve powers of $x$ greater than $Q_{\rm max}$.

We note that increasing $N_f$ increases the reach of a truncation with fixed $p$.
Terms beyond $x^{pN_f}$ are not captured accurately by $\mathcal I_m^{(p)}$. In particular, a purely perturbative index, which may be defined as the $p=0$ truncation, is not generally reliable. It does capture the Higgs limit of the (2,2) specialization: For fixed charges $(Q,h)$, with $h=Q$ corresponding to the Higgs limit, only flux sectors satisfying
\begin{align} \label{eq:ADHM_bare_monopole_contributions}
	\sum_i|\mathfrak m_i|\leq \frac{Q-h}{2N_f}
\end{align}
contribute.
This follows by combining the bare-monopole factor in \eqref{eq:ADHM_bare_monopole_pairwise}, where $c$ is the fugacity for $(Q-h)/2$, with all other letters, which only contribute non-negative powers of $c$. We therefore expect even drastic truncations to remain accurate near the Higgs edge.

We now turn to the factorized form \eqref{eq:factorized_schematic} of the ADHM index. With $k_a$ and $\bar{k}_\mathrm{a}$ denoting the vorticities of the vortex and anti-vortex partition functions, respectively, we have 
\begin{align}
	\mathcal I(x,\xi_\ell,\xi_F,\xi_T)
	&=
	\sum_{ K, \bar{K} \geq 0}
	\xi_T^{K-\bar{K}}
	\mathcal I_{K, \bar{K}}(x,\xi_\ell,\xi_F) \, , 
	& 
	K &=\sum_{\mathrm{a} \in \mathcal{Y}} k_\mathrm{a} \, .
	\label{eq:ADHM_vortex_decomposition}
\end{align}
where $\mathcal I_{K, \bar{K}}$ is independent of $\xi_T$. A truncated version is 
\begin{align}
	\mathcal I_v^{(p)}(x,\xi_\ell,\xi_F,\xi_T)
	&=
	\sum_{K, \bar{K} \geq 0,\,K+\bar{K}\leq p}
	\xi_T^{K-\bar{K}}
	\mathcal I_{K, \bar{K}}(x,\xi_\ell,\xi_F) \, , 
\end{align}
where the subscript $v$ indicates that this truncation is tied to the vorticity sum in the factorized form \eqref{eq:factorized_schematic}.
To obtain exact results up to order $x^{Q_\text{max}}$ terms, it suffices to take
\begin{equation}\label{eq:p-v-rel}
	p\geq Q_\text{max} \, . 
\end{equation}
Some individual ingredients in \eqref{eq:factorized_schematic} involve negative powers of $x$, but rewriting the negative-length Pochhammer symbols using $(a;q)_{-k}=(-q/a)^k q^{k(k-1)/2}/(q/a;q)_k$ and collecting the powers again provides a bound on $p$.
One may define a notion of perturbative index tied to the factorized form as the $p=0$ truncation
\begin{align}
	\mathcal I_v^{(0)}(x,\xi_\ell,\xi_F) = \sum_{|\mathcal{Y}|=N} Z_{\text {pert }}^{\mathcal{Y}}(x,\xi_\ell,\xi_F) \, .
\end{align}

Neither of the perturbative indices $\mathcal I_m^{(0)}$, $\mathcal I_v^{(0)}$ should be confused with the index counting topologically neutral states, which in the form \eqref{eq:ADHM_index} is
\begin{align}
	\mathcal I^{\mathrm{n}}(x,\xi_\ell,\xi_F)
	=
	\sum_{T(\mathfrak m)=0}\mathcal I_{\mathfrak m}(x,\xi_\ell,\xi_F) \, .
	\label{eq:ADHM_semiperturbative_index}
\end{align}
In the factorized form \eqref{eq:factorized_schematic} the same index is
\begin{align}
	\mathcal I^{\mathrm{n}}(x,\xi_\ell,\xi_F)
	=
	\sum_{K=0}^\infty\mathcal I_{K,K}(x,\xi_\ell,\xi_F) \, . 
	\label{eq:ADHM_semiperturbative_index2}
\end{align}
Both expressions count states which are neutral under the topological $\mathrm{U}(1)$ global symmetry by retaining the relevant contributions with non-zero fluxes or vorticities.
 
\paragraph{Comments on ABJM:} The flux decomposition for ABJM is analogous to that of ADHM, with the flux sectors labeled by $(\mathfrak m,\widetilde{\mathfrak m})$ and gauge invariance requiring $\sum_i\mathfrak m_i=\sum_i\widetilde{\mathfrak m}_i$. For ABJM, we can similarly truncate the flux sum to sectors satisfying $\frac{1}{2} \sum_i(|\mathfrak m_i|+|\widetilde{\mathfrak m}_i|) \leq p$. To capture terms up to order $x^{Q{\rm max}}$ accurately, it suffices to take 
\begin{align}\label{eq:p-k-rel}
	p\geq \frac{Q_{\rm max}}{k} \, .
\end{align}
For ABJM in the (2,2) specialization, $\xi_A=\xi_B=y$ and $\xi_T=1$ in \eqref{eq:phys_fug}. The single-letter indices $f^\pm$ depend on $y$ only through $y+y^{-1}$, and the index retains the $y \rightarrow y^{-1}$ reflection symmetry in the truncations for any $p$. This is in contrast with the ADHM index, where the symmetry under $y\rightarrow y^{-1}$, when present, emerges due to non-perturbative contributions beyond the states visible in the UV Lagrangian description.
This also means that there is no analog of the bound \eqref{eq:ADHM_bare_monopole_contributions} for the Higgs limit. Restricting to $p=0$ provides another notion of a purely perturbative index, which for $k=1$ is distinct from $I_m^{(0)}$ and $I_v^{(0)}$ defined in ADHM.

\section{\texorpdfstring{AdS$_4$}{AdS4} black holes, dressing, and entropy}\label{sec:entropy-functions}

In this section we review AdS$_4$ black holes and the phase diagram of supersymmetric configurations for black holes dressed by additional degrees of freedom which carry part of the charge, e.g.\ dual giants \cite{Choi:2024xnv,Choi:2025lck,Jones:2025gno}. The relevant gravity theory dual to ABJM theory is the 4d STU gauged supergravity with $\mathrm{U}(1)^4$ gauge group, where $\mathrm{U}(1)^4$ corresponds to the Cartan of the $\mathrm{SO}(8)$ R-symmetry for $k=1,2$ and to the Cartan of the $\mathrm{SO}(6)$ R-symmetry and a $\mathrm{U}(1)$ flavor symmetry for $k>2$ \cite{Aharony:2008ug}. The solutions can be uplifted to AdS$_4\times S^7/\mathbb{Z}_k$ in M-theory \cite{Cvetic:1999xp}. For ADHM theory with $N_f>1$, the R-symmetry is $\mathrm{SU}(2)\times \mathrm{SU}(2)$ and the relevant theory is a $\mathrm{U}(1)^2$ gauged supergravity which is a subsector of the $\mathrm{U}(1)^4$ theory.

We start from the rotating, electrically charged AdS$_4$ black holes of gauged supergravity, first obtained for $\mathrm{U}(1)^2$ as supersymmetric limits of the non-extremal solutions \cite{Cvetic:2005zi}. The more general $\mathrm{U}(1)^4$ solutions were found in \cite{Hristov:2019mqp}, and their entropy is captured by the entropy functional equations in \cite{Choi:2018fdc}. In a normalization with $\ell=1$ and $Q_I$ denoting the four electric charges, the entropy $S$ and angular momentum $J$ obey 
\begin{subequations} \label{eq:general_four_charge_entropy}
	\begin{align} 
		\frac{1}{16 \pi^4} S^4
		-\frac{\sum_{I<J} Q_I Q_J}{4 \pi^2} S^2
		+Q_1 Q_2 Q_3 Q_4
		&=
		\frac{S^2}{16 \pi^2 G^2_N}
		-\frac{ J^2}{4G^2_N} \, ,
		\label{eq:general_four_charge_entropy_1}
		\\
		-\frac{\sum_I Q_I}{8 \pi^3} S^3
		+\frac{\sum_{I<J<K} Q_I Q_J Q_K}{2\pi}S
		&=
		\frac{ JS}{4 \pi G^2_N} \, .
		\label{eq:general_four_charge_entropy_2}
	\end{align}    
\end{subequations}
In the next two subsections we discuss the two subclasses of these black holes which will be relevant for the index comparisons, and their entropy functions with dressing contributions. As in \cite{Murthy:2020scj,Agarwal:2020zwm,Kim:2023sig,Deddo:2025jrg} we will assume that index specializations with equal chemical potentials are dominated by bulk configurations with the corresponding charges identified.

Our discussion of the phase diagram for composite configurations builds on \cite{Jones:2025gno}, and proceeds at the level of entropy functions without constructing gravity solutions. 
We will not assume that the backreaction of the dressing can be neglected. However, for practical reasons we will assume, without first-principles justification, that the effect of the backreaction on the entropy functions can be fully captured as a change in the effective 4d Newton constant.

\subsection{$(2,2)$ two-charge black holes}
\label{sec:pairwise_equal_slice}

The 4-charge black holes can be specialized to the $\mathrm{U}(1)^2$ gauged supergravity which describes a subsector of 4d $\mathcal N=4$ gauged supergravity with $\mathrm{SU}(2)\times \mathrm{SU}(2)$ gauge group. This is the gravity theory relevant for 3d $\mathcal N=4$ SCFTs, and the results can be applied to theories with no supersymmetry enhancement, such as ADHM with $N_f>1$, and to theories with enhanced supersymmetry.
The solutions are obtained by setting the charges pairwise equal \cite{Cvetic:2005zi},
\begin{align}
	Q_1&=Q_3\,,
	&
	Q_2&=Q_4 \, .
	\label{eq:pairwise_charge_identification}
\end{align}
Supersymmetry imposes, as in \eqref{eq:BPS condition},
\begin{align}
	 E = J+Q_1+Q_2 \, .
	\label{eq:bps_bound_pairwise}
\end{align}
Together with the charge constraint implied by \eqref{eq:general_four_charge_entropy}, this reduces the parameter space for supersymmetric black holes to a two-dimensional sheet in $(J,Q_1,Q_2)$, with $J,Q_1,Q_2 \geq 0$ as noted in Sec.~\ref{sec:tetrahedron}. 
The entropy $S_{\rm BH}$ and angular momentum $J_{\rm BH}$ on the black hole sheet are
\begin{align}
	S_{\text{BH}}(G_N;Q_1,Q_2)
	&=
	\frac{\pi}{2 G_{\text{N}}}
	\left(
	\sqrt{1+16G_{\text{N}}^2 Q_1 Q_2}
	-1
	\right) \, ,
	\label{eq:S_pairwise}
\\
	J_{\text{BH}}(G_N;Q_1,Q_2)
	&=
	\frac{1}{2}\,( Q_1+ Q_2)
	\left(
	\sqrt{1+16G_{\text{N}}^2 Q_1 Q_2}
	-1
	\right) \, .
	\label{eq:J_pairwise}
\end{align}
On the black hole sheet, the entropy vanishes precisely when at least one of the bounds $J,Q_1,Q_2\geq0$ is saturated.

The charges $Q_I$ map to the identically named charges introduced in Secs.~\ref{sec:ABJM_index} and \ref{sec:ADHM_index}. Two-charge black holes are then naturally associated with the (2,2) index \eqref{eq:(2,2) index}.  To facilitate comparisons with index results in Sec.~\ref{sec:results} we introduce $Q=4J+\sum_I Q_I$ and $Q_1-Q_2+Q_3-Q_4$ as independent charges, which the $(2,2)$ index resolves through the $x$ and $y$ fugacities, respectively.
With the identification \eqref{eq:pairwise_charge_identification} these become
\begin{align} \label{eq:pairwise_index_variables}
	Q&= 4J + 2 \left( Q_1+ Q_2 \right) & h&=2 \left( Q_1- Q_2 \right) \, ,
\end{align}
which match the arguments of the index degeneracies $d(Q,h)$ introduced in Sec.~\ref{sec:tetrahedron}. Within this slice, the index leaves one independent charge combination unresolved, which we take as the average
\begin{align}\label{eq:pairwise_Qbar}
	\bar Q&=\frac{Q_1+Q_2}{2} \, .
\end{align}
With the charge identification \eqref{eq:pairwise_charge_identification}, the $(Q,h)$-index locus\eqref{eq:pairwise_index_line} reduces to a $(Q,h)$-index line
\begin{align}\label{eq:pairwise_index_line_id}
2(Q_1-Q_2)&=h \,, & Q_1,Q_2&\geq0 \,, & Q_1+Q_2&\leq \frac{Q}{2} \,.
\end{align}
Along it, $\bar Q$ and $J$ vary at fixed $Q=4J+4\bar Q$ and $h$. States on \eqref{eq:pairwise_index_line} with $(Q_1,Q_2)\neq (Q_3,Q_4)$ contribute to $d(Q,h)$ but are not described by the two-charge black holes. The black hole sheet picks out one point on each index line, where $Q_{\text{BH}}=4J_{\text{BH}}+4\bar Q$ with $J_{\text{BH}}$ given in \eqref{eq:J_pairwise}.
\begin{figure}
	\centering
	\subfigure[][]{\label{fig:3D_black_hole_sheet}
	\includegraphics[width=0.42\linewidth]{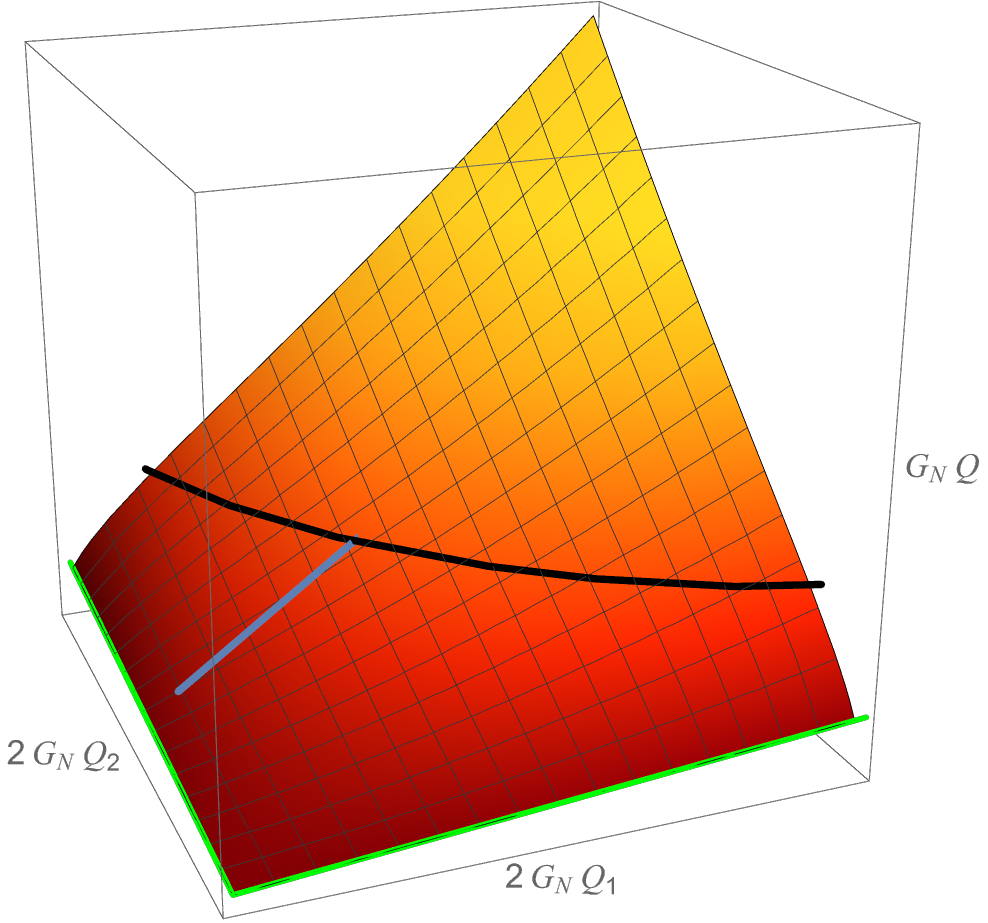}
	}
	\subfigure[][]{\label{fig:phasediagram1}
		\includegraphics[width=0.42\linewidth]{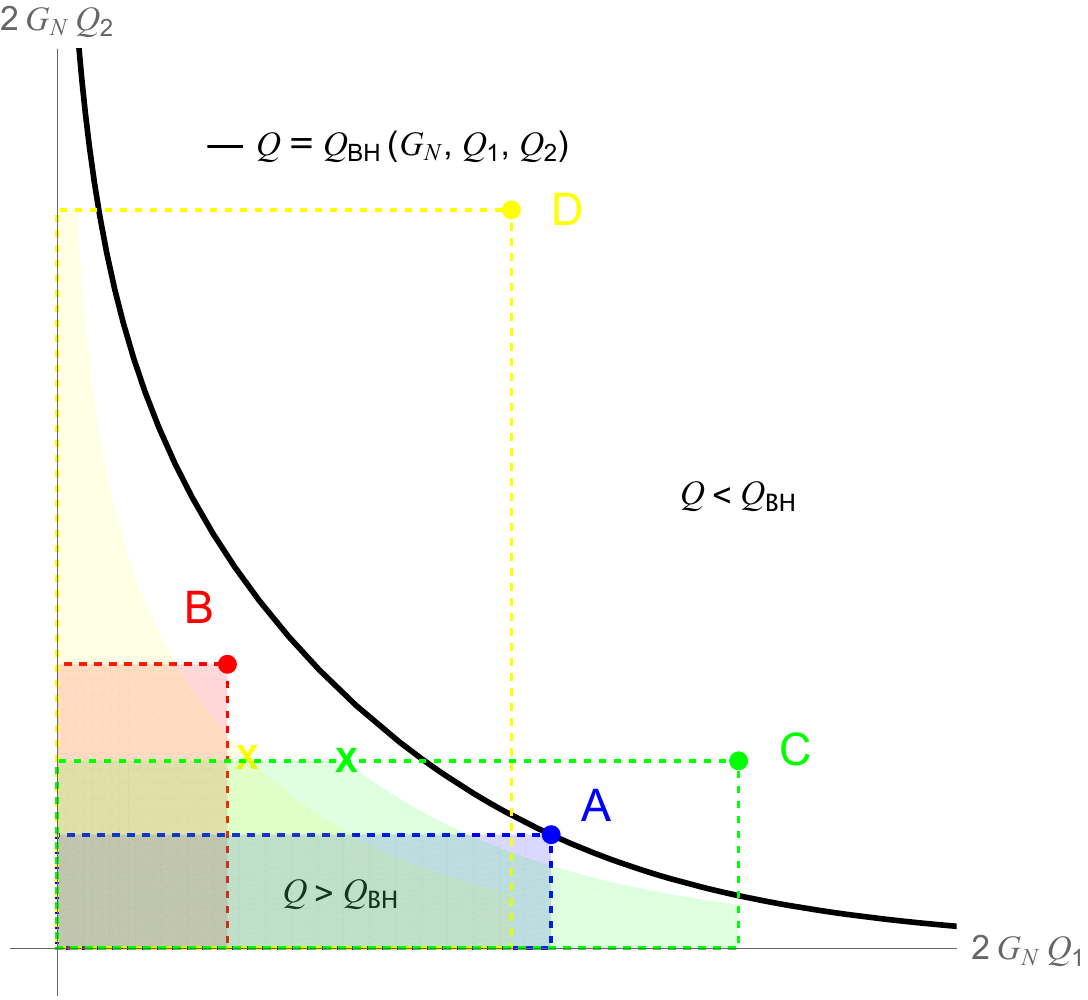}
	}
	\caption{Left: The $(2,2)$ two-charge black hole sheet $Q=Q_{\text{BH}}(G_N;Q_1,Q_2)$, with representative index line at fixed $Q$ and fixed $h=2(Q_1-Q_2)$. At the green lines $J=0$.
	Right: Constant-$Q$ section of Fig.~\ref{fig:3D_black_hole_sheet}. Point $A$ has $Q=Q_{\text{BH}}(G_N;Q_1,Q_2)$. Point $B$ has $Q>Q_{\text{BH}}(G_N;Q_1,Q_2)$ and the dressing only carries angular momentum. Points $C$ and $D$ have $Q<Q_{\text{BH}}(G_N;Q_1,Q_2)$ and require the dressing to carry part of the total electric charge.}
\end{figure}
In terms of $Q$, $h$,
\begin{align}
	S_{\text{BH}}(G_N;Q,h)
	=\frac{\pi}{2G_N}\left(\frac{Q}{4\bar Q_{\rm BH}}-1\right) \, ,
	\label{eq:resentropy_pairwise}
\end{align}
with the averaged charge $\bar Q_{\rm BH}$ fixed by the angular momentum constraint \eqref{eq:J_pairwise} 
\begin{align}
	\bar Q_{\rm BH}(G_N;Q,h)
	=
	\frac{1}{4 \sqrt{2} G_N}
	\sqrt{
		 G^2_N h^2-1
		+
		\sqrt{4 G^2_NQ^2+\left(G^2_N h^2-1\right)^2}
	} \, .
	\label{eq:qbar_charge_constraint}
\end{align}
The entropy vanishes when $|h|=Q$, which corresponds to $J=0$, and is maximal for $h=0$.

\subsubsection{Dressing and entropy maximization}
\label{sec:pairwise_phase_diagram}
We now turn to composite configurations where the total charges $(J,Q_1,Q_2)$ are carried by two components: a core BPS black hole and additional BPS matter outside the black hole. We refer to the composite configurations as dressed black holes (DBH) and to the additional matter as dressing.
The entropy of the dressing is assumed to be subleading, so that the total entropy is obtained by maximizing the entropy of the core black hole over the available charges.
We further assume that the remaining charge-reduced black hole can still be described by the entropy-functions introduced above.

We start with a supersymmetric composite configuration with $(E,J,Q_1,Q_2)$ satisfying the BPS condition \eqref{eq:bps_bound_pairwise} and $J,Q_1,Q_2\geq0$, which follows from $\{\mathcal Q,\mathcal Q^\dagger\}\geq0$. We assume that both components preserve the same supercharge and remain in the $(2,2)$ charge slice. These assumptions could be relaxed, but turn out to be sufficient for the comparison with the index. Distinguishing the black hole and dressing charges by subscripts, the total charge can be decomposed into black hole and dressing contributions as
\begin{align}
	(J,Q_1,Q_2) \ = \ (J_{\rm bh},Q_{1,{\rm bh}},Q_{2,{\rm bh}}) \ + \ (J_{\rm dr},Q_{1,{\rm dr}},Q_{2,{\rm dr}})\,.
\end{align}
We assume that the black hole angular momentum $J_{\rm bh}$ is fixed in terms of the black hole electric charges $Q_{1/2,{\rm bh}}$ by the same relation \eqref{eq:J_pairwise} as in the pure black hole discussion before, but with an effective Newton constant $\tilde G_N$ which may differ from the parameter in the gravitational action. That is, with $J_{\rm BH}$ defined in  \eqref{eq:J_pairwise},
\begin{align}
	J_{\rm bh}&=J_{\rm BH}(\tilde G_N;Q_{1,{\rm bh}},Q_{2,{\rm bh}})\,.
\end{align}
This extends the prescription of \cite{Jones:2025gno} by allowing the effective Newton constant of the core black hole to differ from $G_N$. Here, we leave the nature of the dressing unspecified and treat $\widetilde G_N$ as an independent parameter. For dressings carrying different types of charges we will allow independent $\widetilde G_N$. Positivity of $Q_{1,2},J$ for both components individually implies
\begin{align}\label{eq:purecharges_pairwise}
	0\leq & Q_{1/2,{\rm bh}}\leq Q_{1/2}\,,
	&
	4J_{\rm BH}(\tilde G_N;Q_{1,{\rm bh}},Q_{2,{\rm bh}}) &\leq Q - 2 \left(Q_1 +Q_2\right)\,.
\end{align}
The leading entropy at fixed total charge and for fixed $\tilde G_N$ is then
\begin{align}
	S(\tilde G_N;Q,Q_1,Q_2)
	=
	\underset{Q_{1,{\rm bh}},Q_{2,{\rm bh}}}{\operatorname{max}}\,
	\left[S_{\text{BH}}(\tilde G_N;Q_{1,{\rm bh}},Q_{2,{\rm bh}})\right] \, ,
	\quad
	\text{subject to \eqref{eq:purecharges_pairwise}.}
	\label{eq:fullentropy_pairwise}
\end{align}
Configurations without dressing are pure black holes and are evaluated with $G_N$.

Scenarios for composite configurations are illustrated in Figure~\ref{fig:phasediagram1}, where we fixed $\widetilde G_N=G_N$ for simplicity. The figure shows points labeled $A,B,C,D$ representing different charges $(Q,Q_1,Q_2)$. The rectangles attached to the points indicate the allowed electric charges for the core black hole in composite configurations. The positivity constraint for the angular momentum $J$ further restricts the allowed charges to the shaded regions.
Since the black hole entropy increases with increasing electric charges, the dominant configuration is obtained by maximizing the charges within the allowed region. Point $A$ in Figure \ref{fig:phasediagram1} lies on the black hole sheet and represents an undressed black hole, for which the total charges and black hole charges coincide. For point $B$, $Q_{1/2,{\rm bh}}=Q_{1/2}$ and the dressing only carries excess angular momentum. For points $C$ and $D$ the total electric charges are too large for a pure black hole at the chosen value of $Q$, and part of the electric charge must be assigned to the dressing. In this case, the black hole with maximal entropy leaves no angular momentum to the dressing, $J_{\rm dr}=0$. For general $\widetilde G_N$, the black hole charges are then determined by the saturated angular-momentum constraint $4J_{\rm BH}(\tilde G_N;Q_{1,{\rm bh}},Q_{2,{\rm bh}}) = Q - 2\left(Q_1 +Q_2\right)$.

\begin{figure}
	\centering
	
	\subfigure[][]{\label{fig:phasediagram2}
	\includegraphics[width=0.40\linewidth]{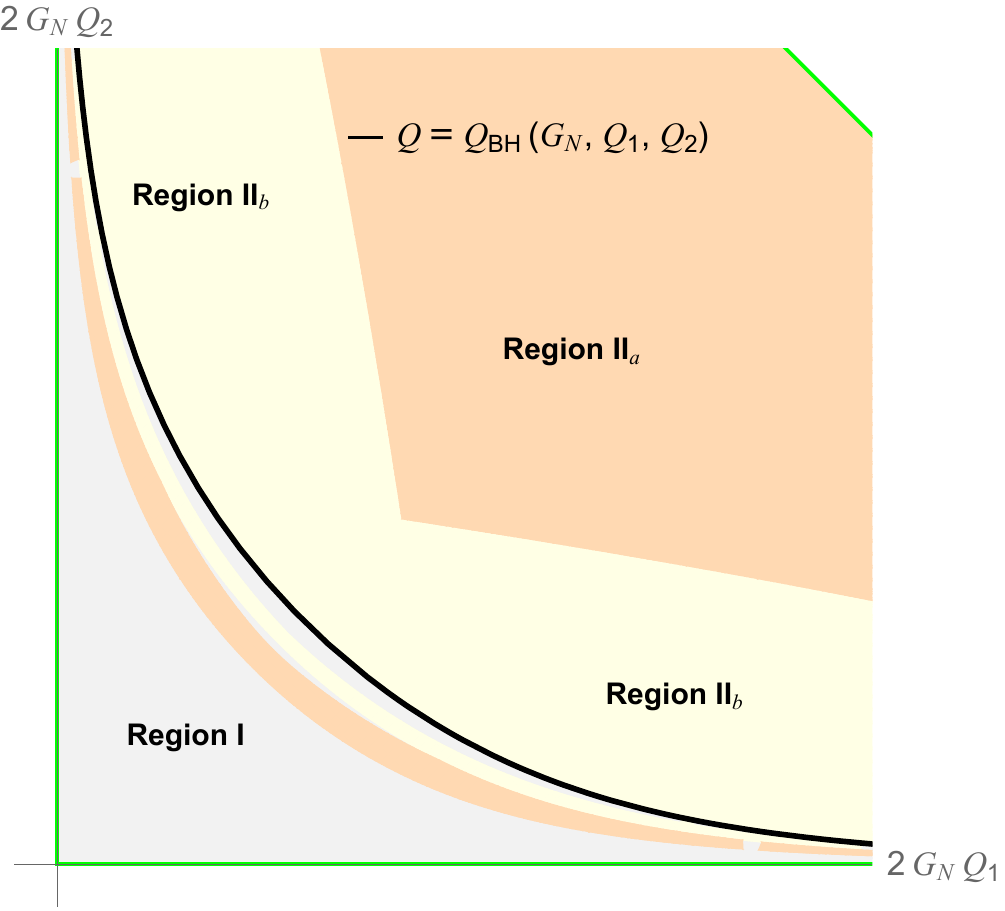}
	}
	\qquad
	\subfigure[][]{\label{fig:phasediagram3}
	\includegraphics[width=0.43\linewidth]{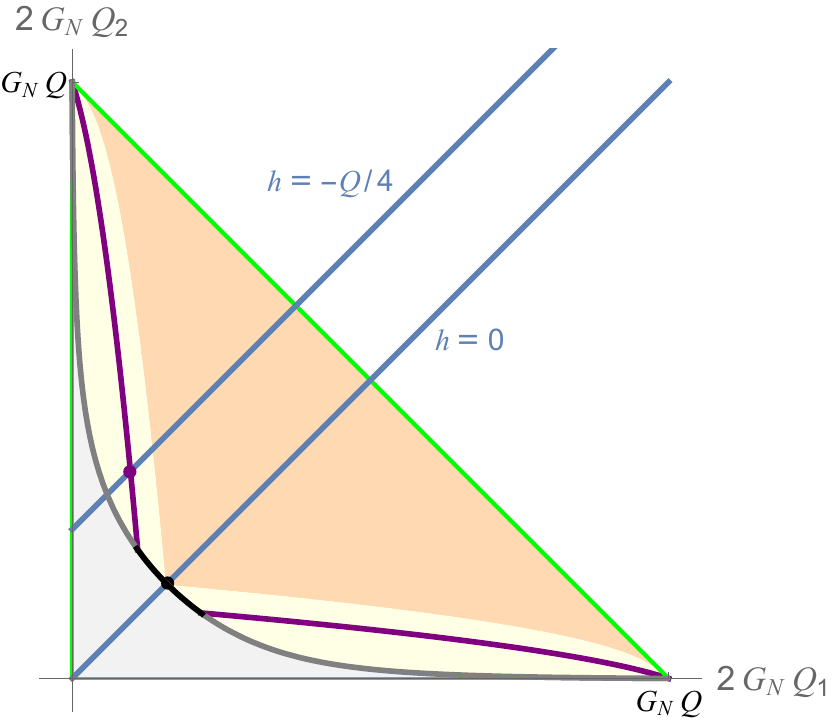}
	}
	
	\caption{Left: Constant-$Q$ phase diagram in the (2,2) slice. The pure black hole curve is shown in black. In region I the dressing carries only angular momentum. In regions II$_b$ and II$_a$ the dressing carries respectively one or both pairwise-equal electric-charge pairs. For the illustration, regions II$_b$ and II$_a$ use $\widetilde G_N/G_N=(15/13)^{3/2}$ and $(15/11)^{3/2}$, respectively. Right: Phase diagram with all $\widetilde G_N = G_N$.  The $(Q,0)$-index and $(Q,-Q/4)$-index lines are represented by blue lines. The black and purple curves are the loci of maximal entropy along these lines, corresponding to pure black holes and $II_b$ dressed black holes, respectively.}
\end{figure}

When $\widetilde G_N\neq G_N$ the entropy extremization becomes slightly more involved. An example phase diagram is shown in Figure \ref{fig:phasediagram2}. In region $I$ the dressing carries only angular momentum. On the black hole curve there is no dressing. Region II$_b$ contains configurations for which the dressing carries one of the two electric charges, and region II$_a$ contains configurations for which it carries both.\footnote{A dressing carrying all three charges $J_{\rm dr}$, $Q_{1,\rm dr}$, $Q_{2,\rm dr}$ would leave the angular momentum constraint unsaturated. Such phases can only dominate with favorable $\tilde G_N$. We do not include them here, as this turns out to be sufficient for the comparison with the index.}
For $\tilde G_N>G_N$, region II can extend below the black hole curve, as shown in Figure~\ref{fig:phasediagram2}. A core with larger $\tilde G_N$ carries more angular momentum at the same electric charges, so that its black hole curve lies below the one for $G_N$. At fixed charges its entropy is also larger, and above the $\tilde G_N$ black hole curve a dressing carrying part of the electric charge leaves a core with larger entropy than the region I core.

To connect this discussion to the superconformal index we note that the index does not probe $(Q,Q_1,Q_2)$ separately, but rather counts states on the $(Q,h)$-index locus \eqref{eq:pairwise_index_line}, which reduces to the index line \eqref{eq:pairwise_index_line_id} in the $(2,2)$ slice. At fixed $Q$ these index lines are the straight blue lines in the phase diagram of figure~\ref{fig:phasediagram3}. Including dressed configurations generalizes the pure black hole entropy \eqref{eq:resentropy_pairwise} to the maximal entropy at fixed $(Q,h)$, obtained by maximizing along the remaining charge direction 
\begin{equation}
	S^{\rm grav}_\mathcal{I}(Q,h)
	=
	\underset{r}{\operatorname{max}}\,
	S(\widetilde G_N;Q,r+h/2,r) \, .
	\label{eq:indentr_pairwise}
\end{equation}
Since the entropy is determined by the black hole, the maximum along an index line only depends on which black holes can be dressed so that the total charges lie on the line. In the limiting case of all $\widetilde G_N \rightarrow G_N$, the maximum lies on the black hole sheet for $G_N|h|\leq1$ and agrees with the pure black hole entropy in \eqref{eq:resentropy_pairwise}. Outside this window it moves onto the dressed branch, specifically region II$_b$ in the (2,2) phase diagram, see figure \ref{fig:phasediagram3}. Along the index lines, this branch takes the closed form
\begin{align}\label{eq:S_IIb}
    S_{{\rm II}_b}(\widetilde G_N;Q,h) = \frac{\pi}{2\widetilde G_N}\left(\sqrt{1+\widetilde G_N(Q-|h|)}-1\right) \, ,
\end{align}
which for $\widetilde G_N=G_N$ agrees with the pure black hole entropy at $G_N |h|=1$.  For $G_N Q \gg 1$ at fixed $h/Q$ and $\widetilde G_N/G_N$, it scales as $S_{{\rm II}_b}\sim(Q/\widetilde G_N)^{1/2}$.

Allowing for $\widetilde G_N\neq G_N$ again modifies this picture: the window $G_N|h|\leq1$, and the dominance of region II$_b$ outside it, are no longer guaranteed. In principle, each type of dressing can lead to a different effective Newton constant and thus to an additional parameter. However, the entropy obtained by extremizing along an index line, which will be relevant for Sec.~\ref{sec:results}, is only sensitive to the effective Newton constant in the region where the maximum is attained. Matching it still bounds the effective Newton constants in the other regions.

\begin{figure}[h!]
	\centering
	\subfigure[][]{\label{fig:BPSentropyfiniteN}
    \includegraphics[width=0.47\linewidth]{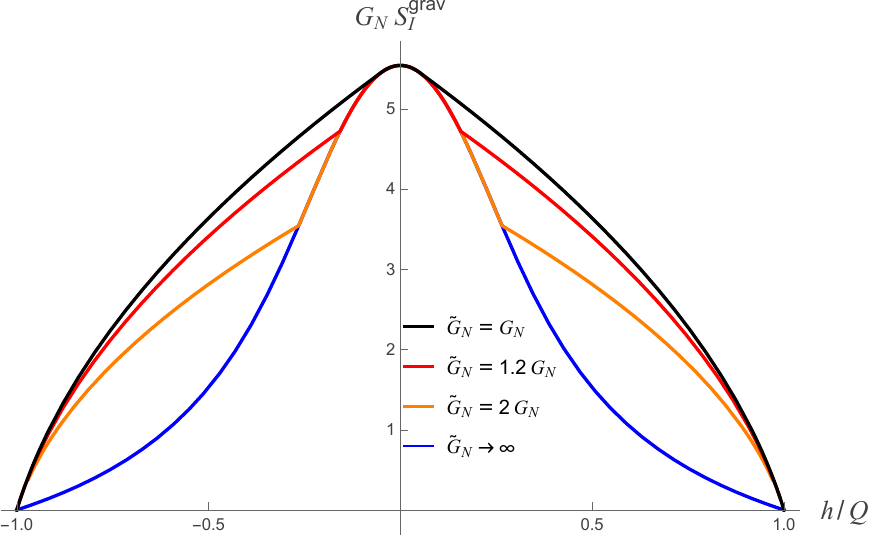}
    }
    \subfigure[][]{\label{fig:BPSentropyfiniteN_trans}
	\includegraphics[width=0.45\textwidth]{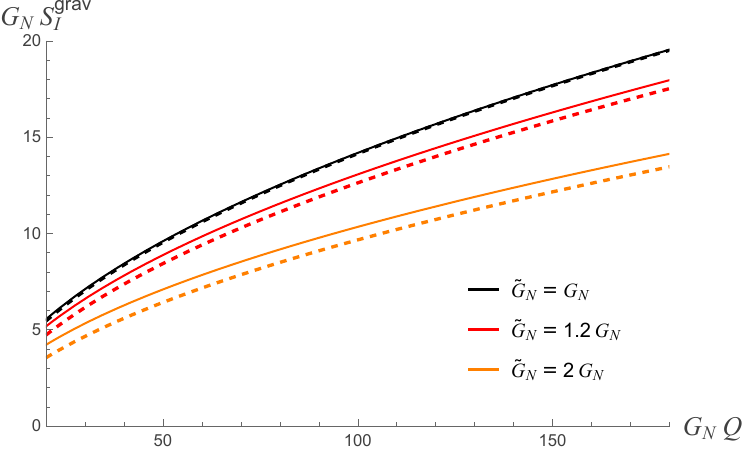}
    }
	\caption{Left: Entropy $G_N S^{\rm grav}_{\mathcal I}(Q,h)$ at fixed $G_N Q=20$, obtained by maximizing along index lines in the (2,2) slice for different $\widetilde G_N$. The central segment is governed by pure black holes and independent of $\widetilde G_N$. The outer segments represent region II$_b$ dressed branches. Right: Values of $G_N S_{\mathcal I}$ at the phase transition between pure black hole and region II$_b$ dressed branches, as a function of $G_N Q$. The solid curves show $G_N S_{\text{BH}}(\widetilde G_N;Q,0)$, whose leading large charge behavior agrees with the entropy at this boundary.}
\end{figure}

Figure \ref{fig:BPSentropyfiniteN} illustrates the entropy obtained from \eqref{eq:indentr_pairwise} for several choices of the effective Newton constants of the dressed branches, including $\widetilde G_N=G_N$.  Increasing $\widetilde G_N$ suppresses the dressed branches and extends the central region governed by pure black holes to larger $|h|$. For $\widetilde G_N\to\infty$ the entropy curve approaches the black hole curve. The transition point from black hole branch to dressed branch is determined by equality of the pure black hole and region II$_b$ entropies \eqref{eq:resentropy_pairwise} and \eqref{eq:S_IIb}, and generally depends on $Q$ as well as $\widetilde G_N/G_N$. Fig.  \ref{fig:BPSentropyfiniteN_trans} shows the entropy at this phase boundary as a function of $G_N Q$. For $G_N Q\gg1$ at fixed $\widetilde G_N/G_N$, \eqref{eq:S_IIb} implies that the transition entropy scales as $(Q/\widetilde G_N)^{1/2}$, as illustrated by the solid curves in Fig.~\ref{fig:BPSentropyfiniteN_trans}.

\subsection{$(1,3)$ two-charge black holes}
\label{sec:three_equal_charges}

The second class of black hole solutions we consider is obtained by setting three of the four charges equal. Since the entropy functions are symmetric under permutations of the $Q_I$, all choices of which charges to identify are equivalent and we set
\begin{align}
	Q_2=Q_3=Q_4\,.
	\label{eq:three_equal_charge_identification}
\end{align}
The BPS condition \eqref{eq:BPS condition} becomes $E = J+\frac{1}{2}(Q_1+3Q_2)$,
with $J,Q_2\geq 0$ and $Q_1 \geq - Q_2$, see Sec.~\ref{sec:tetrahedron}. The charge constraint implied by \eqref{eq:general_four_charge_entropy} again reduces the 3-dimensional space $(J,Q_1,Q_2)$ to a 2-dimensional black hole sheet.
When specializing \eqref{eq:general_four_charge_entropy} to \eqref{eq:three_equal_charge_identification}, one can eliminate $J$ and directly express entropy and angular momentum in terms of two independent electric charges,
\begin{align}
	S_{\text{BH}}(G_N;Q_1,Q_2)
	&=
	\frac{\pi}{2G_N}
	\sqrt{
		\frac{
			\mathcal{C}^2
			\left(
			\mathcal{C}-1
			\right)^2
			-
			16G_N^4Q_2^2(Q_1-Q_2)^2
		}{
			1+G_N^2(Q_1+3Q_2)^2
		}
	} \, ,
	\label{eq:threeequal_entropy}
	\\
	J_{\text{BH}}(G_N;Q_1,Q_2)
	&=
	\frac{
		(Q_1+3Q_2)
		\mathcal{C}^2
		\left(
		\mathcal{C}-1
		\right)
		-
		4G_N^2Q_2(Q_1-Q_2)^2
	}{
		4+4G_N^2(Q_1+3Q_2)^2
	} \, ,
	\label{eq:threeequal_J}
\end{align}
where
\begin{align}
    \mathcal{C} = \sqrt{ 1+8G_N^2Q_2(Q_1+Q_2)} \, .
\end{align}
With $Q_2 \geq 0$, the entropy is non-negative only when $Q_1\geq Q_{1,\min}(G_N;Q_2)$, where
\begin{equation}
    Q_{1,\min}(G_N;Q_2)=-\frac{\left(\sqrt{1+12G_N^2Q_2^2}-1\right)^2}{36G_N^2Q_2} \, .
    \label{eq:threeequal_Q1min}
\end{equation}
For $Q_2>0$, this lower bound satisfies $0 >Q_{1,\min}>-Q_2$, so the BPS black hole configurations realize negative $Q_1$ compatible with $Q_1\geq-Q_2$ but not the entire allowed range. On the black hole sheet the entropy vanishes for $Q_2=0$ or for $Q_1 = Q_{1,\min}$. One can check that the black holes satisfy $J_{\rm BH}+Q_1>0$ wherever $S_{\rm BH}>0$, so they lie inside the region \eqref{eq:tetrahedron_J}.

As for the $(2,2)$ case before, we define the charges $Q$ and $h$ appearing in the index \eqref{eq:(1,3) index} associated with the $x$ and $y$ fugacities, respectively, as 
\begin{align}
	Q&=4J+Q_1+3Q_2\,,
	&
	h&=2(Q_1-Q_2)\,.
	\label{eq:threeequal_index_variables}
\end{align}
The tetrahedral index support discussed in section \ref{sec:tetrahedron} restricts the index comparison to
\begin{align}
	-\frac{2Q}{3}\leq h\leq 2Q \, .
\end{align}
Using \eqref{eq:threeequal_index_variables}, this is equivalent to \eqref{eq:tetrahedron_J} with $Q_2=Q_3=Q_4$, i.e.\ $J+ Q_1 \geq 0$ and $J+ Q_2 \geq0$. Since $J,Q_2\geq 0$ already follows from the superalgebra, the additional restriction is $Q_1\geq-J$.

\subsubsection{Dressing and entropy maximization}

We now apply the composite prescription of subsection \ref{sec:pairwise_phase_diagram}. We again require $J\geq0$ and $Q_2\geq0$, which is implied by $\{\mathcal Q,\mathcal Q^\dagger\}\geq0$, for black hole and dressing individually. The constraint on $Q_1$ is $Q_1\geq-Q_2$, but we impose the additional bound $J+Q_1\geq0$. This is obeyed by the core black holes, as noted below \eqref{eq:threeequal_Q1min}. It is also obeyed by known BPS states in AdS$_4$: the graviton index of \cite{Kim:2009wb} has support only in this region, and the spinless dual giant gravitons in  \cite{Bhattacharyya:2007sa,Nishioka:2008ib} satisfy it.
Combined with $Q_{1,{\rm dr}}\geq-Q_{2,{\rm dr}}$, it gives $Q_{1,{\rm dr}}\geq-\min(J_{\rm dr},Q_{2,{\rm dr}})$, so that the core black hole charges are constrained by
\begin{align}
	&Q_{1,\min}(\tilde G_N;Q_{2,{\rm bh}})\leq Q_{1,{\rm bh}}\leq Q_{1}+\min\!\left(J-J_{\text{BH}}(\tilde G_N;Q_{1,{\rm bh}},Q_{2,{\rm bh}}),\,Q_2-Q_{2,{\rm bh}}\right)\,, \nonumber \\
	&0\leq Q_{2,{\rm bh}}\leq Q_{2}\,,
	\quad 
	4J_{\text{BH}}(\tilde G_N;Q_{1,{\rm bh}},Q_{2,{\rm bh}})
	\leq
	Q-(Q_1+3Q_2)\,.
	\label{eq:threeequalpurecharges}
\end{align}
The entropy for given total charge is again obtained by maximizing the core black hole entropy,
\begin{align}
	S(\tilde G_N;Q,Q_1,Q_2)
	=
	\underset{Q_{1,{\rm bh}},Q_{2,{\rm bh}}}{\operatorname{max}}
	\left[
	S_{\text{BH}}(\tilde G_N;Q_{1,{\rm bh}},Q_{2,{\rm bh}})\right] \, ,
	\quad\text{subject to \eqref{eq:threeequalpurecharges}.}
	\label{eq:threeequalfullentropy}
\end{align}

\begin{figure}[h!]
	\centering
	\subfigure[][]{\label{fig:phasediagramthreeequal}
		\includegraphics[width=0.41\linewidth]{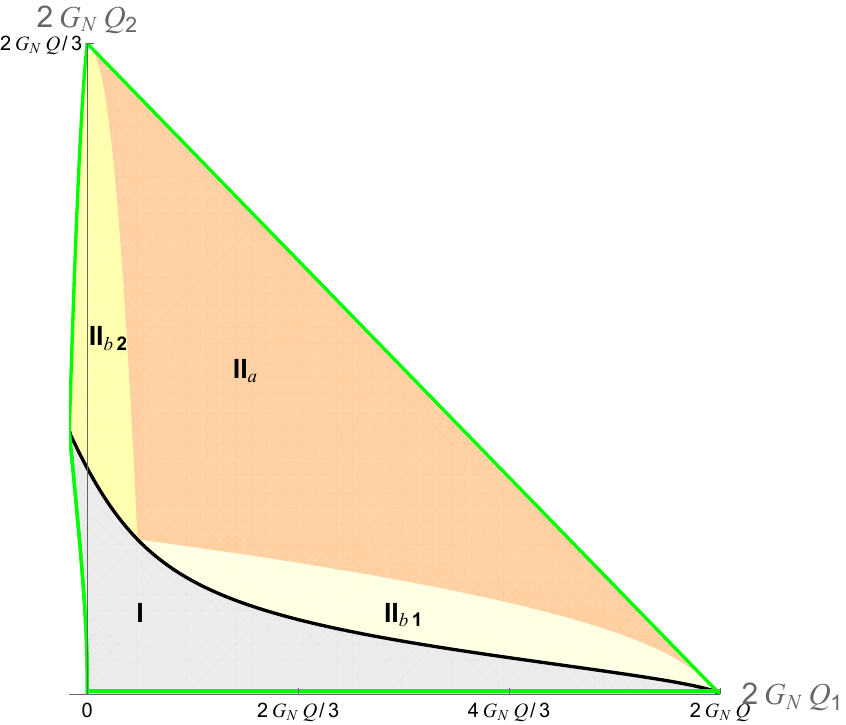}
	}
	\quad
	\subfigure[][]{\label{fig:entropythreeequal}
		\includegraphics[width=0.48\textwidth]{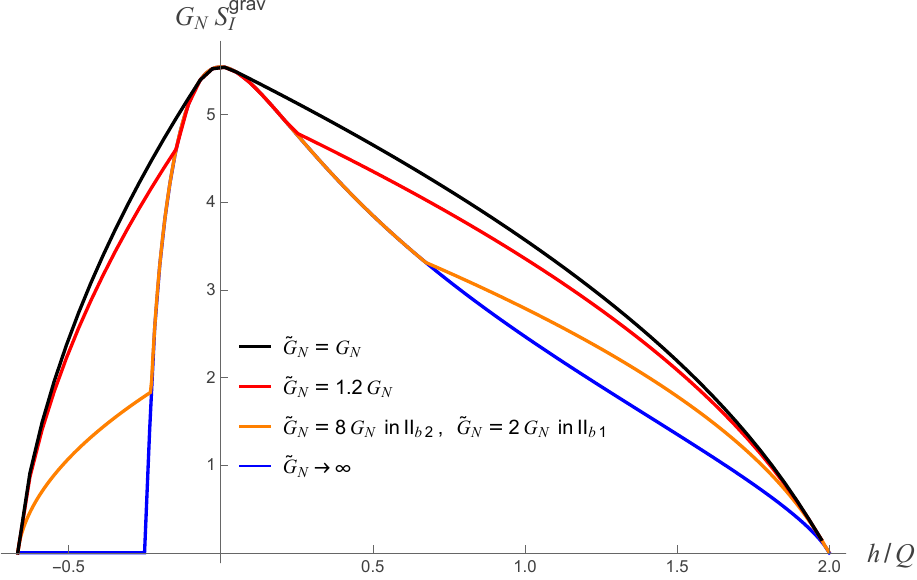}
	}
	\caption{Left: Constant-$Q$ phase diagram in the (1,3) slice.
		The pure black hole curve is shown in black. In region I the dressing carries only angular momentum. In region II$_{b1}$ it carries $Q_1$ charge, in region II$_{b2}$ the three equal
		$Q_2=Q_3=Q_4$ charges, and in region II$_a$ both. 
		Right: Entropy $G_N S^{\rm grav}_{\mathcal I}(Q,h)$ at $G_N Q=20$, obtained by maximizing along index lines in the (1,3) slice. The central segment represents pure black holes, the outer segments represent dressed configurations from region II$_{b2}$ for $h<0$ and II$_{b1}$ for $h>0$. The orange curve uses different effective Newton constants for these two dressed branches. The asymmetry between the two sides reflects that $Q_2$ represents the three equal charges $Q_2=Q_3=Q_4$.  }
\end{figure}

We discuss the phase diagram with the additional assumption that the dressing charges are non-negative; while including all phases is in principle straightforward, we find this sufficient for our purposes.\footnote{%
As before, the dominant core black hole saturates the upper bounds on $Q_{1,{\rm bh}}$ in \eqref{eq:threeequalpurecharges}, i.e.\ $Q_{1,{\rm dr}}=-\min(J_{\rm dr},Q_{2,{\rm dr}})$. Negative $Q_{1,\rm dr}$ therefore forces non-zero $J_{\rm dr}$ and $Q_{2,\rm dr}$, so the dressing carries all three charges.
Unlike in the (2,2) case, such a dressing increases the charge available to the core black hole at fixed total charge, so these phases can dominate without strongly favorable effective Newton constants. This in principle adds 3 regions to Fig.~\ref{fig:phasediagramthreeequal}, which differ in whether the bound is saturated by the first argument, the second argument, or both. This adds 3 effective Newton constants, and the phase diagram in principle depends on all. However, as before we find agreement with the index results without including these phases.}
The resulting phase diagram is shown in Fig.~\ref{fig:phasediagramthreeequal}. The labeling of the regions is analogous to the (2,2) case. In region I the dressing carries only angular momentum. In Region II$_{b1}$ the dressing carries only $Q_1$ charge, in region II$_{b2}$ only the three equal $Q_2=Q_3=Q_4$ charges, and in region II$_a$ it carries both types of electric charge. The charged branches are no longer symmetric due to the asymmetric charge identification.

With an eye towards comparisons to the index, we again define the entropy obtained by maximizing along the remaining charge direction of the index line at fixed $(Q,h)$ in \eqref{eq:threeequal_index_variables},
\begin{align}
	S^{\rm grav}_{\mathcal I}(Q,h)
	=
	\underset{r}{\operatorname{max}}\,
	S(\tilde G_N;Q,r+h/2,r)\,.
	\label{eq:threeequal_indicial_entropy}
\end{align}
Example curves are shown in Figure \ref{fig:entropythreeequal}. The central regime is governed by the pure black hole branch, while the outer parts probe dressed branches. Unlike in the (2,2) case, there is no symmetry under $h\rightarrow -h$ and we allow for independent effective Newton constants in regions II$_{b2}$ and II$_{b1}$. This results in the asymmetric outer branches in Figure \ref{fig:entropythreeequal}.
As commented in the (2,2) case before, while the phase diagram depends on all effective Newton constants, the result of the maximization along the index line is only sensitive to the Newton constant in the region where the maximum is attained. Reproducing this maximum therefore only bounds the other effective Newton constants.

\section{Superconformal index results}\label{sec:results}

In this section we present our main results, which are exact evaluations of the $(2,2)$ and $(1,3)$ indices introduced in Section~\ref{sec:3dCFTs} for ABJM theory and ADHM theory. Details on the implementation can be found in Appendix~\ref{app:computational}. We include all contributions that are required to capture the index exactly up to the claimed charge unless indicated otherwise.\footnote{The approximation used in \cite{Hwang:2025hfs} to push the single-fugacity index to large charge (a cut-off on the vorticity sum which does not retain all contributing terms) does not capture the qualitative behavior of the two-fugacity indices we are interested in, as we show explicitly in appendix \ref{app:pert_vort}.}

To ground the discussion, we start with the single-fugacity index in Subsection~\ref{subsec:singlecharge}, and compare the degeneracies to the entropies of isolated black holes. As mentioned in Sec.~\ref{sec:entropy-functions}, we compare index state counts with fugacity identifications to black holes with the corresponding charge identifications, as in \cite{Murthy:2020scj,Agarwal:2020zwm,Kim:2023sig,Deddo:2025jrg}.
Our results will extend those obtained for the unrefined index \eqref{eq:single_fugacity_index} in \cite{Hwang:2025hfs} to a substantially larger parameter range.
We then turn to the two-fugacity indices. Subsection \ref{sec:(2,2) results} presents results for the (2,2) specialization with $k=1$ / $N_f=1$. We use the ADHM matrix integral to compute the index for $N=2$ and $3$, and also present partial results for $N=4$ obtained from a truncation of the factorized index. Subsection \ref{sec:(2,2) results k>1} extends the analysis to ABJM with $k>1$ and ADHM with $N_f>1$. 
We discuss the (1,3) specialization for  ABJM with $k=1$ and $k>1$ in Subsection \ref{sec:(1,3) results}.

Throughout this section we emphasize the index results, and defer a more detailed discussion of the connection with black hole physics to Section \ref{sec:comparison}.

\subsection{Single-charge index and black hole entropy}\label{subsec:singlecharge}

The large-$N$ limit of the 3d index was previously matched to the black hole entropy in \cite{Choi:2019zpz,Nian:2019pxj}. More recently, \cite{Hwang:2025hfs} computed the single-charge ADHM index at finite-$N$, with exact results for $N=2,3$ and approximate results for $N=4,5$ up to $Q=100$. As in previous finite-$N$ studies in ${\cal N}=4$ SYM \cite{Agarwal:2020zwm,Murthy:2020scj}, close agreement with the black hole entropy was observed.
As a consistency check for our evaluation, we first (re)compute the single-charge ADHM index. 

As discussed in Sec.~\ref{sec:3dCFTs}, $\mathrm{U}(N)$ ADHM theory with $N_f=1$ possesses a decoupled sector which is equivalent to $\mathrm{U}(1)$ ADHM theory with $N_f=1$. In the classification of monotone/fortuitous states \cite{Chang:2024zqi}, this free sector consists of monotone operators. Since our focus is on black hole microstates, we remove these states from the counting by dividing out the decoupled sector from the index, and plot $\mathcal{I}_N/\mathcal{I}_1$ in all figures for $N_f=1$. 

\begin{figure}[h]
    \centering

    \begin{minipage}{0.48\linewidth}
        \centering
        \includegraphics[width=\linewidth]{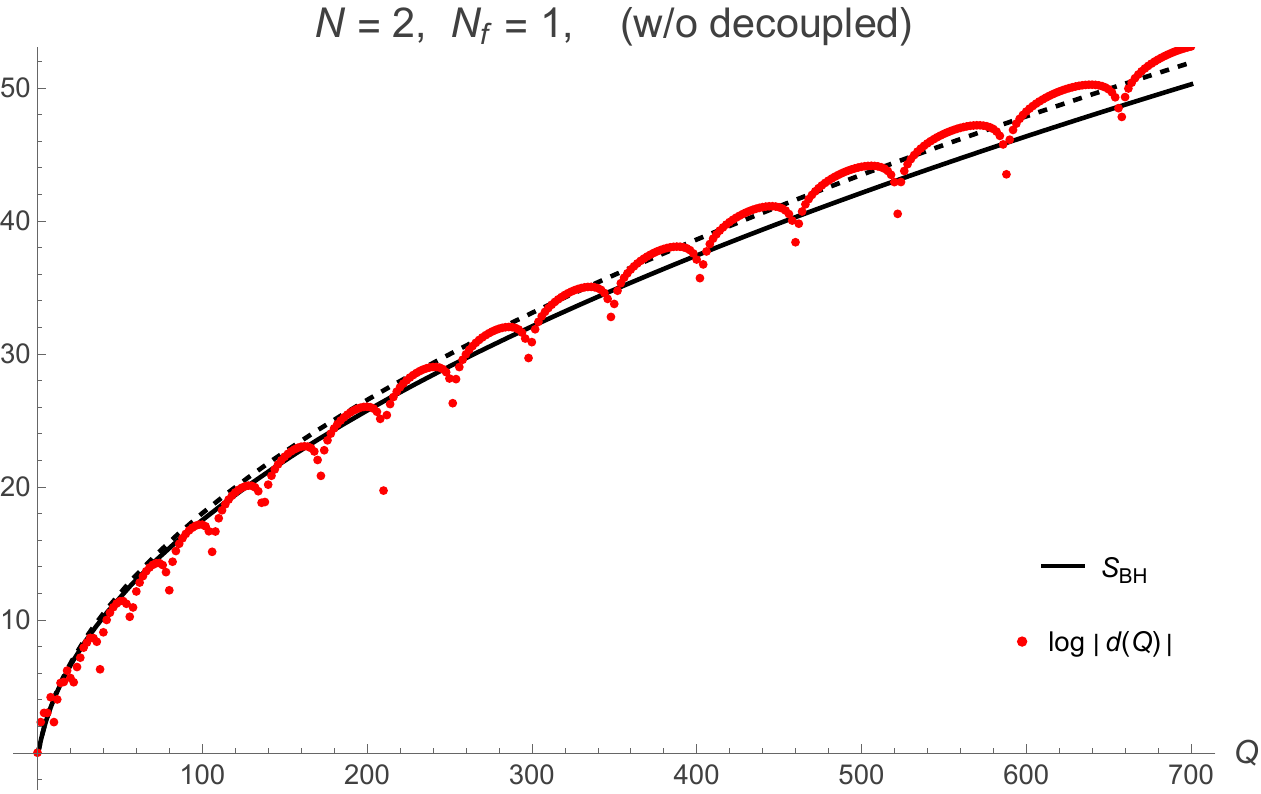}
    \end{minipage}
    \hfill
    \begin{minipage}{0.48\linewidth}
        \centering
        \includegraphics[width=\linewidth]{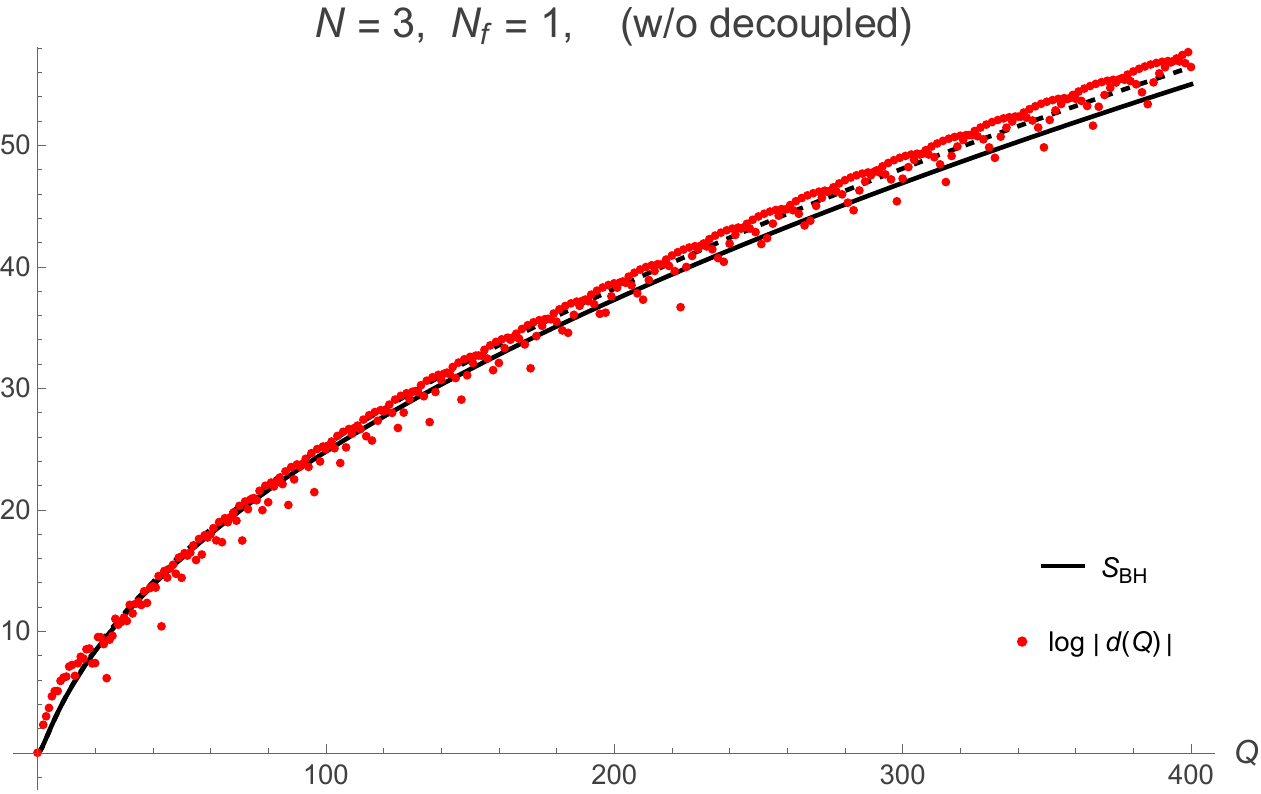}
    \end{minipage}

    \caption{The single-fugacity index with $N=2$ and $N=3$. Two black hole curves are shown, using the finite-$N$ free energy (solid) and the large-$N$ free energy (dashed).}
    \label{fig:ADHM_single_fug}
\end{figure}

We obtained explicit degeneracies up to $Q=700$ and $Q=400$ for $N=2$ and $N=3$, respectively.
Results up to $Q=100$ for $N=2$ and $N=3$ were previously given in \cite{Hwang:2025hfs}, and for this range we find exact agreement.
From these degeneracies we define the indicial entropy
\begin{align}
	S_{\mathcal I}(Q) = \log|d(Q)| \,,
\end{align}
and Fig.~\ref{fig:ADHM_single_fug} shows these entropies as red dots for $N=2$ and $N=3$.
The oscillatory modulations of the underlying slowly-varying behavior reflect transitions in the signs of the degeneracies $d(Q)$.
For comparison, two black hole curves with different effective 4d Newton constant $G_N$ are shown as well. They are obtained by relating the 4d Newton constant to the SCFT free energy $F$ using the large-$N$ or semi-classical relation
\begin{align}
	\frac{\pi}{2G_{\text{N}}} &=F \, ,
	\label{eq:Newton}
\end{align}
where the AdS radius has been set to one.
The dashed curve is computed with the large-$N$ approximation to the free energy \eqref{eq:large-N F}, evaluated for finite $N$, as was done in \cite{Hwang:2025hfs}. The solid curve instead uses the exact finite-$N$ free energy \eqref{eq:ZS3_ADHM_exact}. After the contributions of decoupled sectors are removed, both curves match the index reasonably well. For the following plots we use the finite-$N$ free energy, noting that for the values of $N$ considered here, the leading large-$N$ result can differ substantially from the exact finite-$N$ result.

\subsection{(2,2) specialization $k=1$ ABJM / $N_f=1$ ADHM} \label{sec:(2,2) results}

\begin{figure}[h]
	\centering
	
	\begin{minipage}{0.48\linewidth}
		\centering
		\includegraphics[width=\linewidth]{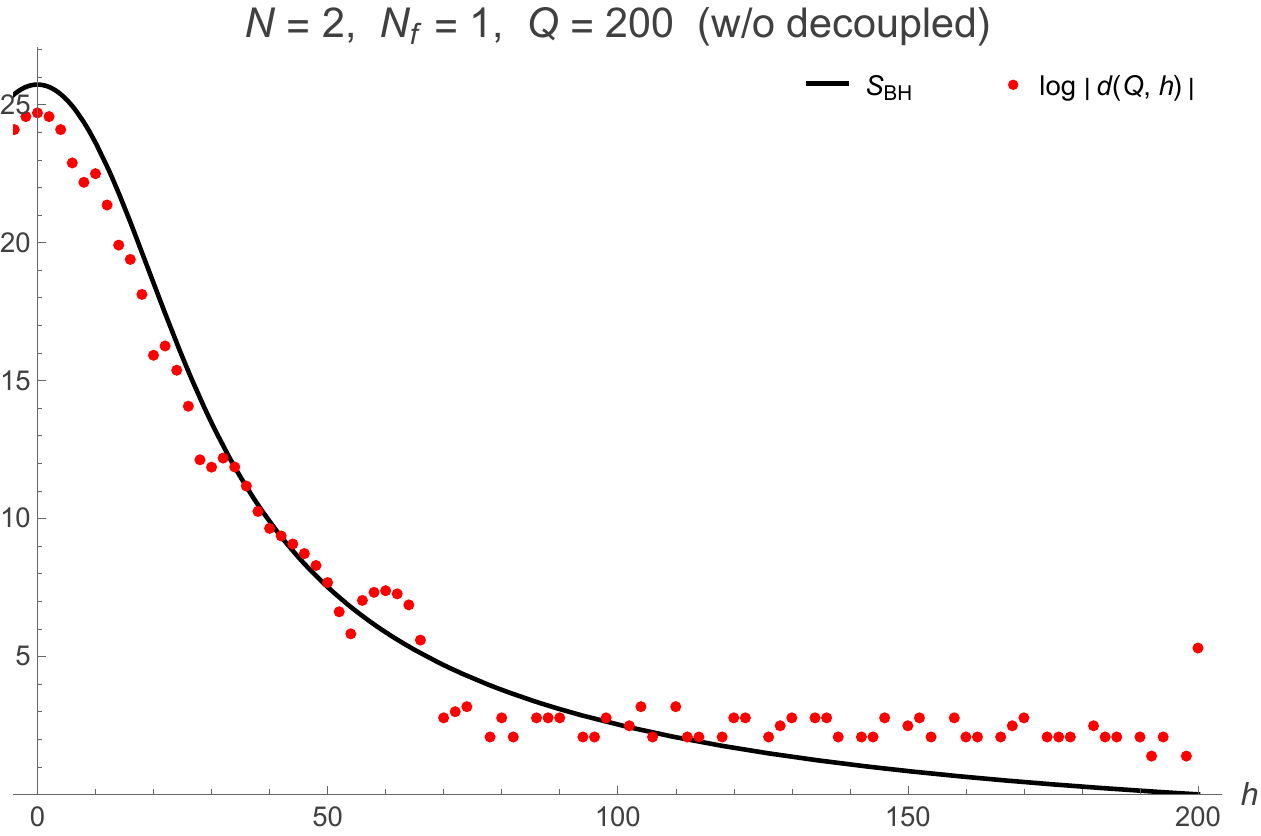}
	\end{minipage}
	\hfill
	\begin{minipage}{0.48\linewidth}
		\centering
		\includegraphics[width=\linewidth]{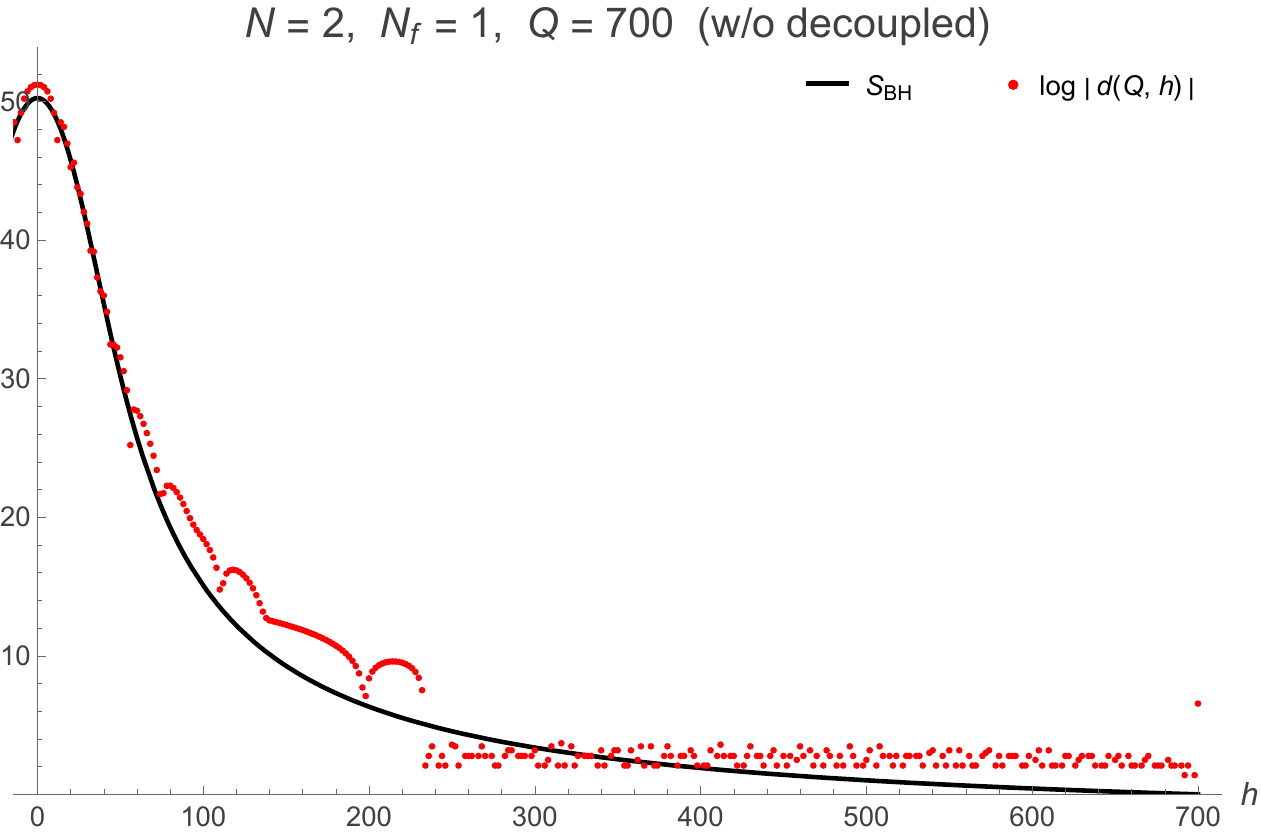}
	\end{minipage}
	
	\caption{Indicial entropies for $N=2$ with $Q=200$ and $700$. The curves show a central peak, which is well approximated by the entropy of isolated black holes, and a plateau at large $h$ which decreases relative to the central peak as $Q$ is increased.}
	\label{fig:ADHM_N2_y2}
\end{figure}

We now turn to two-fugacity indices, starting with the (2,2) specialization defined in Section \ref{sec:ABJM-index-slices}.
We define the associated refined indicial entropy as
\begin{align} \label{eq:refined_indicial_entropy}
	S_{\mathcal I}(Q,h) = \log|d(Q,h)| \, .
\end{align}
Figure \ref{fig:ADHM_N2_y2} shows the indicial entropies for $N=2$, with the decoupled sector removed, as function of $h$ for two fixed values of $Q$. The degeneracies for fixed $Q$ can be obtained exactly from the truncation discussed in Section \ref{sec:flux-vortex}.
The index has nonzero degeneracy in the regime $|h|\leq Q$ and is symmetric under $h\to-h$, so we restrict the plots to $h\geq 0$.
Qualitatively, the indicial entropies show a central peak at small $h$ which morphs into an approximately constant plateau as $h$ approaches $Q$. We find that the central peak grows approximately like $\sqrt{Q}$, as expected from \eqref{eq:resentropy_pairwise}. In contrast, the endpoints at $h=Q$, which can be taken as proxy for the plateau height, only grow like $\log Q$. See App. \ref{app:scalings} for further details. The logarithmic growth of the endpoints is expected from the Higgs and Coulomb Hilbert series \eqref{eq:large_hilbert_series}, although the precise coefficients differ due to the removal of the decoupled sector.
Since the endpoints have enhanced supersymmetry they do not correspond to black hole microstates.
Similar tails were observed in the SCI of 4d $\mathcal N=4$ SYM and conjectured to correspond to graviton states \cite{Choi:2025lck}.
Black hole entropy curves, obtained as described in section \ref{subsec:singlecharge}, are shown as well. They provide an increasingly accurate fit as $Q$ is increased. They match the support of the indicial entropies and provide an accurate fit of the central peak, which increasingly dominates the indicial entropy as $Q$ is increased.

\begin{figure}[h]
	\centering
	
	\begin{minipage}{0.48\linewidth}
		\centering
		\includegraphics[width=\linewidth]{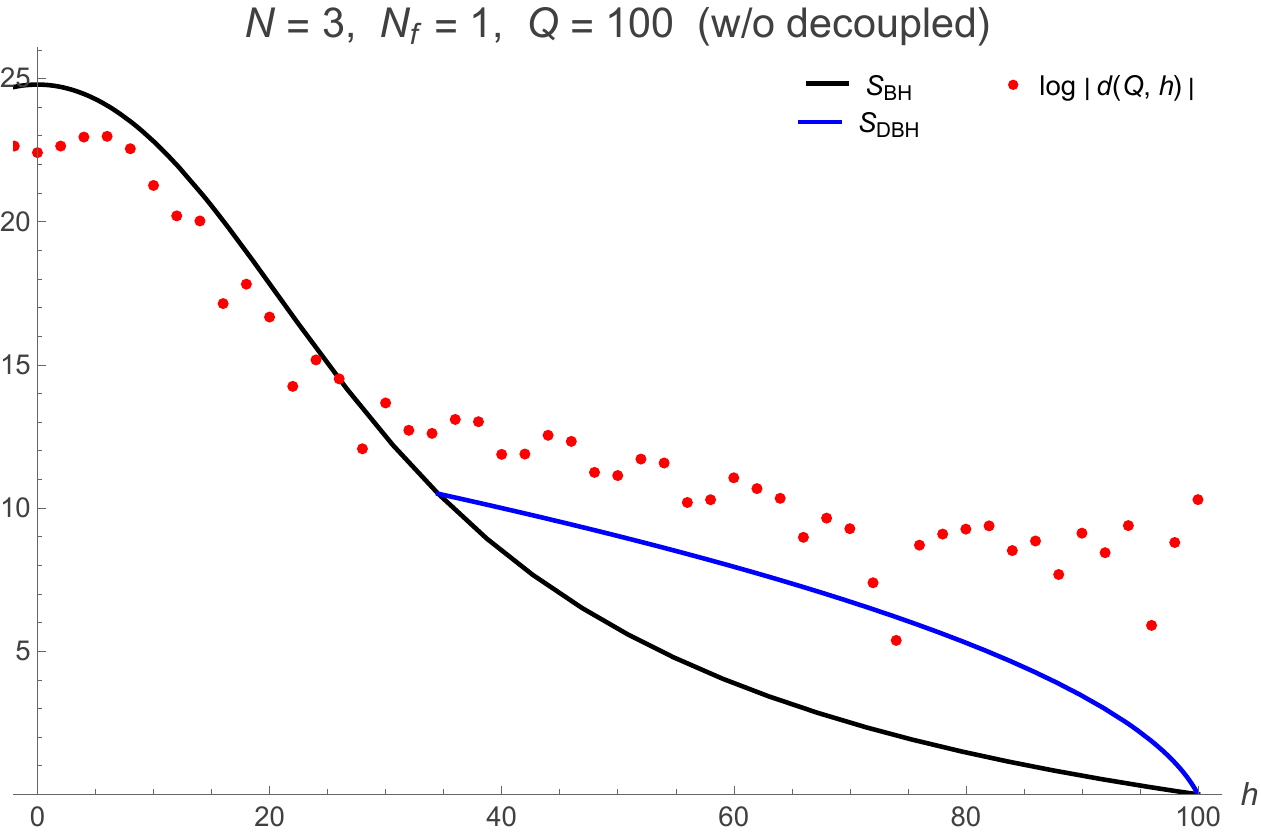}
	\end{minipage}
	\hfill
	\begin{minipage}{0.48\linewidth}
		\centering
		\includegraphics[width=\linewidth]{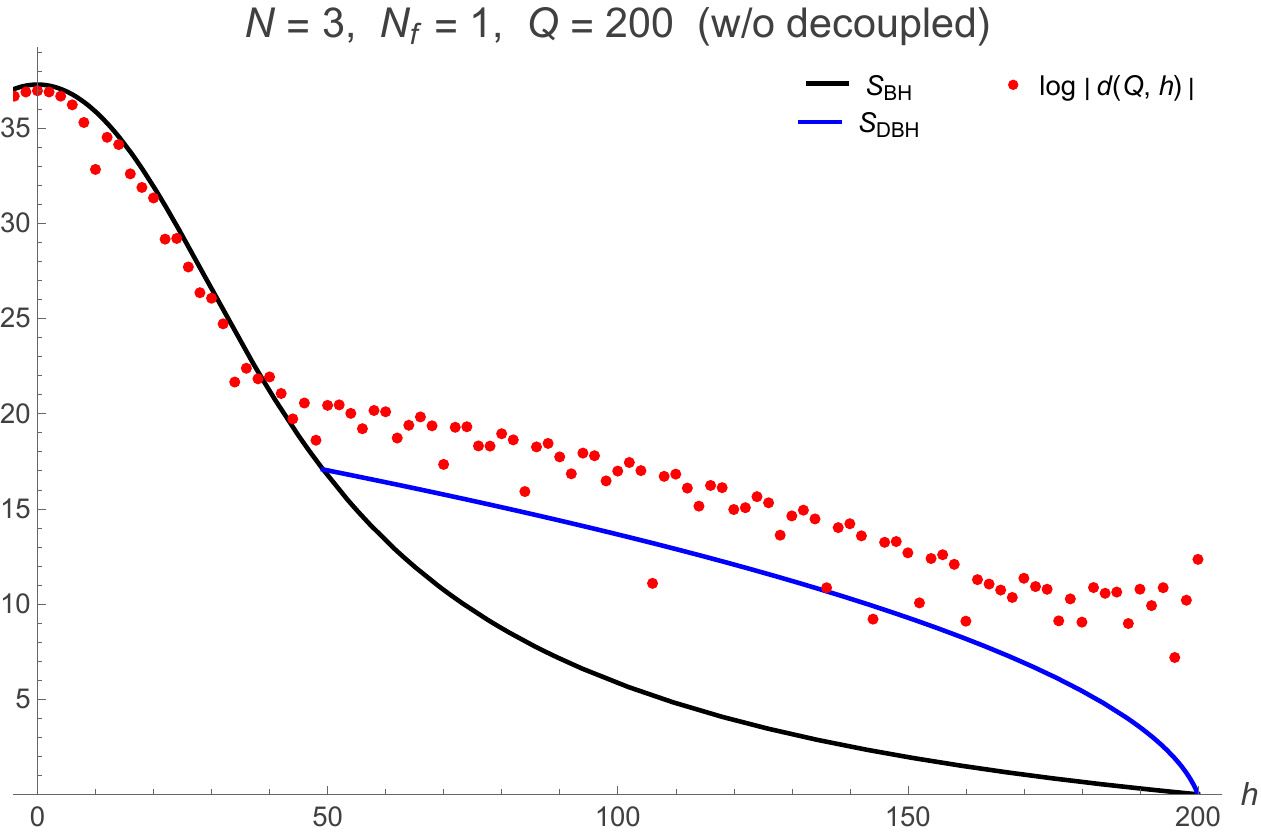}
	\end{minipage}
	\hfill
	\begin{minipage}{0.48\linewidth}
		\centering
		\includegraphics[width=\linewidth]{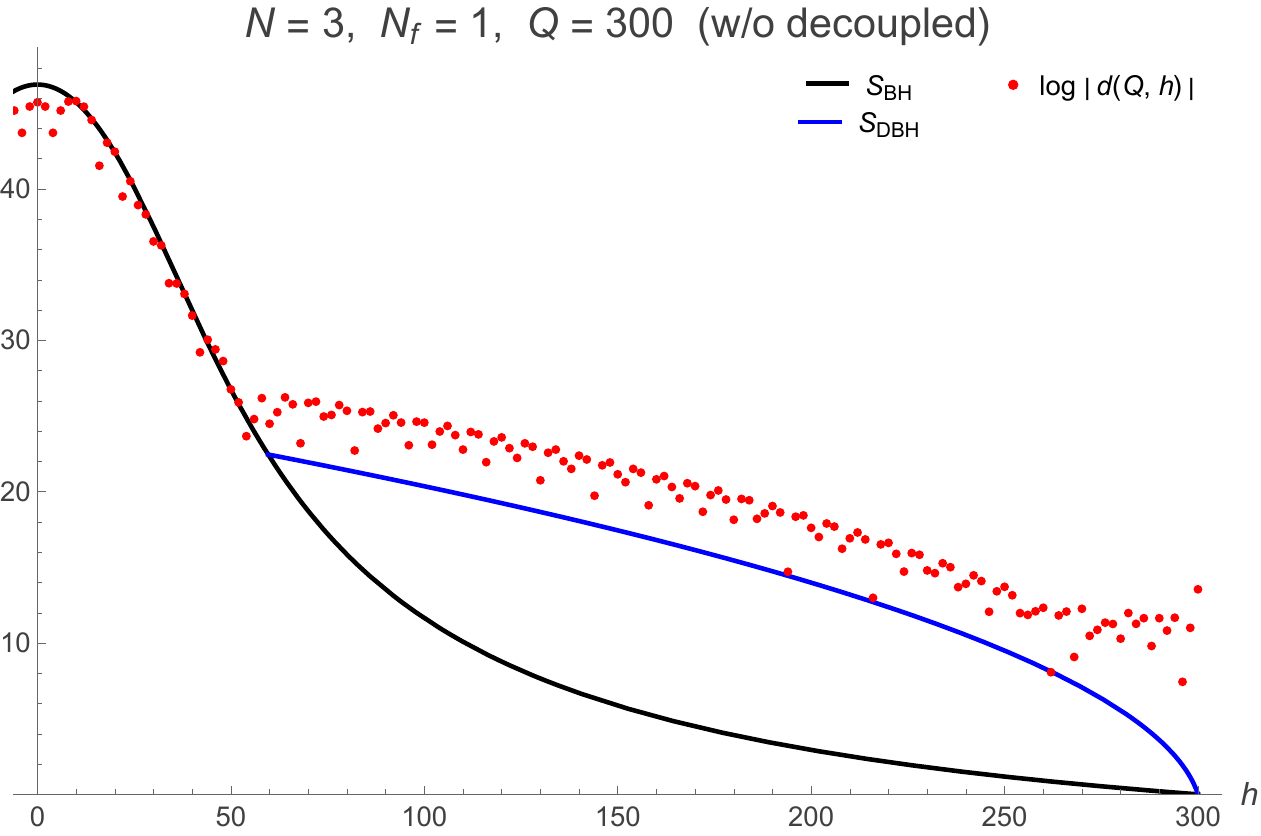}
	\end{minipage}
	\hfill
	\begin{minipage}{0.48\linewidth}
		\centering
		\includegraphics[width=\linewidth]{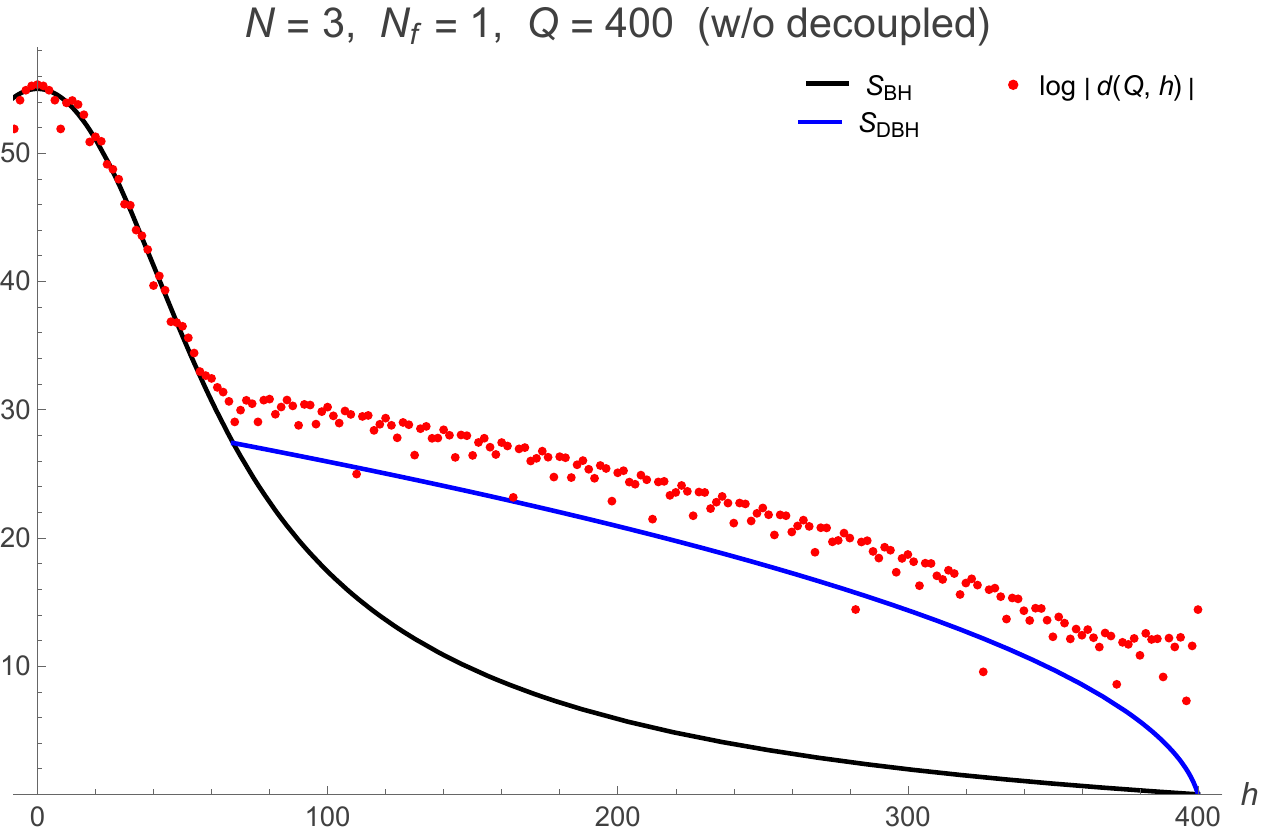}
	\end{minipage}
	
	\caption{The $N=3$ index for a range of $Q$. In contrast to $N=2$, DBH-like behavior is clearly visible. Best-fit DBH entropy curves are shown in blue.}
	\label{fig:ADHM_N3_y2}
\end{figure}

The results for $N=3$, shown in Fig.~\ref{fig:ADHM_N3_y2} for four different values of $Q$, are qualitatively different. 
The indicial entropies are still well approximated by isolated black hole entropies for small $h$, but sharply transition into an arc above the black hole curves at moderate $h$, before tapering out into a plateau near $h=Q$.
For $Q=100$, these regimes correspond roughly to $h<30$, $30<h<75$, and $h>75$ respectively. 
Unlike the plateaux, the arc-shaped excesses do not decrease relative to the central peak as $Q$ is increased, and instead show the same approximate $\sqrt{Q}$ scaling as the pure black hole entropy. This behavior is expected for dressed black holes, whose entropy \eqref{eq:S_IIb} grows like $\sqrt{Q}$ at fixed $h/Q$, as discussed in the text surrounding Fig.~\ref{fig:BPSentropyfiniteN_trans}.

The blue curves show semi-classical entropies for black holes realizing dressed phases, as discussed in section~\ref{sec:entropy-functions}. The effective Newton constants $\widetilde G_N$ which govern the core black holes in the dressed phases of region II$_b$, i.e.\ the parameter in \eqref{eq:S_IIb}, are fitted. In these fits, a constant offset was included as a second free parameter. This offset is not shown in the fitted curves. Additional details on the fitting ansatz can be found in App.~\ref{app:fitting}. As Fig.~\ref{fig:ADHM_N3_y2} shows, the index results align increasingly well with the semi-classical DBH entropy curves as $Q$ is increased. The relative offset shrinks and the qualitative behavior matches increasingly well. While the fitting ansatz in the gravitational description is empirical rather than derived from first principles, we find the agreement convincing and use it throughout as an efficient way to capture the index results.

\begin{figure}[h]
	\centering
	\includegraphics[width=0.48\linewidth]{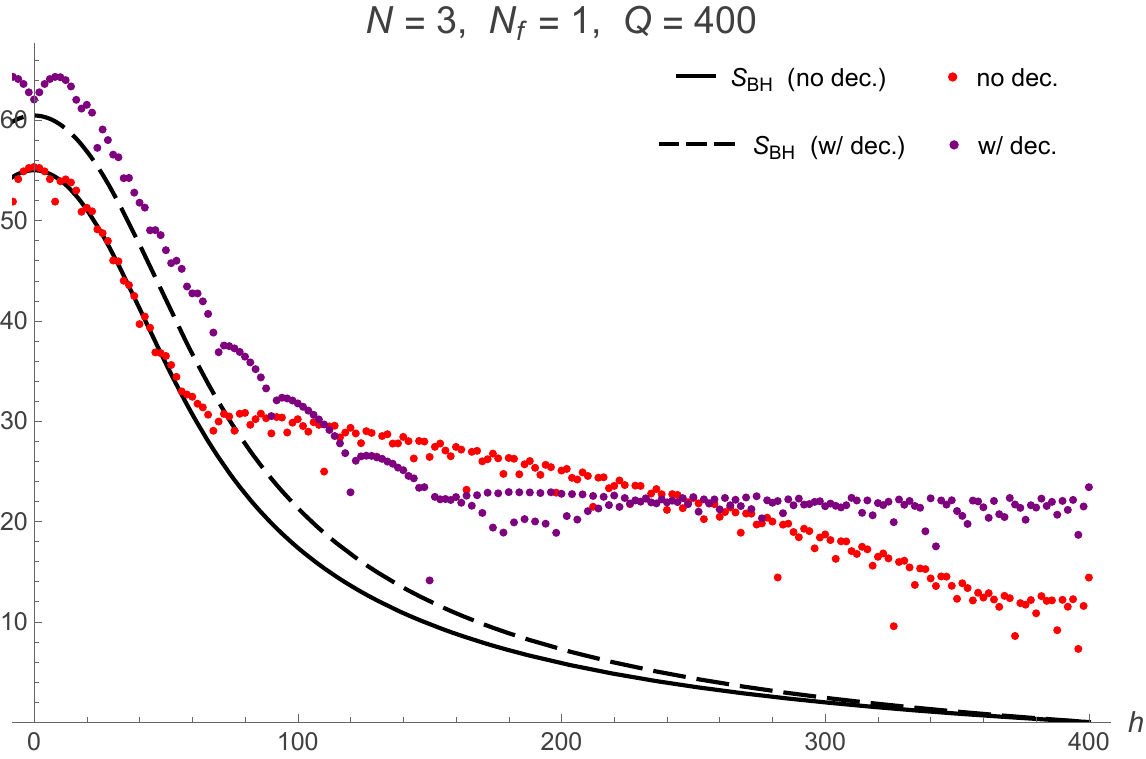}
	\caption{The full $N=3$ index with the decoupled sector included (purple) compared to the index without decoupled sector. To provide an entropy comparison for the full index, the dashed black curve is computed directly from the free energy of \eqref{eq:ZS3_ADHM_exact}. The solid black curve removes the contribution \eqref{eq:Z_decoupled} of the decoupled sector as before.}
	\label{fig:ADHM_y2_3_1_400_dnd_comp}
\end{figure}

In the definition of the indicial entropies we removed the decoupled sector, which is not expected to correspond to bulk degrees of freedom in the large-$N$ gravitational description. One may ask how strong the impact is at finite $N$. To address this point for $N=3$, $Q=400$, we compare in Figure~\ref{fig:ADHM_y2_3_1_400_dnd_comp} the indicial entropies extracted from the full index, including the decoupled sector, to those of the reduced index, with decoupled sector removed.
We find that the indicial entropy of the full index (purple) shows no convincing evidence of DBH behavior while the indicial entropy of the reduced index (as before in red) clearly does.
Curiously, the arc-shaped excess in the reduced indicial entropy rises above the plateau in the full indicial entropy.
We conclude that the contributions of the decoupled sector are not simply masking the DBH shoulder additively, and instead induce cancellations in the full index. We leave the question of whether DBH-like behavior emerges upon further increasing $Q$ for the future.

Exactly computing the factorized index \eqref{eq:factorized_schematic} for larger $N$ becomes increasingly difficult, since the number of independent integers $k_\mathrm{a}$ in the vorticity sums grows with $N$. For $N=4$, we did not reach large enough $Q$ to diagnose DBH behavior.  However, we recall that \cite{Hwang:2025hfs} found success in approximating the $N=4$ and $N=5$ single-fugacity index through truncations of the vorticity sums well below the exactness threshold \eqref{eq:p-v-rel}. We present similar explorations in App.~\ref{app:vort_cutoff} for the (2,2) two-fugacity index. Although, as noted earlier, this approximation fails to accurately capture the qualitative behavior of the two-fugacity indices in general, especially if applied aggressively, we find the dressing to be remarkably robust.


\subsection{(2,2) specialization $k>1$ ABJM and $N_f>1$ ADHM} \label{sec:(2,2) results k>1}

We now turn to ABJM theory with $k>1$ and ADHM theory with $N_f>1$. Both are computationally more challenging than $k=1$ and $N_f=1$, but as we shall see they compensate by revealing DBH-like behavior for lower values of $N$ and $Q$.

Since ABJM theory with $k>1$ is not equivalent to ADHM theory, the factorized form of the index cannot be used, and we directly evaluate the matrix integral \eqref{eq:ABJM_index}. The product form of the gauge group makes the index in \eqref{eq:ABJM_index} computationally challenging, and we limit the analysis to $N=2$. 
For ADHM theory with $N_f>1$, the factorized form of the index involves colored Young tableaux and becomes considerably more involved, so that we likewise return to the matrix integral form. We note that the number of flux sectors needed for fixed $Q$ in (\ref{eq:p-Nf-rel}) changes favorably as $N_f$ is increased.
With a single gauge group this index is more tractable and we still obtain useful results for $N=2$ and $N=3$.

The nature of the decoupled sectors also changes. For ABJM with $k>1$ there is no decoupled sector, so we compute the full index and the associated indicial entropies. For ADHM with $N_f>1$ there still is a decoupled sector, but it reduces to the $\mathrm{U}(1)$ part of the $\mathcal{N}=4$ adjoint hypermultiplet. We will continue to remove this contribution.

\begin{figure}[h]
	\centering
	
	\begin{minipage}{0.48\linewidth}
		\centering
		\includegraphics[width=\linewidth]{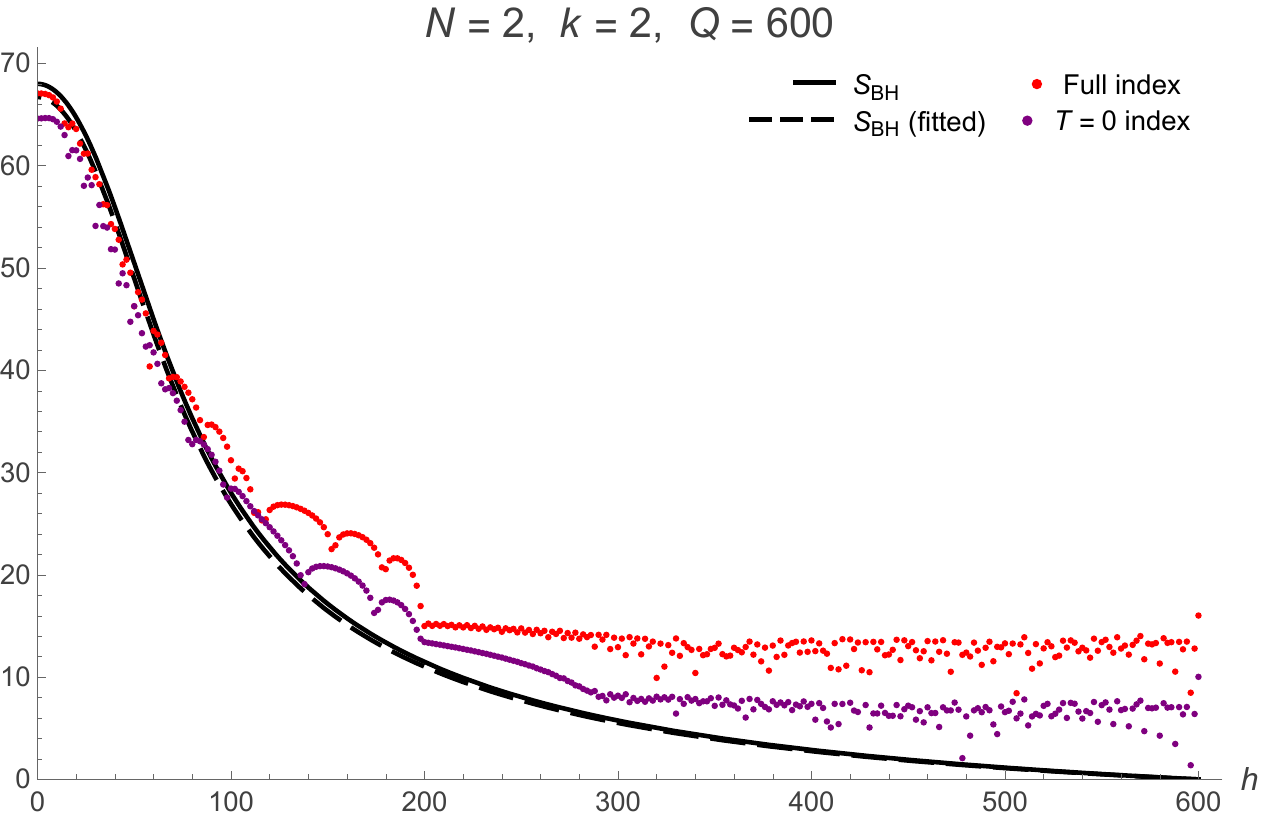}
	\end{minipage}
	\hfill
	\begin{minipage}{0.48\linewidth}
		\centering
		\includegraphics[width=\linewidth]{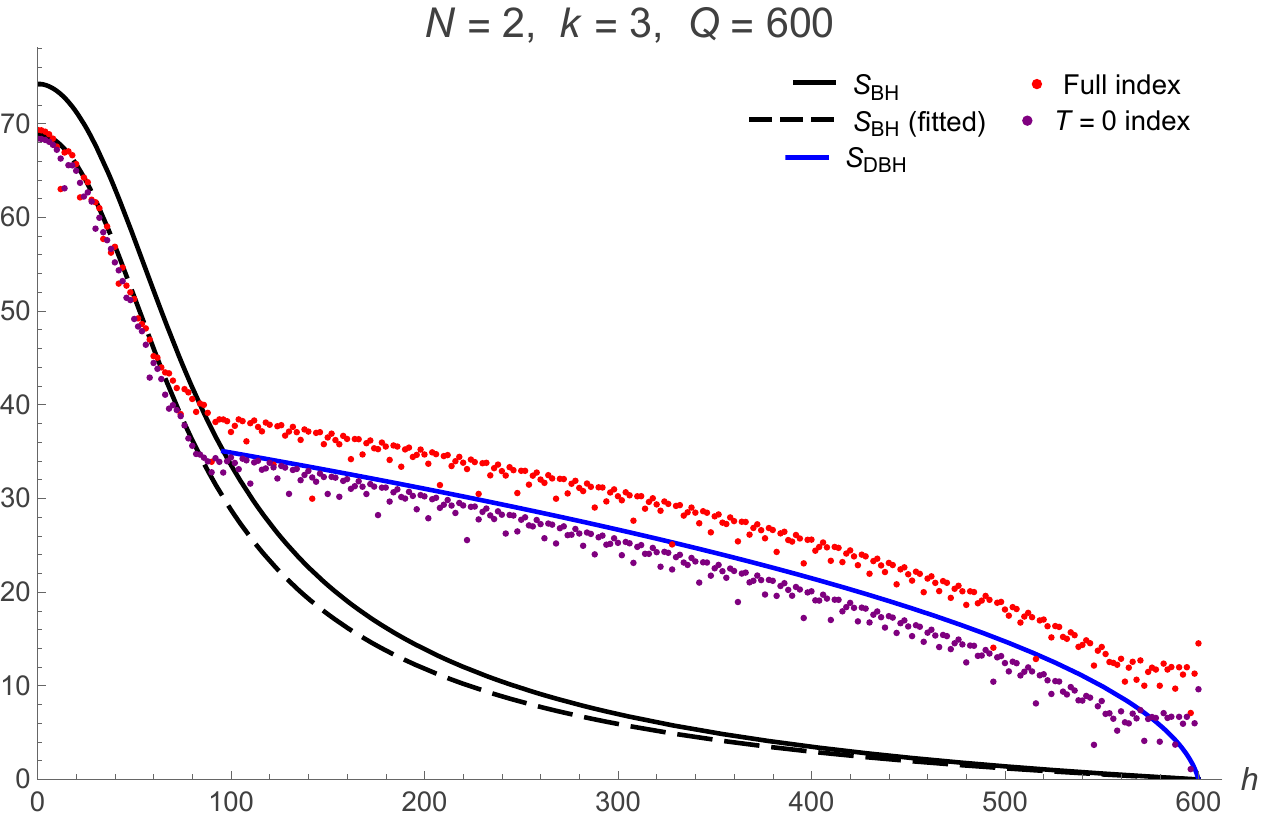}
	\end{minipage}
	\hfill
	\begin{minipage}{0.48\linewidth}
		\centering
		\includegraphics[width=\linewidth]{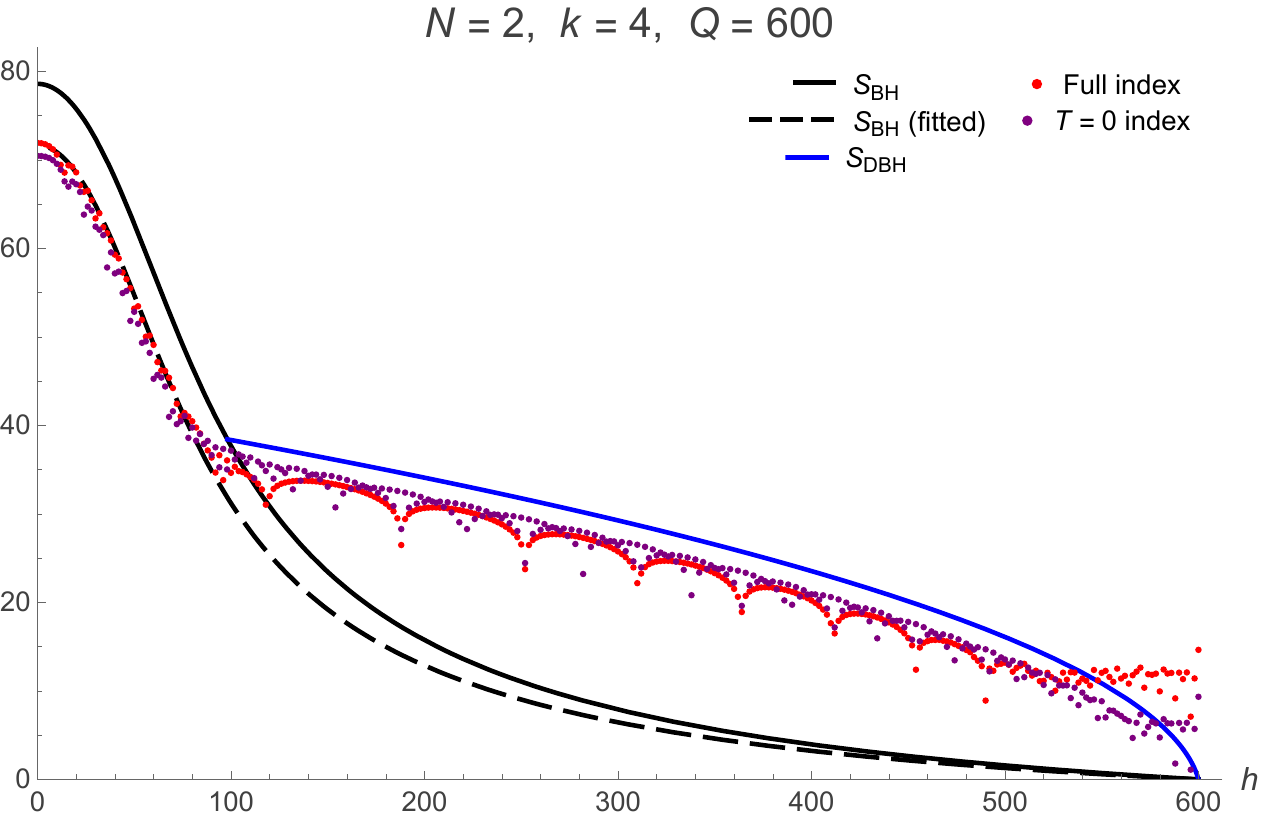}
	\end{minipage}
	\hfill
	\begin{minipage}{0.48\linewidth}
		\centering
		\includegraphics[width=\linewidth]{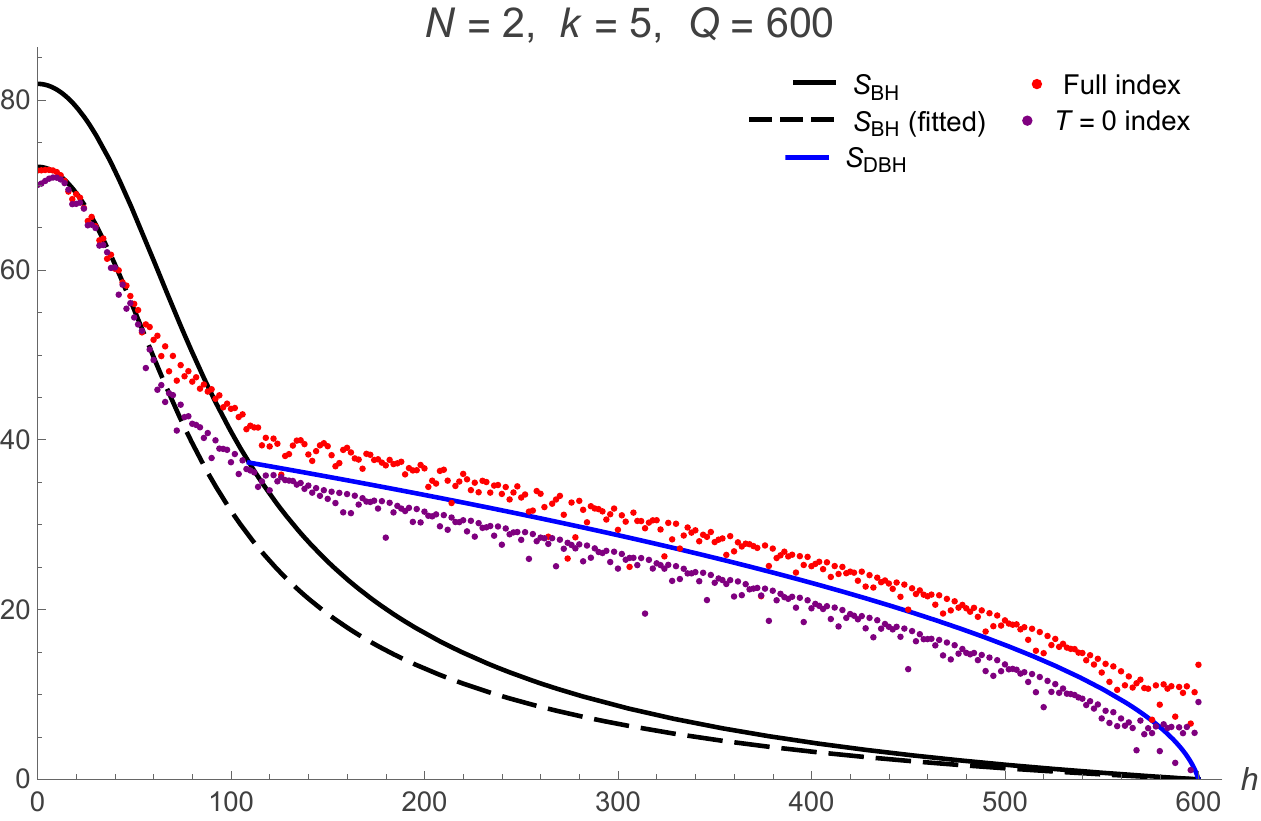}
	\end{minipage}
	
	\caption{The ABJM index with $N=2$ for various $k$. DBH-like behavior appears for $k\geq 3$. In these cases, the DBH entropy curve with best-fit $\tilde G_N$ is shown.}
	\label{fig:ABJM_N2_k_y2}
\end{figure}

\paragraph{$k>1$ ABJM theory:}
We start with the results for $k>1$ ABJM theory. Fig.~\ref{fig:ABJM_N2_k_y2} displays the indicial entropies for $N=2$ with $2\leq k\leq 5$. The $k=2$ results are qualitatively similar to the $k=1$ results in Fig.~\ref{fig:ADHM_N2_y2} at least for the values of $Q$ we obtained results for, with a central peak resembling the entropy curve of an isolated black hole tapering out into plateaux at larger $|h|$ whose prominence decreases with increasing $Q$, and no clear evidence of DBH-like behavior. 
For $k\geq 3$ this changes qualitatively. The indicial entropies for $k\geq 3$ show clear transitions into arc-shaped shoulders before eventually morphing into plateaux, showing similar DBH-like behavior as the $(N,k)=(3,1)$ indicial entropies. We also note that the agreement between the central peak and the isolated black hole entropy obtained through the identification of the Newton constant with the free energy slightly deteriorates as $k$ is increased. The central peaks are nevertheless still captured accurately by black hole curves, as we show through black hole curves with best-fit Newton constants shown as dashed curves.

With rich behavior arising already for $N=2$, we use this case to investigate the difference between fugacity and charge identifications. So far we compared black holes with charge identifications to indices with fugacity identifications. An alternative would be to directly count states realizing the charge identifications imposed on the black holes. While generally more challenging from a computational perspective, we can partially implement this as follows. The (2,2) charge constraint \eqref{eq:pairwise_charge_identification} implies $T=0$ through  \eqref{eq:geo_charges}. This restriction is captured by the index counting the topologically neutral states with $T=0$, using the ABJM analog of \eqref{eq:ADHM_semiperturbative_index}. The results are included in Fig.~\ref{fig:ABJM_N2_k_y2}. In the regime where an isolated black hole dominates, the $T=0$ indicial entropies align closely with the $\xi_T=1$ $(2,2)$ index. In the dressed phase the qualitative behavior also matches, with the $T=0$ indicial entropies closely following the fitted semi-classical entropy curve determined from the $\xi_T=1$ $(2,2)$ index, while almost entirely eliminating the constant shift observed in the $\xi_T=1$ index. This supports our treatment of the fugacity identifications.

\begin{figure}[h]
	\centering
	
	\begin{minipage}{0.48\linewidth}
		\centering
		\includegraphics[width=\linewidth]{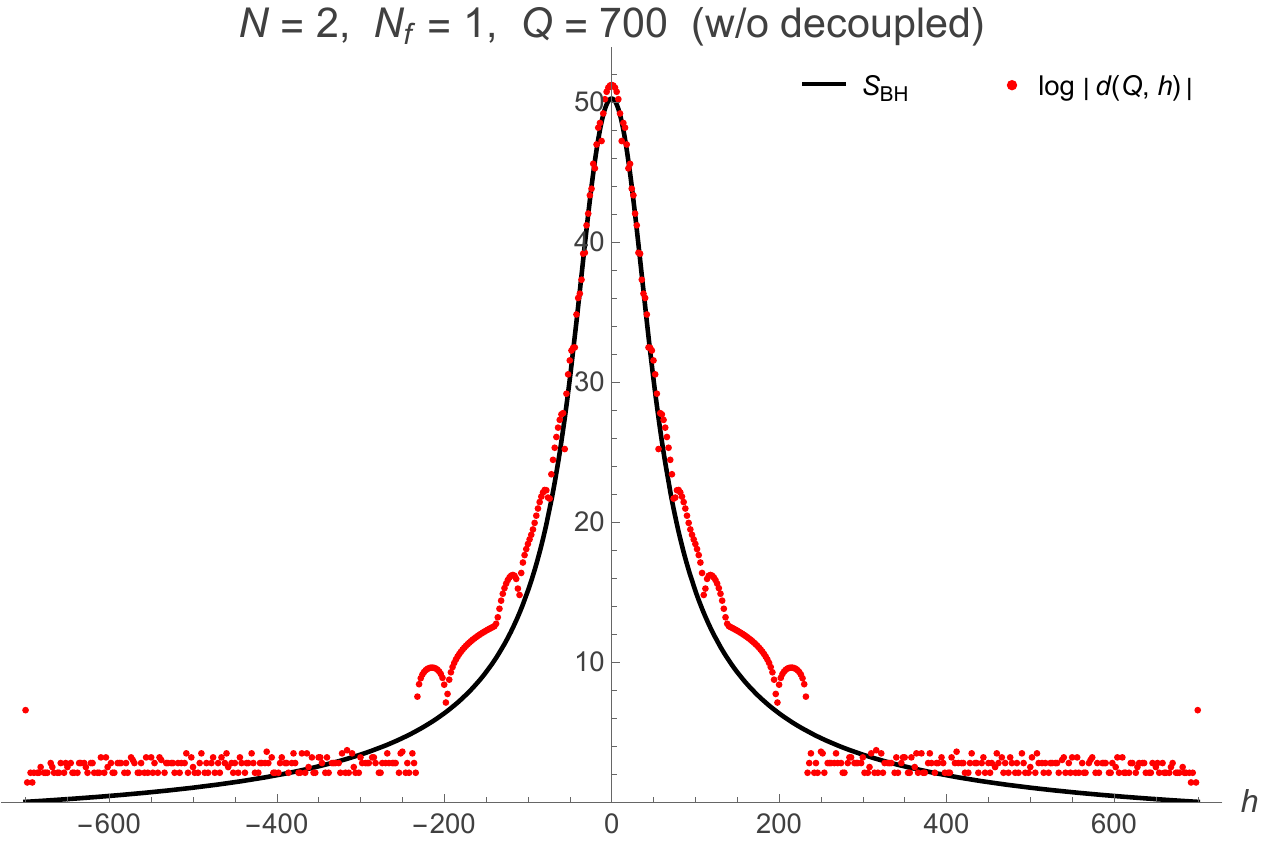}
	\end{minipage}
	\hfill
	\begin{minipage}{0.48\linewidth}
		\centering
		\includegraphics[width=\linewidth]{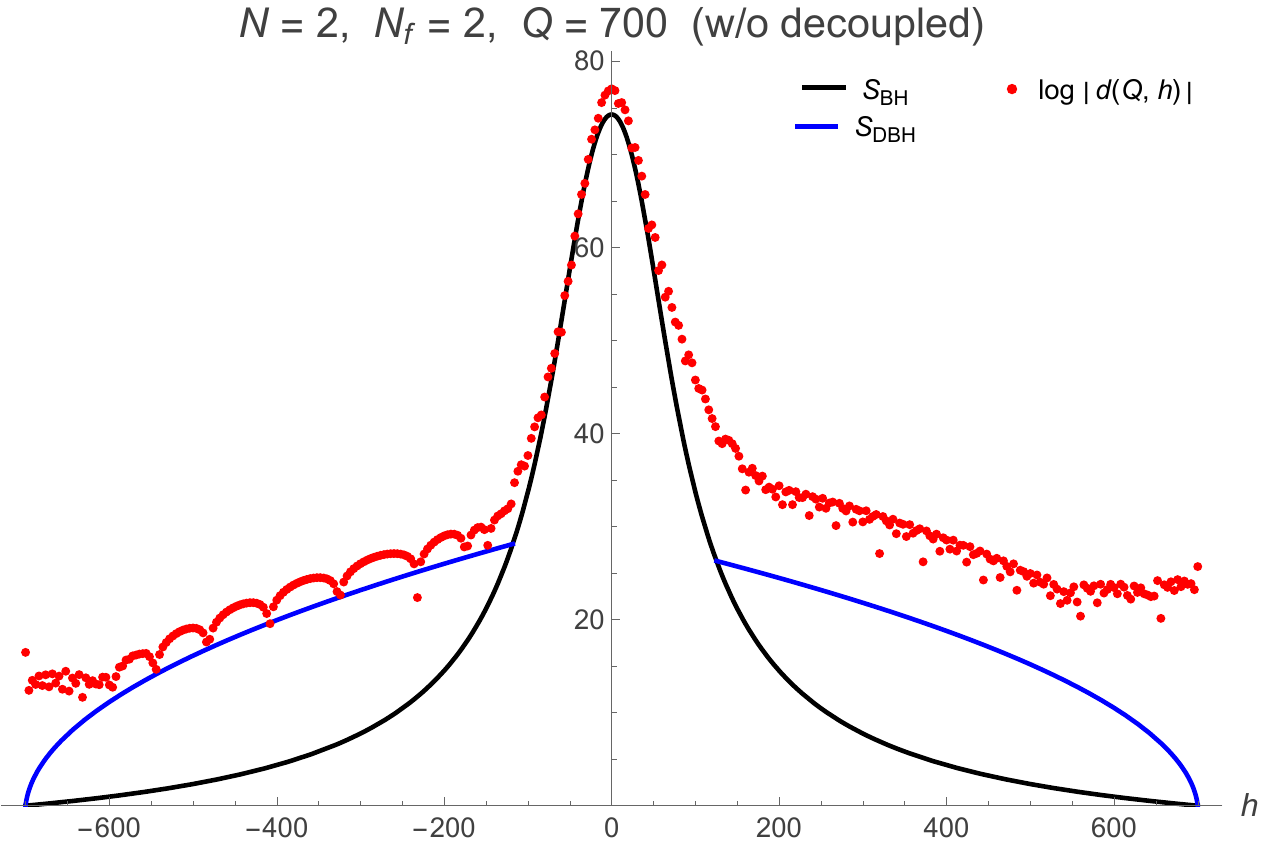}
	\end{minipage}
	\hfill
	\begin{minipage}{0.48\linewidth}
		\centering
		\includegraphics[width=\linewidth]{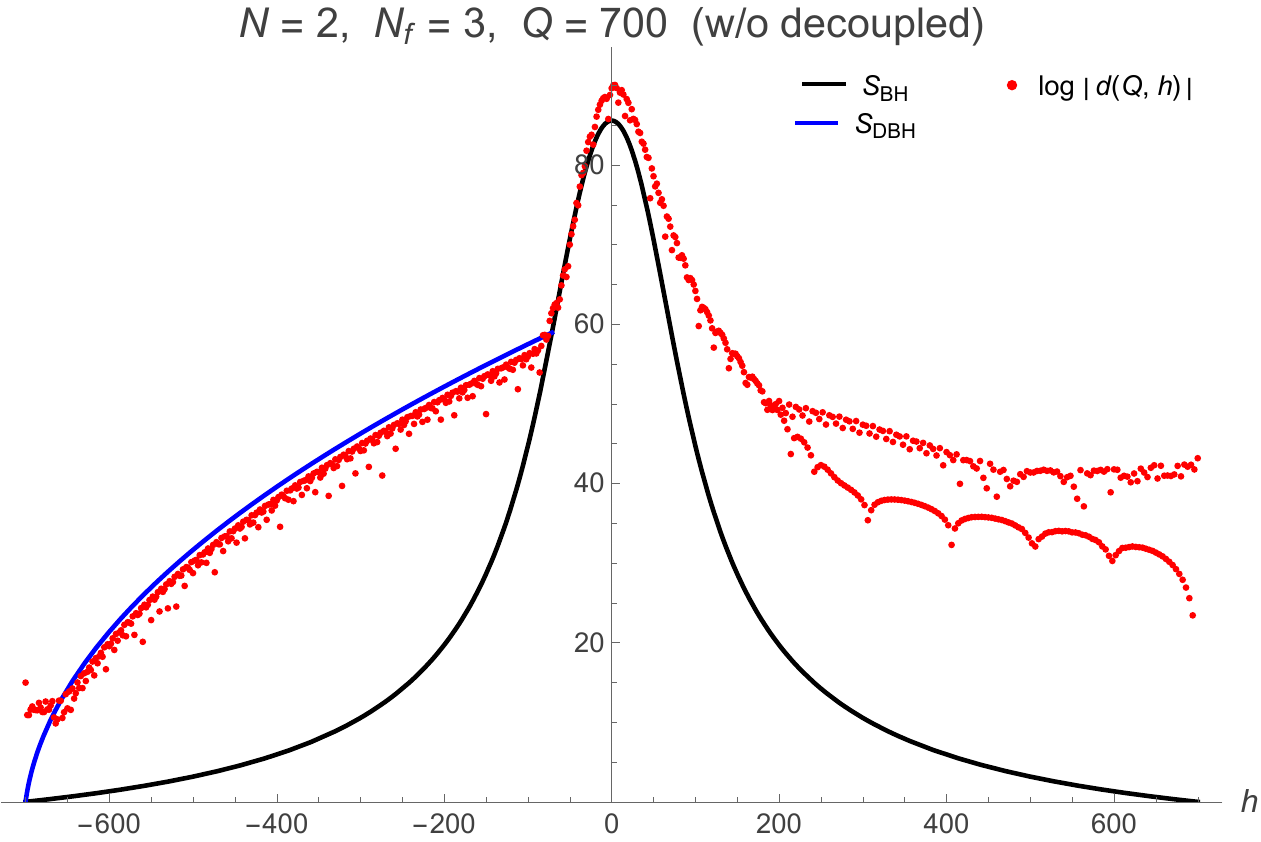}
	\end{minipage}
	\hfill
	\begin{minipage}{0.48\linewidth}
		\centering
		\includegraphics[width=\linewidth]{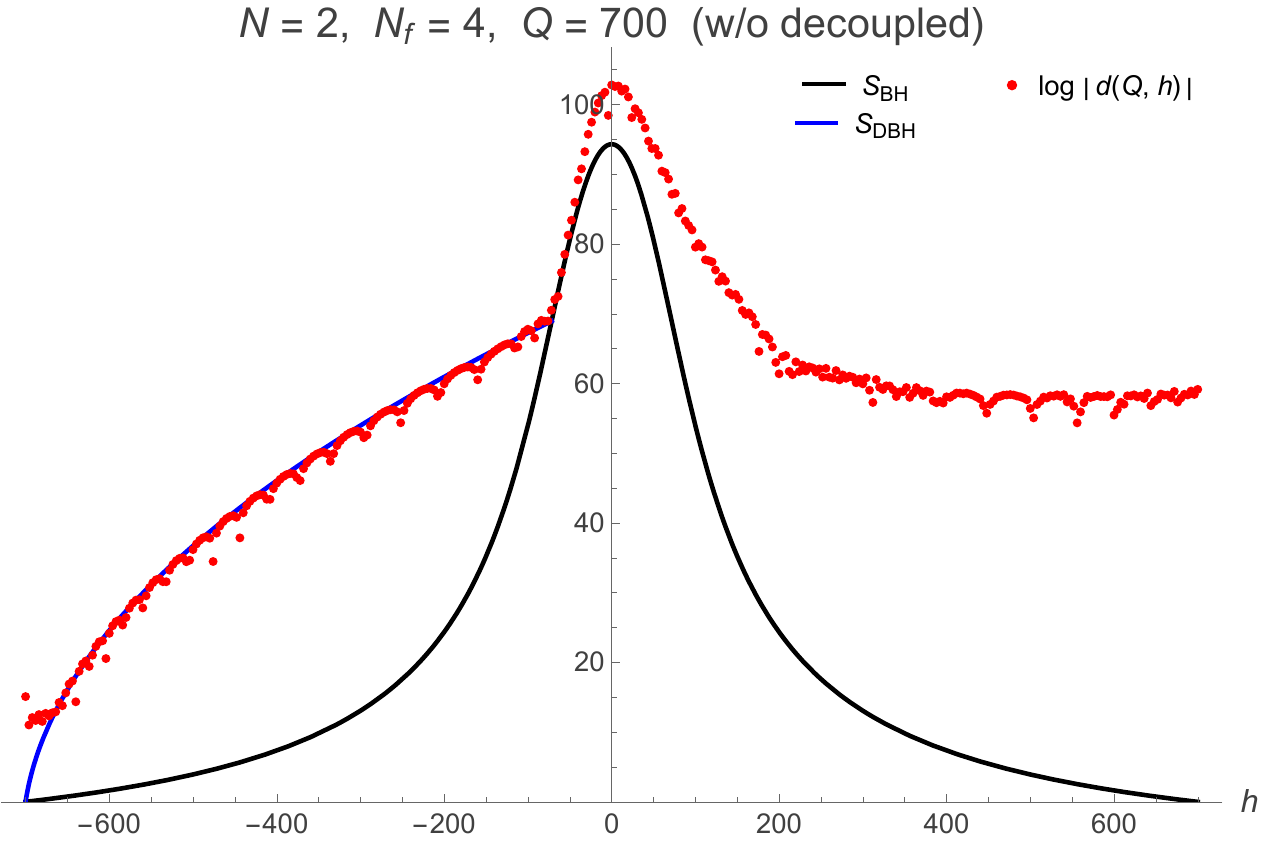}
	\end{minipage}
	
	\caption{The index at $N=2$ with various $N_f$. DBH-like behavior appears for $N_f>1$. In these cases, the DBH entropy curve with best-fit $\tilde G_N$ is shown.}
	\label{fig:ADHM_N2_flavors_y2}
\end{figure}

\begin{figure}[h]
    \centering

    \begin{minipage}{0.48\linewidth}
        \centering
        \includegraphics[width=\linewidth]{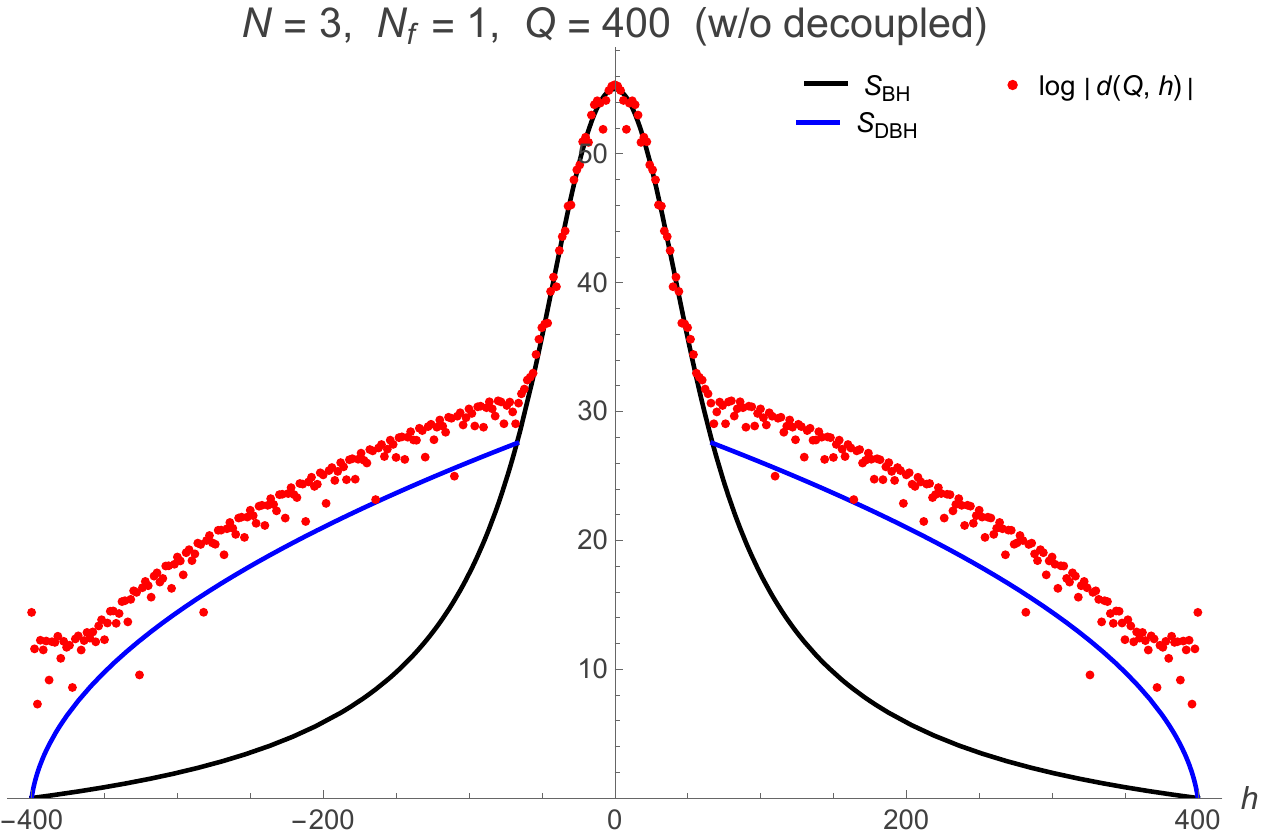}
    \end{minipage}
    \hfill
    \begin{minipage}{0.48\linewidth}
        \centering
        \includegraphics[width=\linewidth]{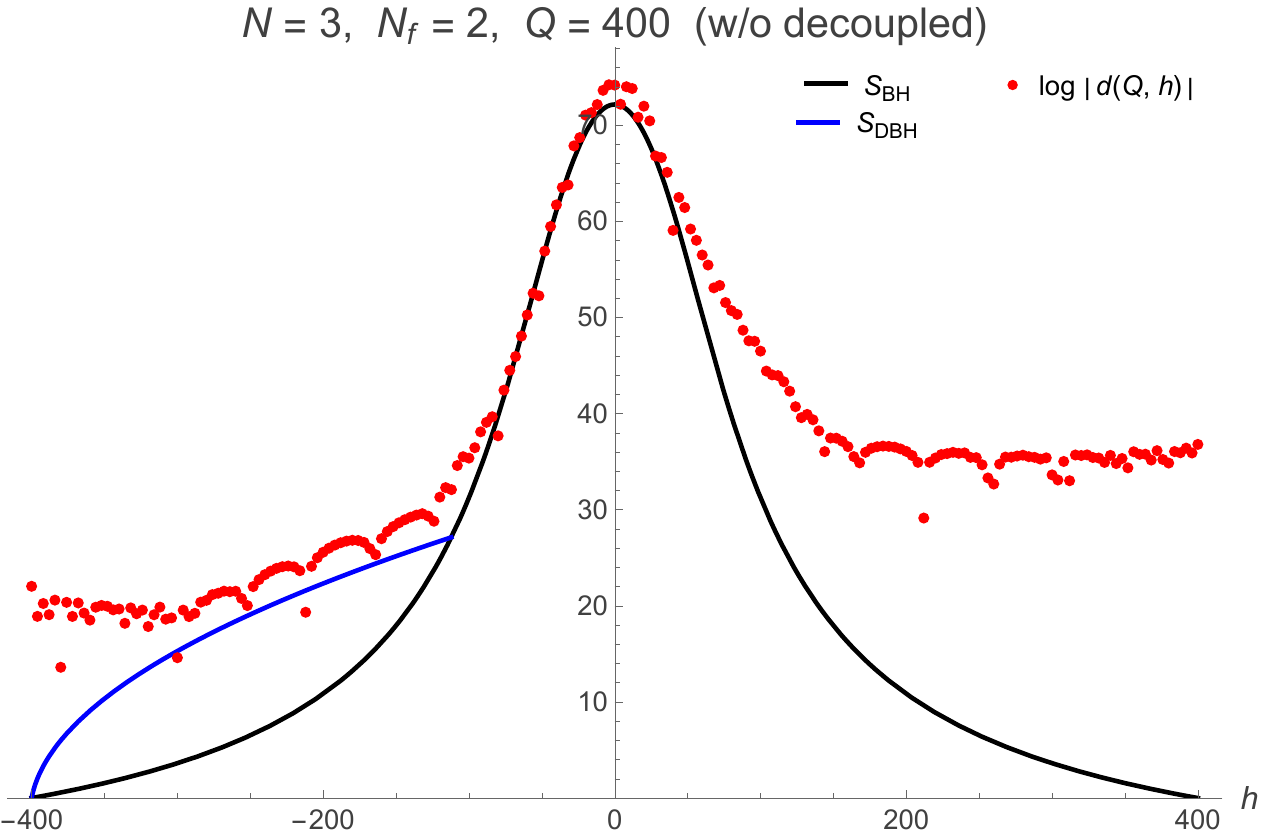}
    \end{minipage}
    \hfill
    \begin{minipage}{0.48\linewidth}
        \centering
        \includegraphics[width=\linewidth]{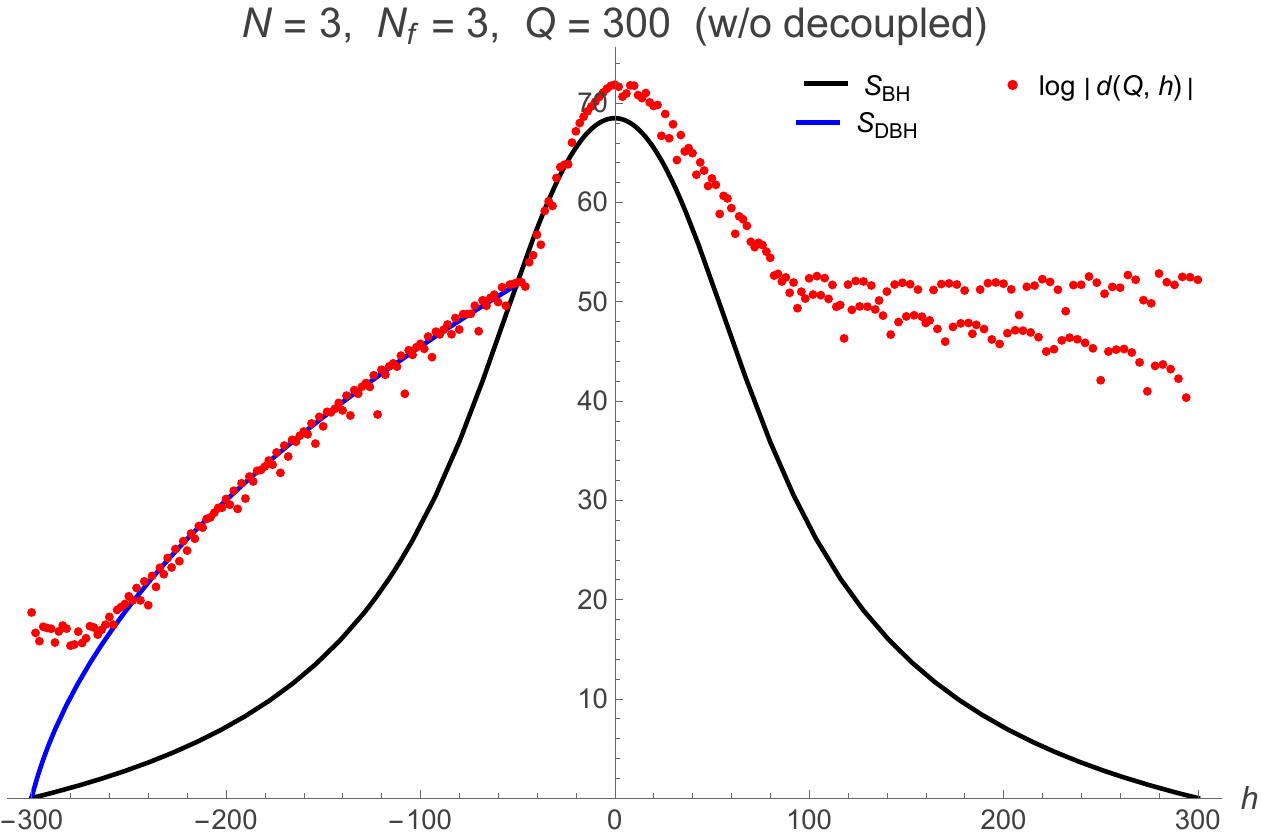}
    \end{minipage}
    \hfill
    \begin{minipage}{0.48\linewidth}
        \centering
        \includegraphics[width=\linewidth]{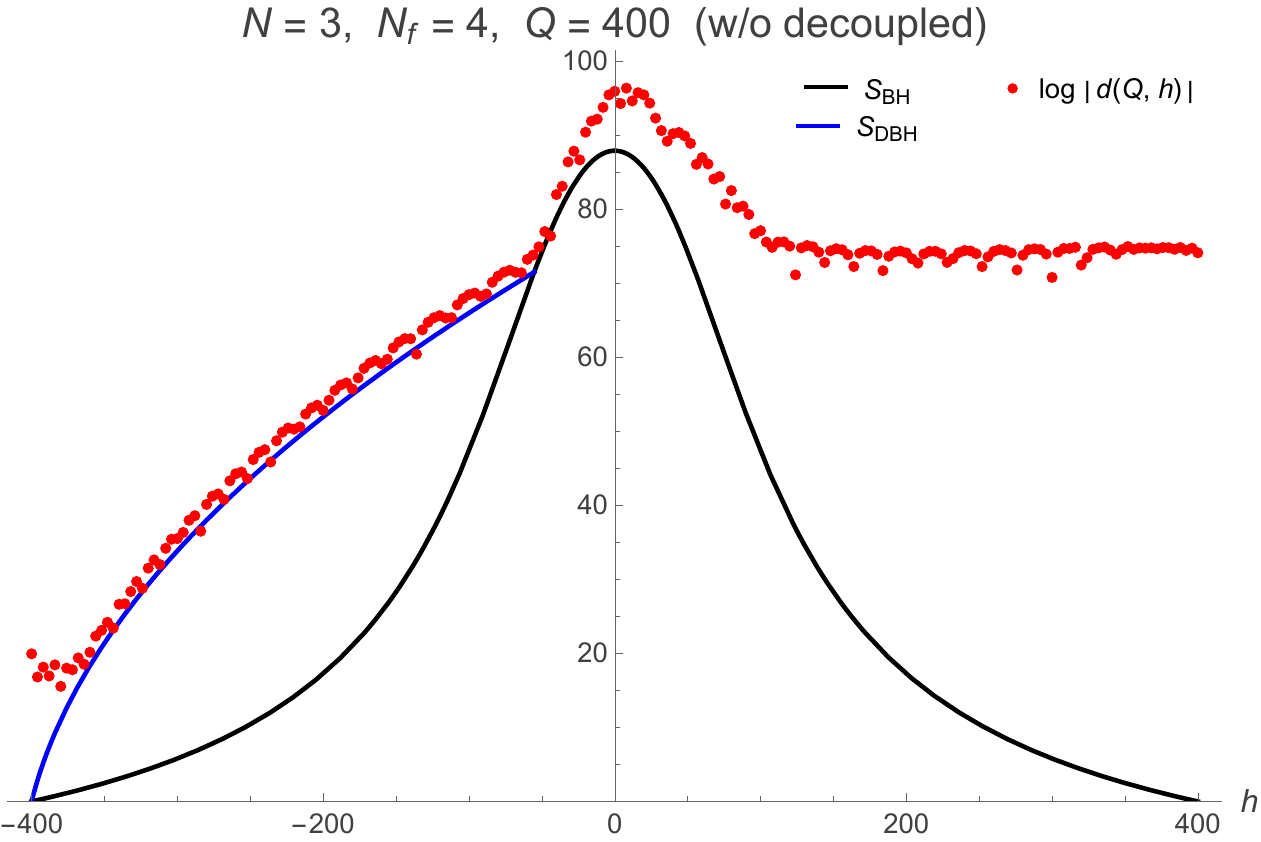}
    \end{minipage}

    \caption{The index at $N=3$ with various $N_f$. Unlike for the $N=2$ index, DBH-like behavior appears on the left side of the plots for all $N_f$.}
    \label{fig:ADHM_N3_flavors_y2}
\end{figure}

\paragraph{$N_f>1$ ADHM theory:}
We now turn to the results for ADHM theory with $N_f>1$. We start with $N=2$, for which the indicial entropies for $1\leq N_f\leq 4$ are shown in Fig.~\ref{fig:ADHM_N2_flavors_y2}, all for $Q=700$. The $N_f=1$ case, which also captures $k=1$ ABJM, shows no DBH-like behavior, in line with Fig.~\ref{fig:ADHM_N2_y2}. 
For $N_f\geq 2$ we observe clear DBH-like behavior. The difference between $h>0$ and $h<0$, however, is qualitative. Clean arc-shaped excesses always appear on the $h<0$ side, while the excesses over the isolated black hole entropy curves on the $h>0$ side are increasingly plateau-like as $N_f$ is increased. The endpoint $h=+Q$ corresponds to the Higgs limit, and the growing degeneracy with increasing $N_f$ is manifest in \eqref{eq:large_hilbert_series} and aligns with the extended Higgs branch flavor symmetry with increasing $N_f$. That these increasing endpoint degeneracies are accompanied by plateaux of similar magnitude is in line with the previous results. In contrast, the plateaux on the $h<0$ side remain low, consistent with the Coulomb limit at $h=-Q$ being independent of $N_f$ in \eqref{eq:large_hilbert_series}. The plateaux on the $h>0$ side for $N_f>1$ rise above the shoulders on the $h<0$ side. So they may hide DBH-like shoulders at the accessible values of $Q$. They also appear to extend into the central region: for $N_f>1$, the isolated black hole curves with Newton constant fixed by the free energy sit slightly below the central peaks.  A curious feature for $N_f=3$ is that the indicial entropy bifurcates beyond $h\sim200$ into two branches between which the indicial entropies alternate. Larger $Q$ would be desirable for $N_f=3,4$ to settle the behavior on the $h>0$ side.
Figure \ref{fig:ADHM_N3_flavors_y2} shows similar plots for $N=3$ at lower $Q$. The index already displays DBH behavior for $N_f=1$ and continues to do so on the $h<0$ side with $N_f>1$, while the results on the $h>0$ side are inconclusive.

\subsection{(1,3) specialization for ABJM} \label{sec:(1,3) results}

We now turn to the $(1,3)$ specialization for ABJM theory, where this specialization is universal as discussed in section \ref{sec:ABJM-index-slices}. For $k=1$ this covers $N_f=1$ ADHM theory, and we still use the ADHM description in this case as we find it more efficient. 
The support of the indicial entropies is $-2/3Q\leq h\leq 2Q$ as discussed below (\ref{eq:Qh-def}) and $h\rightarrow -h$ is not a symmetry.

\begin{figure}[h]
	\centering
	
	\begin{minipage}{0.48\linewidth}
		\centering
		\includegraphics[width=\linewidth]{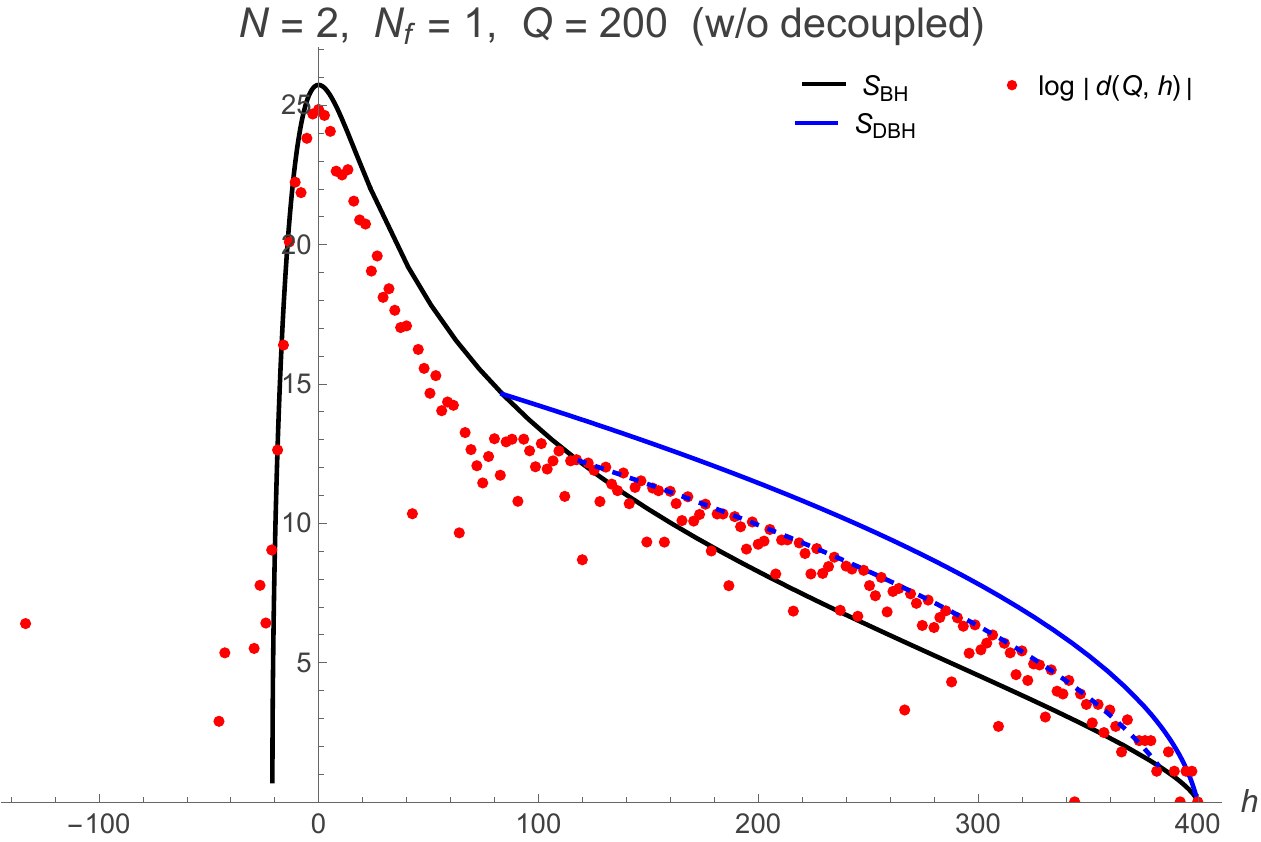}
	\end{minipage}
	\hfill
	\begin{minipage}{0.48\linewidth}
		\centering
		\includegraphics[width=\linewidth]{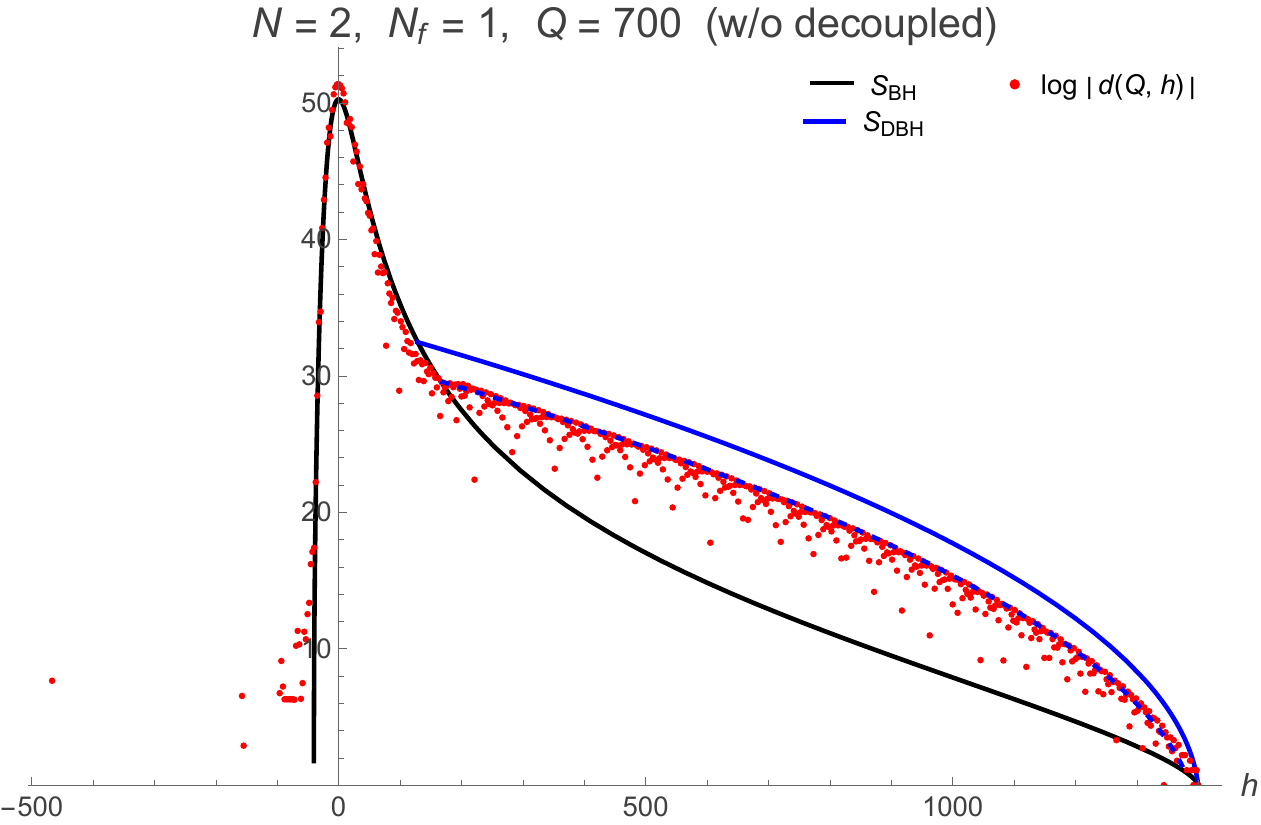}
	\end{minipage}
	
	\begin{minipage}{0.48\linewidth}
		\centering
		\includegraphics[width=\linewidth]{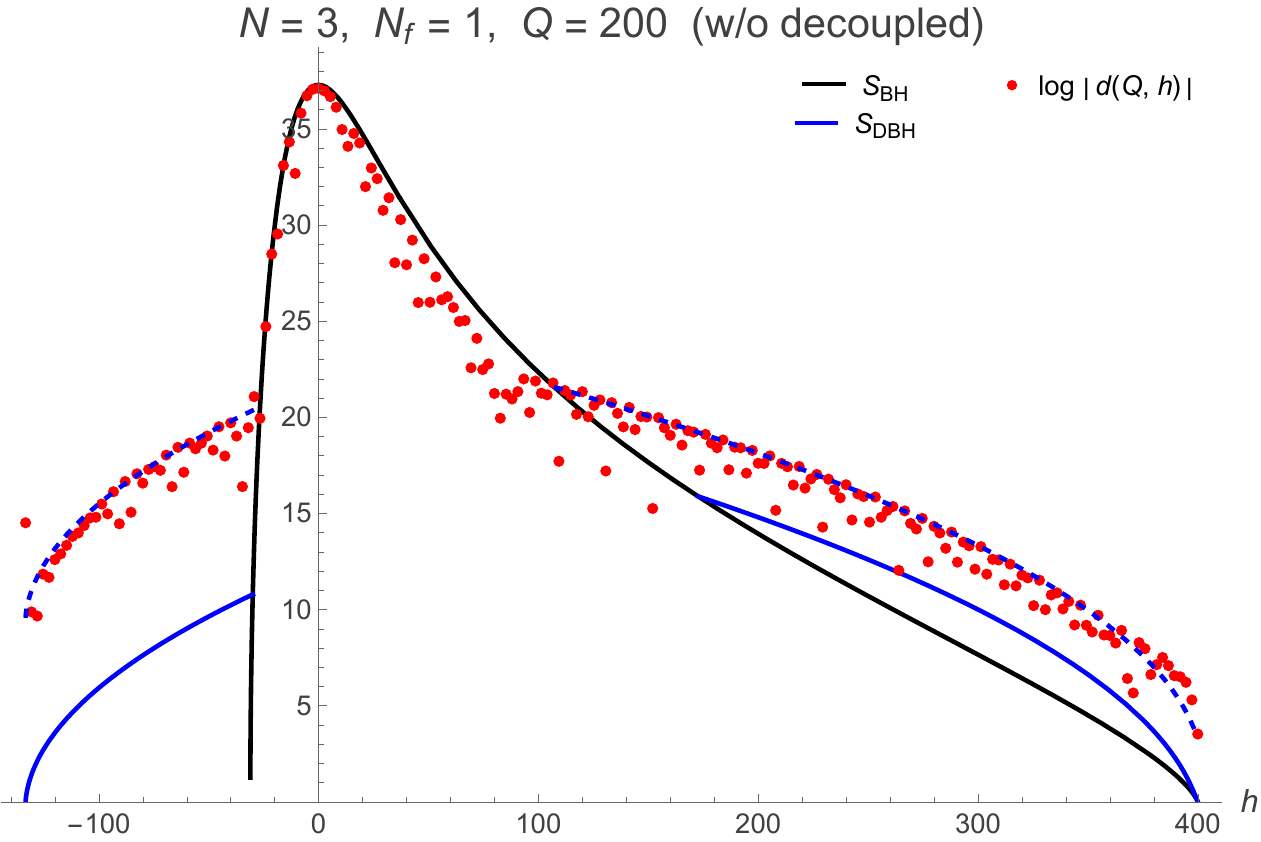}
	\end{minipage}
	\hfill
	\begin{minipage}{0.48\linewidth}
		\centering
		\includegraphics[width=\linewidth]{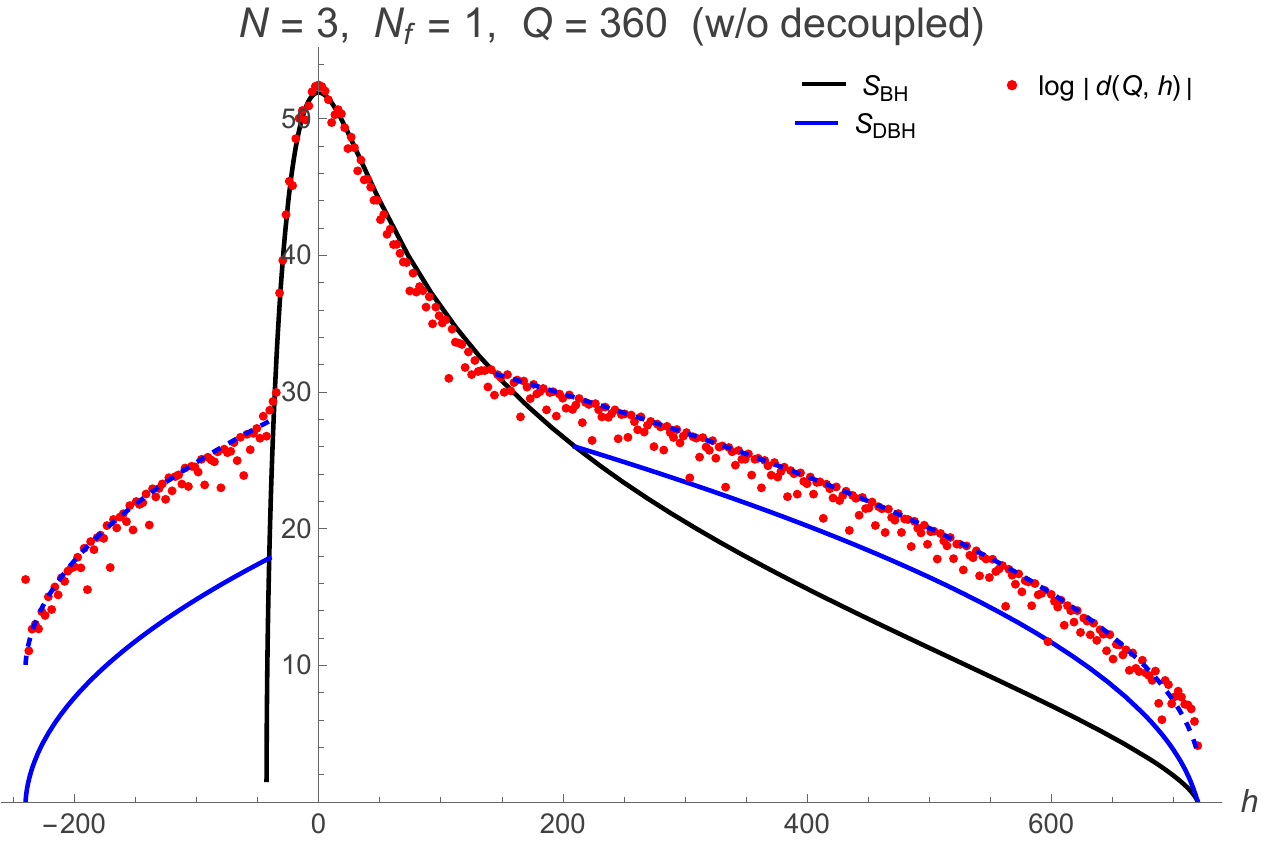}
	\end{minipage}
	
	\includegraphics[width=0.48\linewidth]{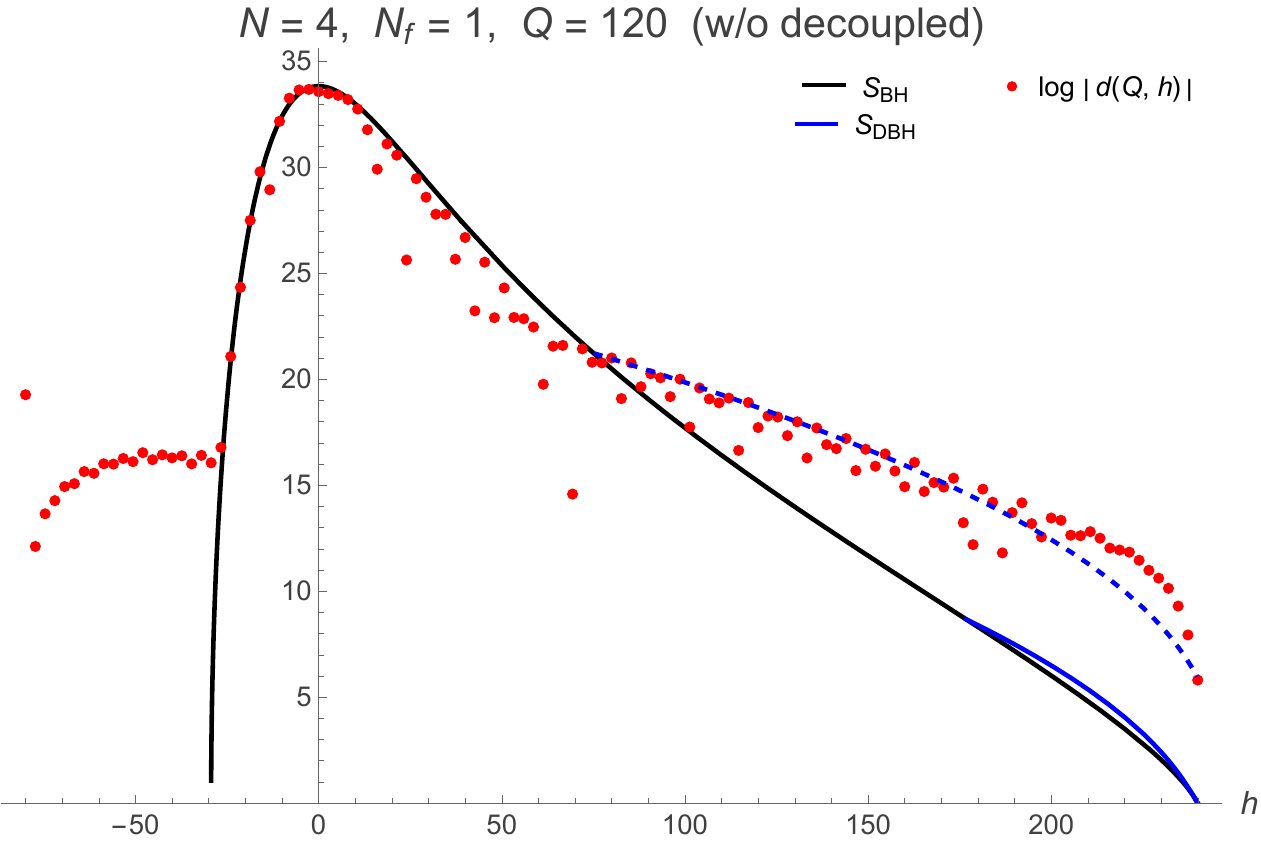}
	\hfill
	\includegraphics[width=0.48\linewidth]{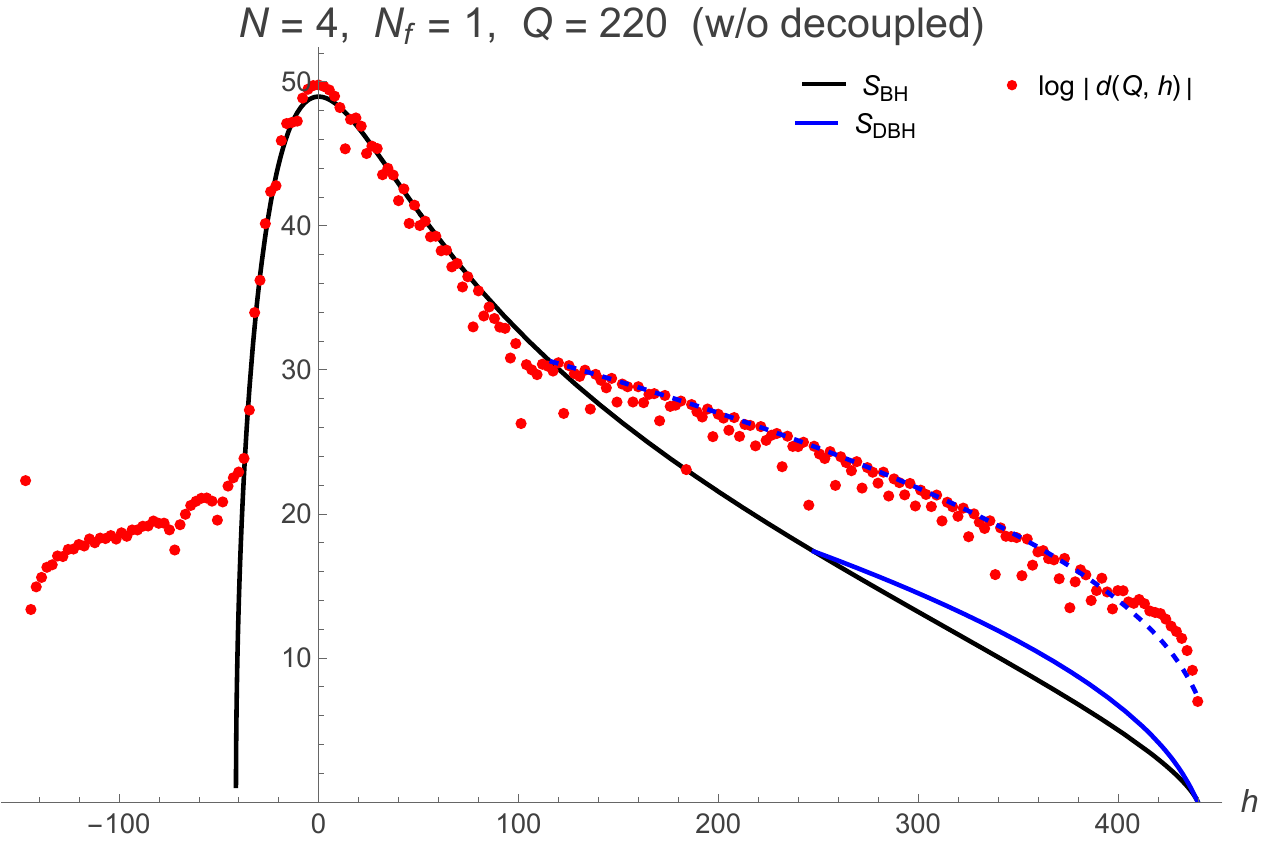}
	
	\caption{Top row: Indicial entropies for $N=2$ with $Q=200$ and $700$, with DBH behavior for $h>0$. 
		Middle row: $N=3$ with $Q=200$ and $360$, which shows DBH behavior for $h>0$ and $h<0$. 
		The dashed blue lines show the DBH curves shifted by the best-fit offsets.
		Bottom row: $N=4$ with comparatively small $Q=120$ and $Q=220$.
	}
	\label{fig:ADHM_N234_y}
\end{figure}

We start with $k=1$.
The indicial entropies for $N=2,3,4$ and two values of $Q$ for each $N$ are shown in Fig.~\ref{fig:ADHM_N234_y}. 
For $N=2$ the indicial entropy traces the black hole entropy in the central region, as in earlier figures. There is a clear excess on the $h>0$ sides, especially for the larger value of $Q=700$, already at $N=2$. The plateaux are suppressed strongly relative to the remaining features. 
On the $h<0$ sides, by contrast, the slowly-varying part of the indicial entropies for $N=2$ terminates well before the endpoints, which appear completely disconnected. There is no clear shoulder-like excess over the black hole entropy curve and no plateaux connecting the termination point of the black hole curve to the endpoint of the indicial entropy at $h=-2Q/3$.
For $N=3,4$, the endpoint of the indicial entropies at $h=-2Q/3$ remains disconnected from the slowly-varying part of the indicial entropies, but a shoulder fills in the space between the endpoint of the black hole curve and the endpoint of the indicial entropy. For $N=3$, where relatively large values of $Q$ could be captured, this leads to well-matching fits to DBH curves on both sides of the plots, i.e.\ for $h>0$ and $h<0$. Though the offsets are still non-negligible, especially on the $h<0$ sides, the qualitative behavior of the shoulders is captured well by the fits once the offset is accounted for. We added the entropy curves including the offset as dashed lines. The $N=4$ results are limited to lower $Q$. 
On the $h>0$ side, the indicial entropies are well captured by a DBH fit once the remaining offsets are taken into account (dashed lines), although towards the right end at $h=2Q$ the indicial entropy rises and then drops more sharply than the fitted curve. To a lesser extent this can also be observed for $N=3$. This slight discrepancy appears to decay as $Q$ is increased and we only used the first half of the index points in the DBH regime for the fits. The left side of the plots may show emerging evidence of DBH but calls for larger $Q$.

\begin{figure}[h]
    \centering

    \begin{minipage}{0.48\linewidth}
        \centering
        \includegraphics[width=\linewidth]{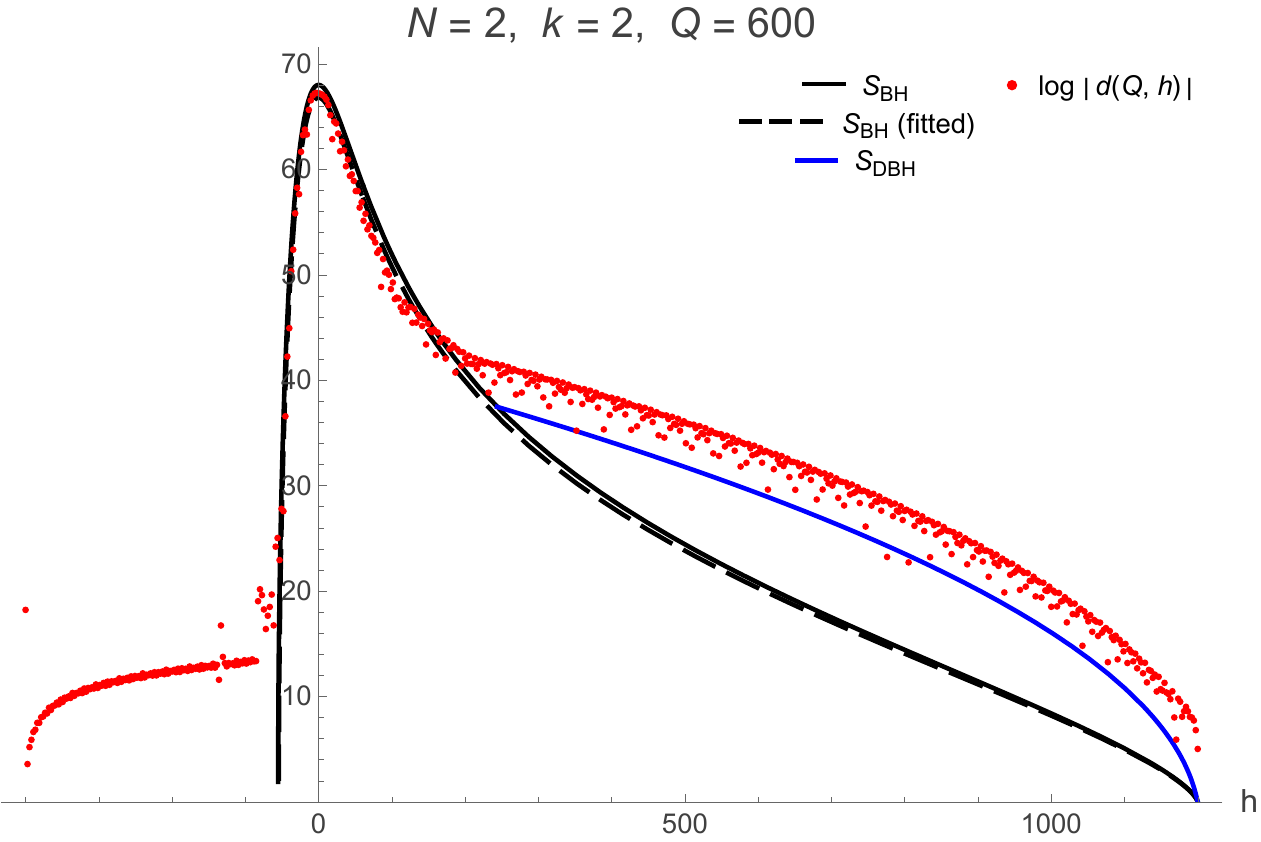}
    \end{minipage}
    \hfill
    \begin{minipage}{0.48\linewidth}
        \centering
        \includegraphics[width=\linewidth]{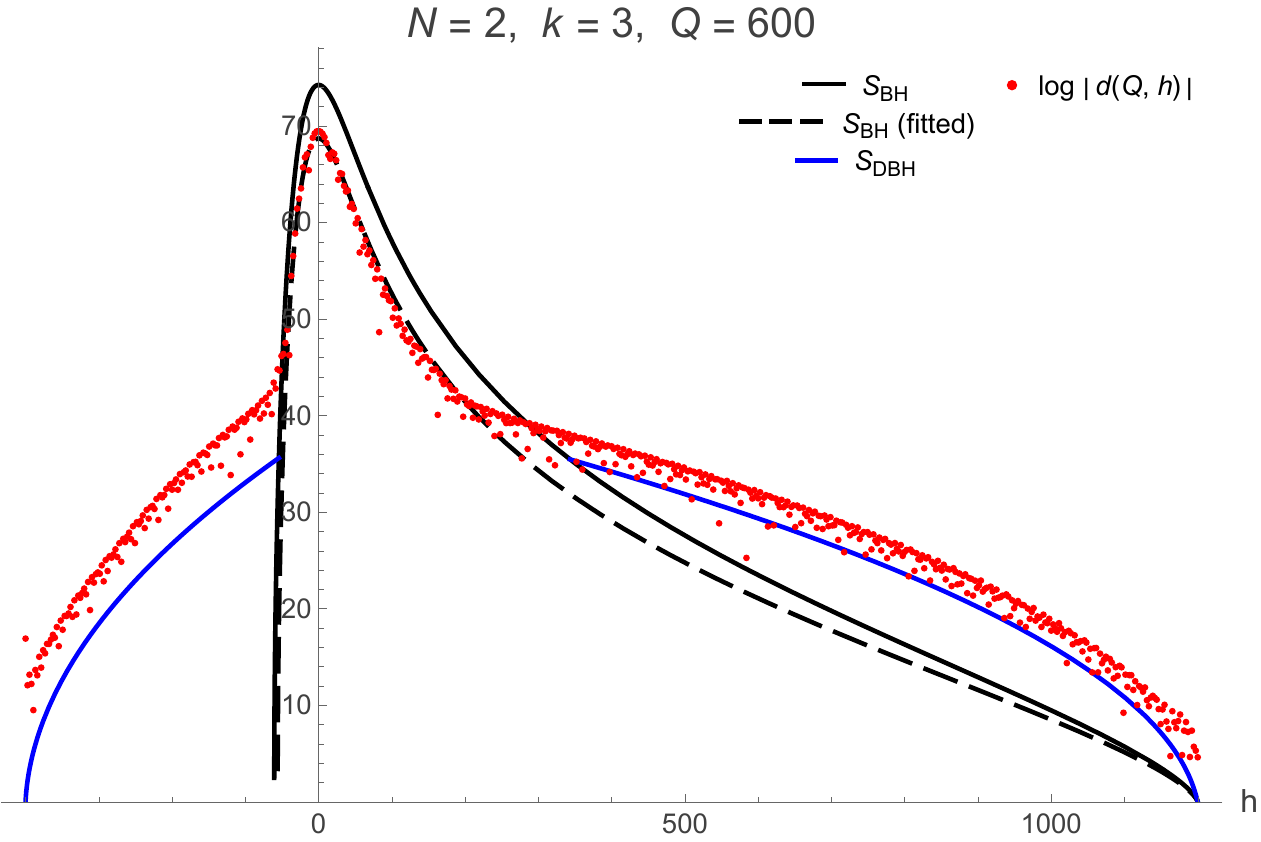}
    \end{minipage}
    \hfill
    \begin{minipage}{0.48\linewidth}
        \centering
        \includegraphics[width=\linewidth]{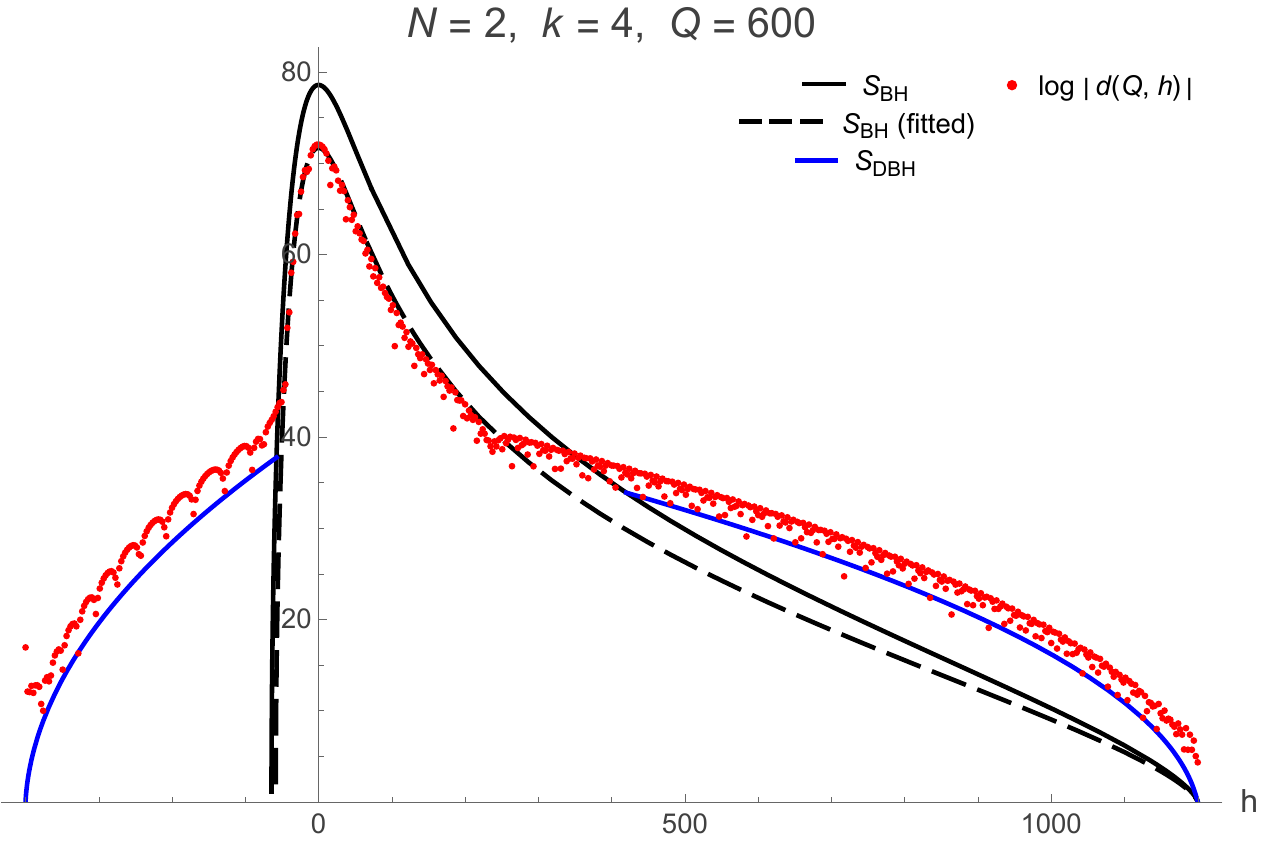}
    \end{minipage}
    \hfill
    \begin{minipage}{0.48\linewidth}
        \centering
        \includegraphics[width=\linewidth]{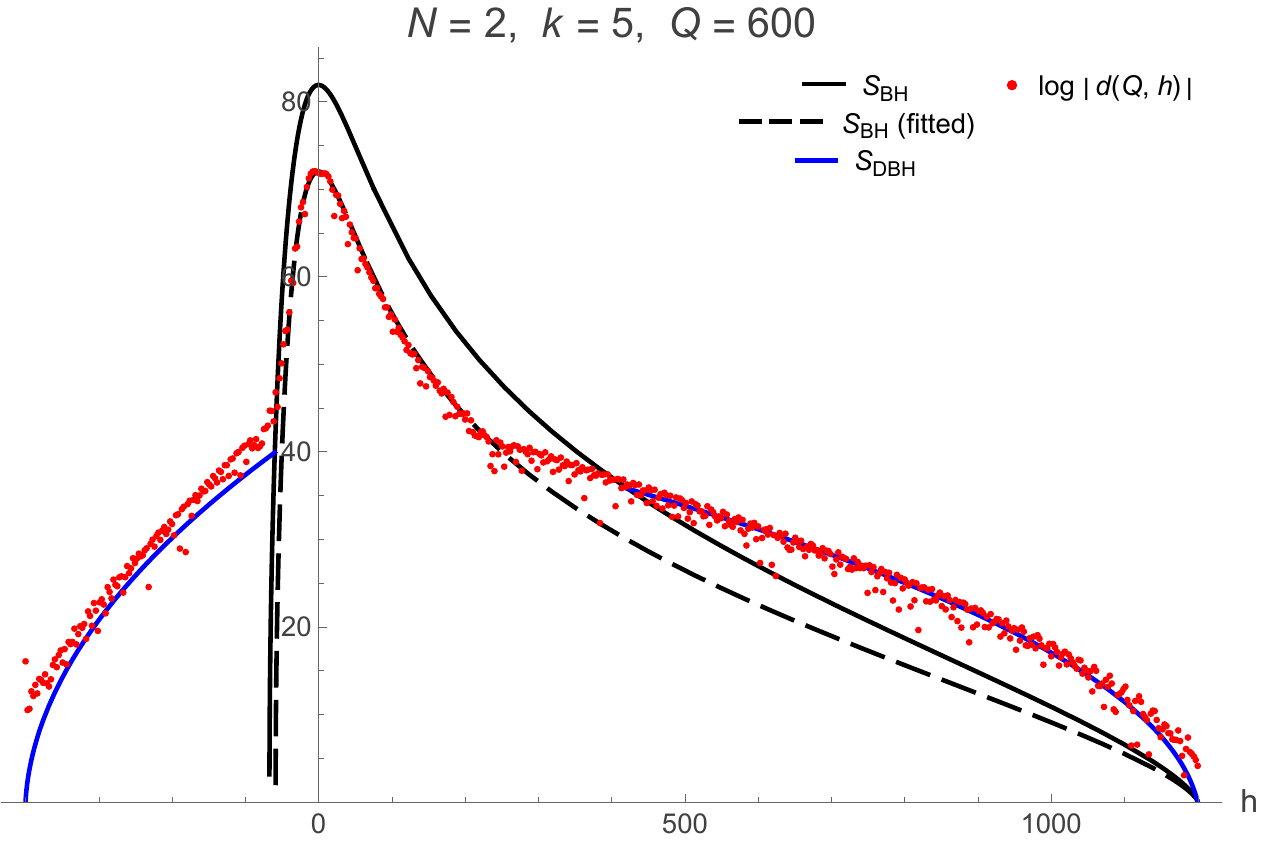}
    \end{minipage}

    \caption{The ABJM index at $N=2$ with $k=2,3,4,5$. The solid black hole curve uses the free energy from \eqref{eq:ZS3_ABJM_exact}, and the dashed black hole curve is fitted with $G_N$ as a free parameter.}
    \label{fig:ABJM_N2_k_y}
\end{figure}

The results for $k>1$ are shown in Fig.~\ref{fig:ABJM_N2_k_y} and cover $k=2,3,4,5$ at $Q=600$.
In the central region of the plots we note a slight increase in the discrepancy between the solid black curve, showing the entropy for an isolated black hole with Newton constant fixed by the free energy, and the indicial entropy as $k$ is increased. This parallels the discussion for the $(2,2)$  specialization with $k>1$. The mismatch can be eliminated by fitting the Newton constant, with the resulting dashed curves fitting the indicial entropy very well in the central region.
We also note that the shoulders on the $h<0$ side, which for $k=1$ only appeared at $N=3$, now appear already at $N=2$ for $k \geq 3$. This once again parallels the earlier onset of DBH-like behavior as $k$ is increased for the $(2,2)$ specialization.

\section{Discussion}\label{sec:comparison}

We computed the exact signed degeneracies of states captured by the superconformal index in ADHM and ABJM theories at finite $N$, for two specializations, dubbed $(2,2)$ and $(1,3)$, which each keep the dependence on two charges explicitly. For ADHM with $N_f=1$ or ABJM with $k=1$ we obtained results up to $N=3$ and $N=4$ in the $(2,2)$ and $(1,3)$ specializations, respectively. For ADHM with $N_f>1$ we obtained results up to $N=3$ and for ABJM with $k>1$ up to $N=2$. We compared the degeneracies to entropy curves motivated by large-$N$ holographic descriptions involving central black holes dressed by additional matter.

We find black hole behavior to emerge for large charge in the single-fugacity index even at the smallest values of $N$. For the single-fugacity index in $N_f=1$ ADHM / $k=1$ ABJM this was previously shown in \cite{Hwang:2025hfs}, whose results we extended to a substantially larger range of charge. While oscillatory modulations in the indicial entropy leave some freedom in extracting an effective Newton constant from the match to a gravitational description, the results are consistent with the semi-classical relation between free energy and Newton constant (\ref{eq:Newton}).

Our main focus was the extension to two-fugacity indices and a broader class of theories. The more detailed information about the CFT spectrum provided by retaining two fugacities in turn allows for more detailed comparisons to gravitational expectations. With the total charge $Q$ appearing in the BPS relation fixed, we can study the dependence of the (signed) state degeneracies, or indicial entropies, on an independent linear combination of charges, $h$. The $(2,2)$ and $(1,3)$ specializations give access to two different additional charge combinations.

While the indicial entropies as functions of $h$ do not match pure black hole entropies over the entire range of $h$, they do generally match in the region with largest degeneracies. This matching region dominates the single-fugacity index and discrepancies elsewhere are not in conflict with the single-fugacity index matching black hole entropies. In the matching region, an effective Newton constant can be extracted by fitting one free parameter in the black hole entropy to the index. For ABJM with $k\leq 2$ and ADHM with $N_f=1$, this effective Newton constant is consistent with the semi-classical relation between Newton constant and sphere free energy of the 3d SCFT (\ref{eq:Newton}). This relation does not extend beyond $k\leq 2$, with the index saturating at large $k$ in line with the recent discussion in \cite{Belin:2026glc}, see App.~\ref{sec:k-gg-N}.

The discrepancies of the indicial entropies from pure black hole curves came in two qualitatively different forms. The first are plateau-like excesses whose relative strength decreases with increasing $N$ and $Q$. 
These plateaux are expected to correspond to graviton states, as can be made precise for the endpoints of the plateaux.
The second discrepancy are excesses which take the form of `shoulders' to either side of the black hole curve. Over the range we tested these features persisted and sharpened upon increasing $N$ and/or $Q$.

In order to connect the persistent shoulders to a gravitational description, we used a simple entropy extremization ansatz which is motivated by recent discussions of composite configurations comprising black holes dressed by additional matter, without committing to a particular type of dressing. The main assumptions are: 1) While the central black hole can carry reduced charges and see a modified Newton constant, due to the dressing taking away charge, the remaining black hole is still described by the same entropy function as an isolated black hole. 2) The dressing can carry charge but contributes only negligibly to the entropy. We further included a purely empirical constant offset in the fitting ansatz to account for contributions that are expected to become subdominant at large $N$.

To give a quantitative form of the fitted Newton constants $\widetilde G_N$ governing the central black hole in dressed phases, where they were observed, we translate them to an effective rank $N_{\rm eff}$ by inverting the free energy expression
\begin{equation}
    \frac{\pi}{2\widetilde G_N}=F_\text{eff}=-\log Z_{S^3}[N_\text{eff}]+\log Z_{S^3}^\text{(dec)}.
\end{equation}
Since $\widetilde G_N$ was introduced as fit parameter in an effective description, we do not expect $N_{\rm eff}$ to generally be integer. We therefore use the Airy approximation \eqref{eq:Airy_Z} for $Z_{S^3}$, since the exact expressions \eqref{eq:ZS3_ABJM_exact}, \eqref{eq:ZS3_ADHM_exact} are restricted to integer rank. 
As discussed in App.~\ref{app:fitting}, the dependence of $\tilde G_N$ on $Q$ is consistent with convergent behavior as $Q\rightarrow\infty$. In the following tables we simply used the largest $Q$ available. For the $(2,2)$ indices the results are:
\begin{center}
\begin{tabular}{|c|c|c|c|}
	\hline
	\multicolumn{4}{|c|}{ABJM $N=2$}\\
	\hline
	$k$ & $3$ & $4$ & $5$\\
	\hline
	$N_{\rm eff}$ & $0.8$ & $0.8$ & $0.8$\\
	\hline
\end{tabular}
\hfill
\begin{tabular}{|c|c|c|c|}
	\hline
	\multicolumn{4}{|c|}{ADHM $N=2$}\\
	\hline
	$N_f$ & $2$ & $3$ & $4$\\
	\hline
	$N_{\rm eff}^L$ &  $0.7$ & $1.3$ & $1.3$ \\
	\hline
\end{tabular}
\hfill
\begin{tabular}{|c|c|c|c|c|}
	\hline
	\multicolumn{5}{|c|}{ADHM $N=3$}\\
	\hline
	$N_f$ & $1$ & $2$ & $3$ & $4$\\
	\hline
	$N_{\rm eff}^L$ & $1.7$ & $1.0$ & $2.2$ & $2.4$\\
	\hline
\end{tabular}
\end{center}
For ADHM we only included fits for the $h<0$ shoulders, as the $h>0$ shoulders were not consistently reliable. 
For the $(1,3)$ indices where shoulders could be identified clearly:
\begin{center}
	\begin{tabular}{|c|c|c|c|c|c|}
		\hline
		\multicolumn{6}{|c|}{ABJM $N=2$}\\
		\hline
		$k$ & $1$ & $2$ & $3$ & $4$ & $5$\\
		\hline
		$N_{\rm eff}^L$ & $-$ & $-$ & $0.8$ & $0.8$ & $0.7$ \\
		$N_{\rm eff}^R$ & $1.4$ & $1.0$ & $0.9$ &  $0.8$ & $0.8$ \\
		\hline
	\end{tabular}
    \qquad
    \begin{tabular}{|c|c|}
		\hline
		\multicolumn{2}{|c|}{ABJM $N=3$}\\
		\hline
		$k$ & $1$\\
		\hline
		$N_{\rm eff}^L$ & $1.3$\\
		$N_{\rm eff}^R$ & $1.8$\\
		\hline
	\end{tabular}
     \qquad
    \begin{tabular}{|c|c|}
		\hline
		\multicolumn{2}{|c|}{ABJM $N=4$}\\
		\hline
		$k$ & $1$\\
		\hline
		$N_{\rm eff}^R$ & $2.0$\\
		\hline
	\end{tabular}
\end{center}
Reassuringly, we always have $N_{\rm eff}<N$. We take the  qualitative agreement between the gravitational curves for dressed black holes and the finite-$N$ indicial entropies as evidence that multi-component configurations in AdS$_4$ indeed play a role.

Looking forward, it would be desirable to reach larger $N$ and $Q$ to explore the cases our results did not settle conclusively. It would also be interesting to explore other quantities which may afford systematic analytic control, such as the topologically twisted index for which finite-$N$ methods were recently deployed in \cite{Peng:2026udd}, and to broaden the space of theories, e.g.\ to the $\mathcal N=4$ theories discussed in \cite{Hayashi:2022ldo}
or the 3d long-quiver theories discussed in \cite{Coccia:2020wtk}. In a similar spirit, it would be interesting to connect other large-$N$ limits, such as the one considered in \cite{Komatsu:2026ano,Chang:2026yeg,Du:2026lbd}, to finite-$N$ results. Last but not least, it would be tantalizing to connect the small-$N$ black hole behavior we observed to the fortuity program \cite{Chang:2024zqi,Belin:2025hsg}.

\section*{Acknowledgments}
We thank Sunjin Choi, Dongming He, Chiung Hwang, Vineeth Krishna, Finn Larsen,  Eunwoo Lee, Yuezhang Tang, Paolo Vallarino and Alberto Zaffaroni for useful exchanges.
This work is partially supported by the U.S. Department of Energy under grant DE-SC0007859. ED is supported in part by a Leinweber Graduate Fellowship. SG is supported by the VUB Research Council through the Strategic Research Program High-Energy Physics. 

\vspace{.5cm}
\noindent
{\bf AI statement:} Codex was used to optimize our code for the index evaluation, as described in Appendix \ref{app:computational}, and for proofreading the manuscript. All other parts were done by the authors.

\appendix

\section{The 3d $\mathcal{N}=8$ Superconformal Algebra}\label{app:3d_N=8_SCA}
The space of supersymmetric states is constrained by the 3d $\mathcal{N}=8$ superconformal algebra. We work in radial quantization with Hermitian generators for the $\mathrm{SO}(3)$ rotation and $\mathrm{SO}(8)$ R-symmetry groups. In these conventions, the (anti)commutation relations \cite{Ferrara:2000ki} include
\begin{align}
& \left\{\mathcal{Q}_\alpha^i, \mathcal{S}_\beta^j\right\}=\delta^{i j} M_{\alpha \beta}+2 \epsilon_{\alpha \beta}\left(-iT^{i j}+\delta^{i j} D\right), \\
& {\left[D, \mathcal{Q}_\alpha^i\right]=\frac{1}{2} \mathcal{Q}_\alpha^i, \quad\left[D, \mathcal{S}_\alpha^i\right]=-\frac{1}{2} \mathcal{S}_\alpha^i,} \\
& {\left[M_{\alpha \beta}, \mathcal{Q}_\gamma^i\right]=\epsilon_{\gamma \alpha} \mathcal{Q}_\beta^i+\epsilon_{\gamma \beta} \mathcal{Q}_\alpha^i, \quad\left[M_{\alpha \beta}, \mathcal{S}_\gamma^i\right]=\epsilon_{\gamma \alpha} \mathcal{S}_\beta^i+\epsilon_{\gamma \beta} \mathcal{S}_\alpha^i,} \\
& {\left[T^{i j}, \mathcal{Q}_\alpha^k\right]=i\left(\delta^{k i} \mathcal{Q}_\alpha^j-\delta^{k j} \mathcal{Q}_\alpha^i\right), \quad\left[T^{i j}, \mathcal{S}_\alpha^k\right]=i\left(\delta^{k i} \mathcal{S}_\alpha^j-\delta^{k j} \mathcal{S}_\alpha^i\right),}
\end{align}
where $\mathcal{Q}_\alpha^i$ and $\mathcal{S}_\alpha^i$ are the supercharges, $D$ is the dilatation operator, $M_{\alpha \beta}$ are the generators of $\mathrm{SO}(3)$ in spinor notation, and $T^{ij}$ are the $\mathrm{SO}(8)$ generators in the vector representation. Note that the quantum numbers $(J,Q_1,Q_2,Q_3, Q_4)$ correspond to the eigenvalues under $(M_{12}/2,\;T^{12},T^{34},T^{56},T^{78})$. In radial quantization we have $(\mathcal{Q}_\alpha^i)^\dagger=\mathcal{S}^{i\alpha}=\epsilon^{\alpha\beta}\mathcal{S}^i_\beta$. In order to rotate to a basis of supercharges that have definite eigenvalues under the Cartans $T^{i,i+1}$, we define
\begin{equation}
    \mathcal{Q}^{m\pm}_\alpha=\frac12\left(\mathcal{Q}_\alpha^{2m-1}\pm i\mathcal{Q}_\alpha^{2m}\right),\qquad m=1,2,3,4.
\end{equation}
To match the conventions of \cite{Kim:2009wb,Hwang:2025hfs}, we must also convert the supercharges to the spinor representation of $\mathrm{SO}(8)$ via triality. This amounts to transforming the $T^{ij}$ generators as \cite{McRae:2025vvf}
\begin{equation}
\begin{pmatrix}
T^{12}\\
T^{34}\\
T^{56}\\
T^{78}
\end{pmatrix}
\;\longrightarrow\;
\frac12
\begin{pmatrix}
-1 & -1 & 1 & 1 \\
1 & 1 & 1 & 1\\
-1 & 1 & -1 & 1 \\
-1 & 1 & 1 & -1
\end{pmatrix}
\begin{pmatrix}
T^{12}\\
T^{34}\\
T^{56}\\
T^{78}
\end{pmatrix},
\end{equation}
Altogether, the $\{\mathcal Q,\mathcal S\}$ anti-commutation relations schematically take the form
\begin{equation}\label{eq:{Q,Qd}-app}
    \left\{\mathcal{Q}, \mathcal{Q}^{\dagger}\right\}=E \pm J+\frac{1}{2}\left( \pm Q_1 \pm Q_2 \pm Q_3 \pm Q_4\right).
\end{equation}
Each choice of $\mathcal{Q}$ realizes a unique sign combination, with the constraint that there are an even number of plus signs preceding the $Q_I$. 

\section{Computational strategies}\label{app:computational}

The results in section~\ref{sec:results} are produced using the matrix integral forms for the ABJM and ADHM indices in \eqref{eq:ABJM_index} and \eqref{eq:ADHM_index}, respectively, and the factorized form of the ADHM index in \eqref{eq:factorized_schematic}. For $k=1$ ABJM we use the ADHM description because the integration over gauge holonomies for $\mathrm{U}(N)$ is faster than for $\mathrm{U}(N)\times\mathrm{U}(N)$. For the ADHM index, the factorized form was previously used in \cite{Hwang:2025hfs} for the single fugacity index and reported to be faster than the matrix integral. For our calculations we found that the matrix integral, with a blockwise evaluation which we explain below, is faster than the factorized form for $N\leq 3$. The factorized form still is more efficient at larger $N$, and was used to obtain the $N=4$ results. For $k>1$ ABJM we use the matrix integral \eqref{eq:ABJM_index}.

\paragraph{Blockwise evaluation:}
The evaluation of the matrix integrals for the ABJM index \eqref{eq:ABJM_index} and ADHM index \eqref{eq:ADHM_index} in principle proceeds by expanding the integrand in the gauge holonomies and keeping the terms picked out by the contour integrals. This computation can be sped up considerably by generalizing the decomposition methods of \cite{Deddo:2025jrg}. To evaluate the index up to $\mathcal{O}(x^{Q_\text{max}})$, we can expand all plethystic exponentials and then organize terms by the gauge holonomies $z_i$. The integrand for the two-fugacity ADHM indices up to this order takes the schematic form
\begin{equation}\label{eq:block_schematic}
\sum_{\mathfrak{m}}c_\mathfrak{m}(x,y)\left[\prod^N_{\substack{i,j\geq1\\i>j}}\sum_n a_{\mathfrak{m},n}(x,y)\frac{z_i^n}{z_j^n}\right]\prod_{i=1}^N\left[\left(\sum_p f_{\mathfrak{m},p}(x,y)z_i^p\right)\left(\sum_q f_{\mathfrak{m},q}(x,y)z_i^{-q}\right)\right]^{N_f},
\end{equation}
where $a_{\mathfrak{m},n}(x,y)$ and $f_{\mathfrak{m},n}(x,y)$ are associated with the adjoint and (anti)fundamental contributions and $c_\mathfrak{m}(x,y)$ collects terms with no gauge holonomies. The sums over $n$, $p$ and $q$ encompass all terms arising from the expansion of the plethystic exponential to the desired order $Q_\text{max}$. After precomputing the $c$, $a$, and $f$ functions, the integration of \eqref{eq:block_schematic} reduces to  a combinatorial problem of building products of terms that yield no leftover $z_i$'s. This method is very efficient for $N\leq3$ but quickly grows intractable at larger $N$. A similar decomposition is possible for ABJM, though the treatment of the two sets of holonomies $z,\tilde{z}$ is more involved and restricts feasibility to $N\leq 2$. We initially implemented this procedure in Mathematica.

\subsection{Codex implementation}
For numerical evaluation of the ABJM and ADHM indices, we initially wrote Mathematica code. This was used as a seed to produce faster code in C++ using Codex.

We began the AI implementation by pointing Codex to the relevant literature and instructing it to produce a C++ implementation from scratch, including custom support for the manipulation of large series expansions. The results were compared to our own code for validation. For the matrix integral expressions, Codex produced correct code almost immediately. However, it was initially unable to implement the blockwise method satisfactorily. We therefore uploaded our Mathematica code as seed, after which it was able to apply the method successfully. After many rounds of optimization, the Codex implementation eventually became substantially faster than our original code.

For the factorized index, Codex required significantly more guidance. Na\"ively applying the fugacity relations for the (2,2) and (1,3) specializations yields singular results, and a successful implementation needs to correctly identify and remove vanishing factors in \eqref{eq:Z_pert}, \eqref{eq:Z_vort}. To this end, symbolic manipulations in three independent fugacities are initially required before a finite limit can be taken. 

\section{Black hole and endpoint entropy}\label{app:scalings}
\begin{figure}[h]
	\centering
	
	\begin{minipage}{0.48\linewidth}
		\centering
		\includegraphics[width=\linewidth]{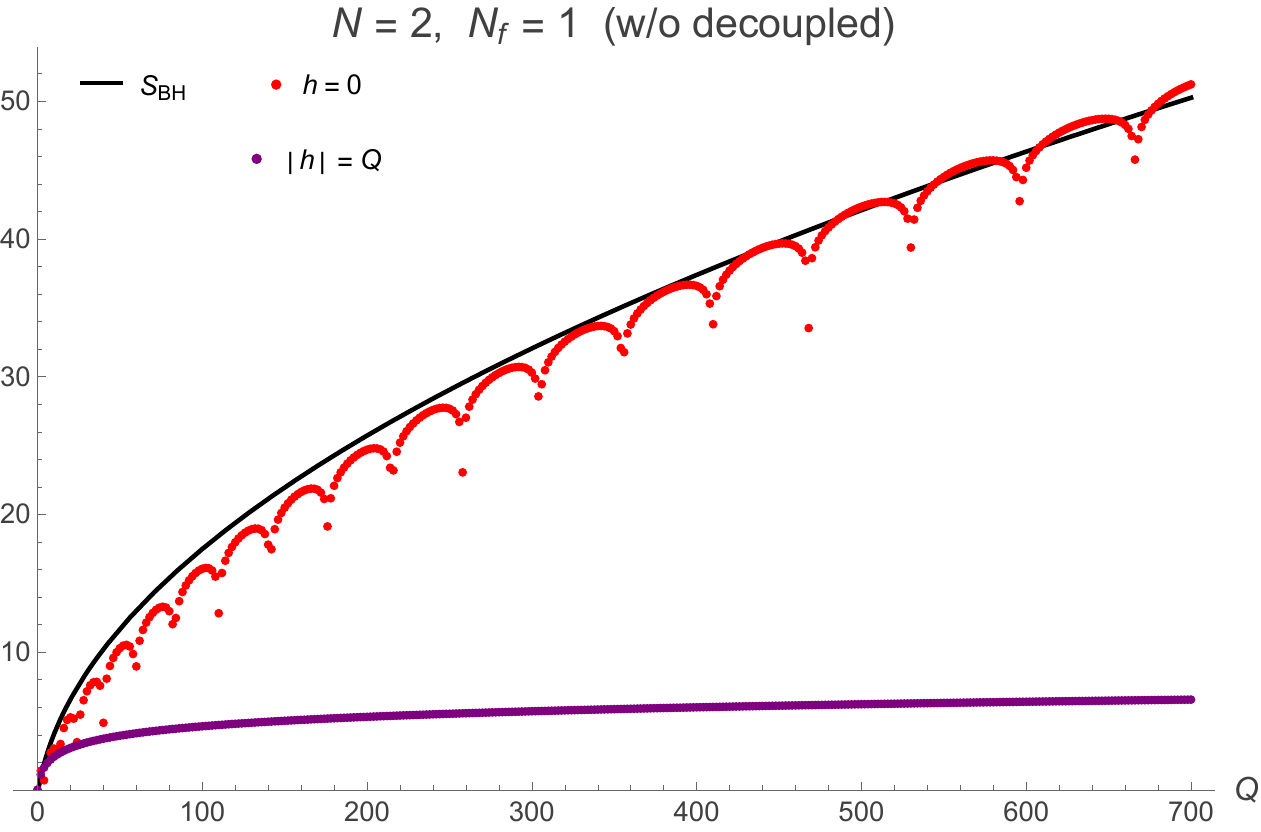}
	\end{minipage}
	\hfill
	\begin{minipage}{0.48\linewidth}
		\centering
		\includegraphics[width=\linewidth]{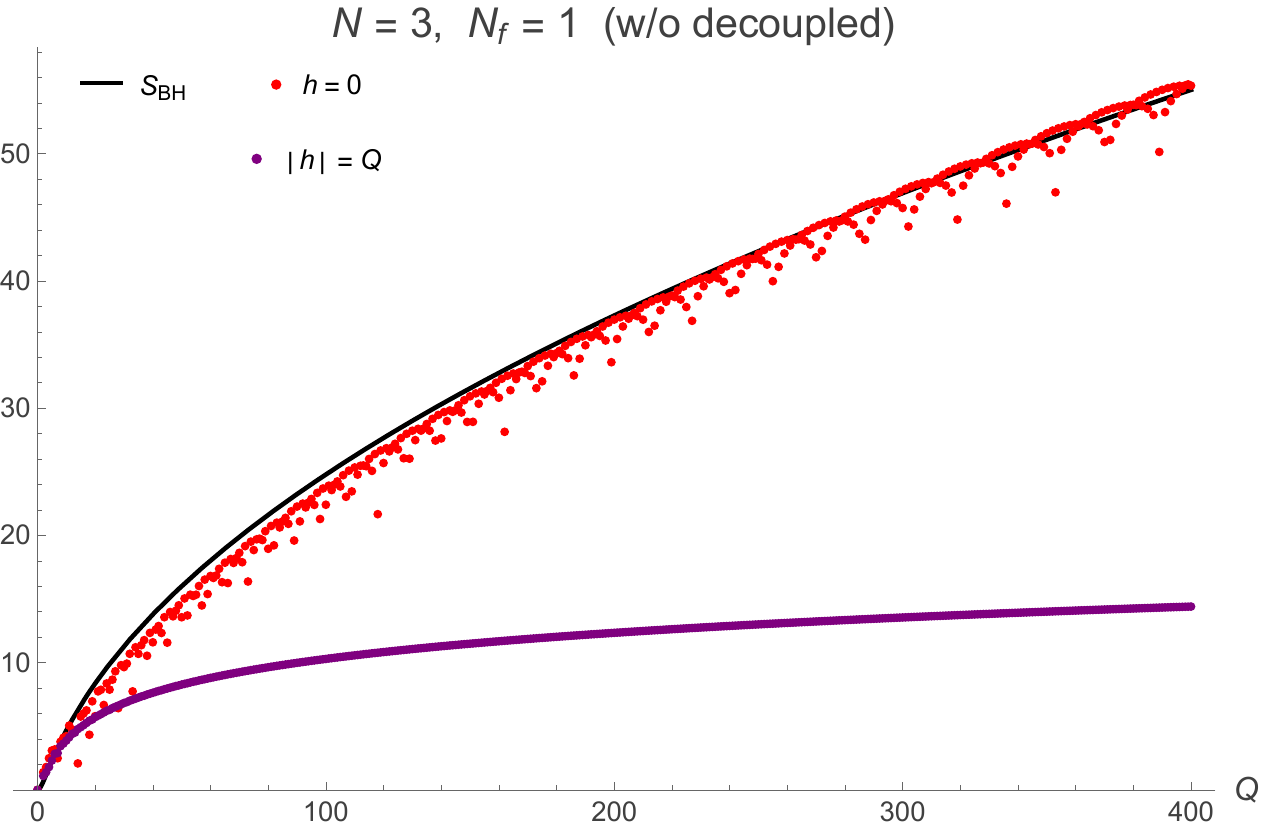}
	\end{minipage}
	
	\caption{The indicial entropy at $h=0$ and $|h|=Q$ for $0\leq Q\leq 700$ for $N=2$ and $0\leq Q\leq 400$ for $N=3$. The $h=0$ entropy tracks the black hole closely, while the $|h|=Q$ entropy shows slower logarithmic growth.}
	\label{fig:ADHM_BH_higgs}
\end{figure}

In the (2,2) index results in Sec.~\ref{sec:(2,2) results}, we observed plateau behavior in the index that deviates from the black hole entropy for sufficiently large $h$. This behavior is most readily visible for $N=2$, $k=1,2$ although it is also apparent close to $|h|\sim Q$ even when the DBH phase exists. To explore the distinction between the plateau and black hole regimes, we plot the indicial entropy $S_I(Q,h)=\log|d(Q,h)|$ at $h=0$ and $|h|=Q$ in Fig.~\ref{fig:ADHM_BH_higgs} for $N=2,3$. Due to the dominance of $h=0$ states in the indicial entropy, the $h=0$ plots are very similar to the single-fugacity plots in Fig.~\ref{fig:ADHM_single_fug} despite the latter summing over all $h$. They therefore maintain the $\sqrt{Q}$ growth characteristic of the black hole. However, the $|h|=Q$ points, which provide an upper bound for the entropy of the plateau, show much slower growth. The exact numerical values at large $Q$ are strongly suggestive of $S_I(Q,Q)=\log Q+\mathcal{O}(1)$ for $N=2$ and $S_I(Q,Q)=3\log Q+\mathcal{O}(1)$ for $N=3$. For the full index with the decoupled sector, the $|h|=Q$ curve would match the Higgs branch Hilbert series exactly with log coefficients given by the exponents in \eqref{eq:large_hilbert_series}. Removing the decoupled sector in Fig.~\ref{fig:ADHM_BH_higgs} maintains the logarithmic shape of the curves, but simply changes the precise coefficients.
The difference in $Q$ scaling for the black hole and plateau phases provides a useful diagnostic for distinguishing black hole and multigraviton phases. Since the shoulder-shaped features of Sec.~\ref{sec:(2,2) results} exhibit growth similar to the black hole, as expected from \eqref{eq:S_IIb},  it supports their identification as a new dressed phase rather than as multigraviton features.

\section{Fitting entropy curves}\label{app:fitting}

To compare the index with the dressed black hole ansatz in Sec.~\ref{sec:results}, we computed $S^{\rm grav}_{\mathcal I}(Q,h)$ from \eqref{eq:indentr_pairwise} and \eqref{eq:threeequal_indicial_entropy} and fit the value of $\tilde{G}_N$ to the index data. 

For the fits, we have to deal with the sign oscillations of the degeneracies $d(Q,h)$. The goal for the fits is to capture an enveloping curve tracing the peak degeneracies. Therefore, only index points that are local maxima and also lie clearly within the arc were used for the fits. As discussed in Sec.~\ref{sec:results}, we further include a constant offset in the fitting ansatz which is intended to parametrize unresolved finite-$N$ effects.

\begin{figure}[h]
	\centering
	\begin{minipage}{0.48\linewidth}
		\centering
		\includegraphics[width=\linewidth]{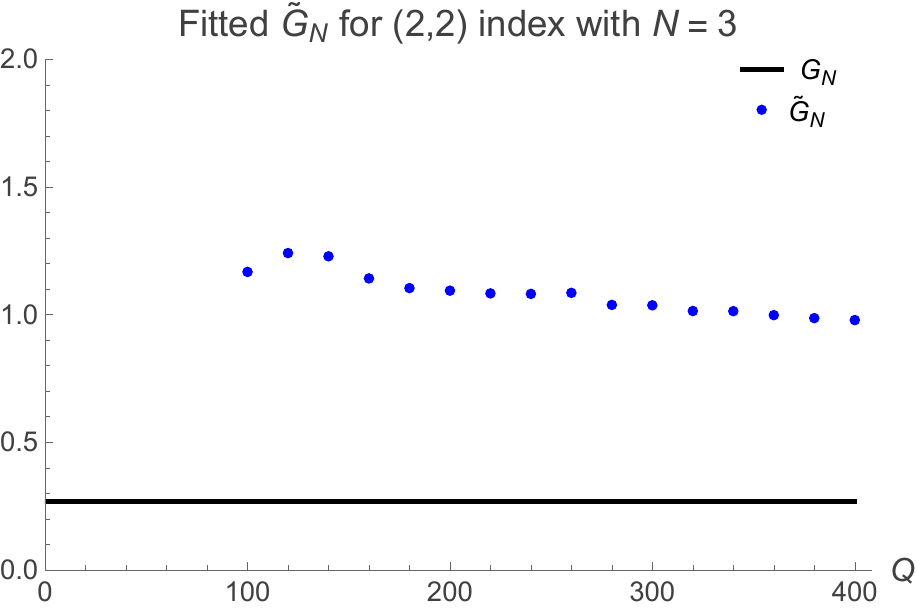}
	\end{minipage}
	\hfill
	\begin{minipage}{0.48\linewidth}
		\centering
		\includegraphics[width=\linewidth]{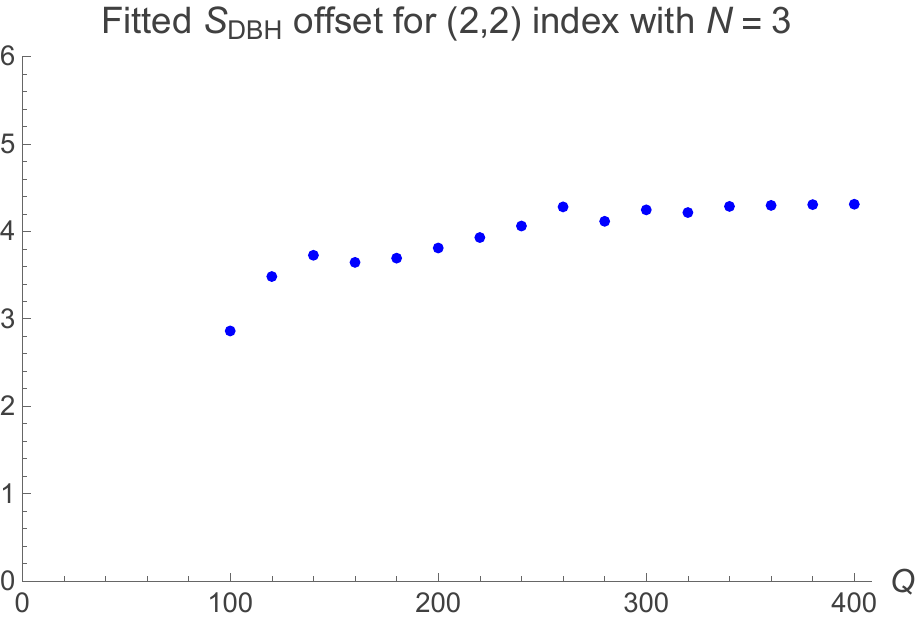}
	\end{minipage}
	
	\begin{minipage}{0.48\linewidth}
		\centering
		\includegraphics[width=\linewidth]{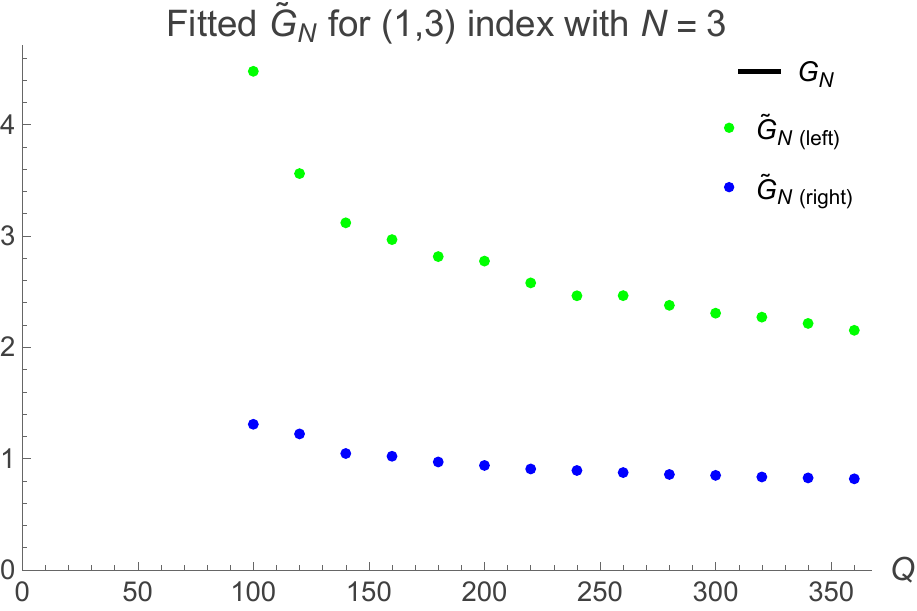}
	\end{minipage}
	\hfill
	\begin{minipage}{0.48\linewidth}
		\centering
		\includegraphics[width=\linewidth]{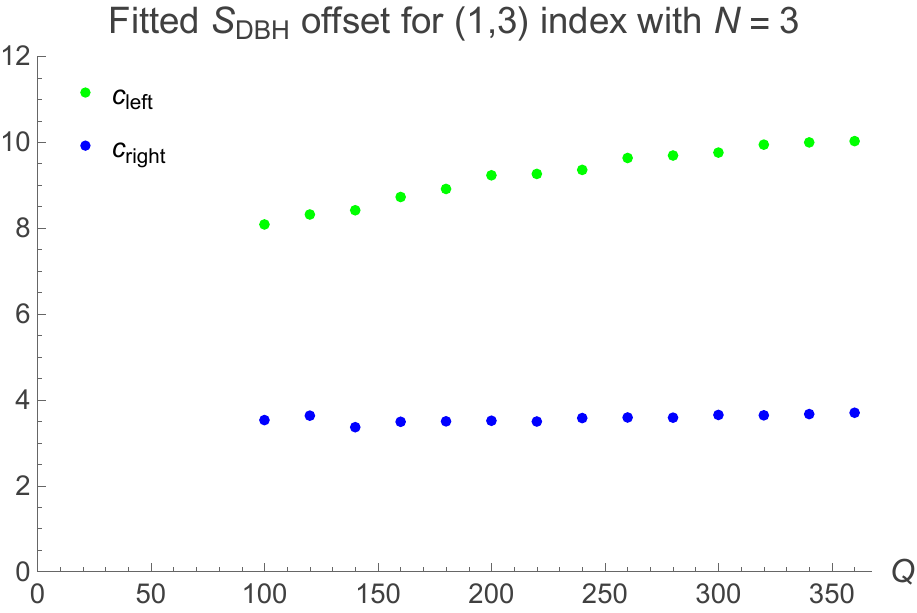}
	\end{minipage}

	\caption{Top: $\tilde G_N$ and vertical offset of the DBH curve fitted to the (2,2) index for a range of $Q$.
	Bottom: $\tilde G_N$ and vertical offset for the (1,3) index for a range of $Q$. Because this index is not symmetric, there are two independent values for the $h<0$ and $h>0$ regimes. 
	}
	\label{fig:ADHM_N3_y2_fit_params}
\end{figure}

One may ask how the best-fit value of $\tilde{G}_N$ at finite $N$ depends on $Q$. As an example, we consider the $(2,2)$ index for ADHM with $N_f=1$ and $N=3$. The fitted values of $\tilde{G}_N$ are shown for a range of $Q$ in the top row in Fig.~\ref{fig:ADHM_N3_y2_fit_params}. At small $Q$, $\tilde{G}_N$ begins large but decreases into a more constant trend for high $Q$. The bottom row shows similar data for the $(1,3)$ index. In both cases the values at the largest $Q$ shown appear reasonably well converged.


\section{Perturbative indices and vorticity cutoffs}\label{app:pert_vort}

In this section, we study to what extent perturbative sectors of the indices alone may demonstrate black hole and DBH behavior.

\paragraph{The zero-flux sector}\label{app:zero_mono_sector}

\begin{figure}[h]
    \centering
    \begin{minipage}{0.48\linewidth}
        \centering
        \includegraphics[width=\linewidth]{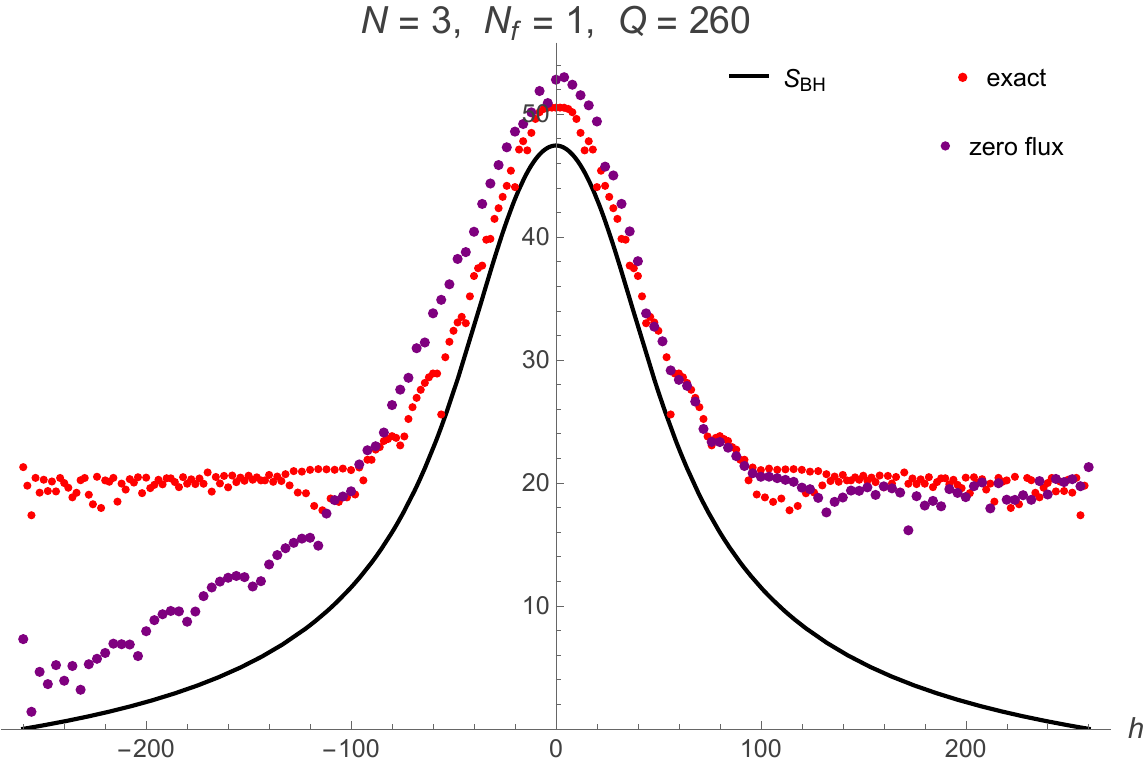}
    \end{minipage}
    \hfill
    \begin{minipage}{0.48\linewidth}
        \centering
        \includegraphics[width=\linewidth]{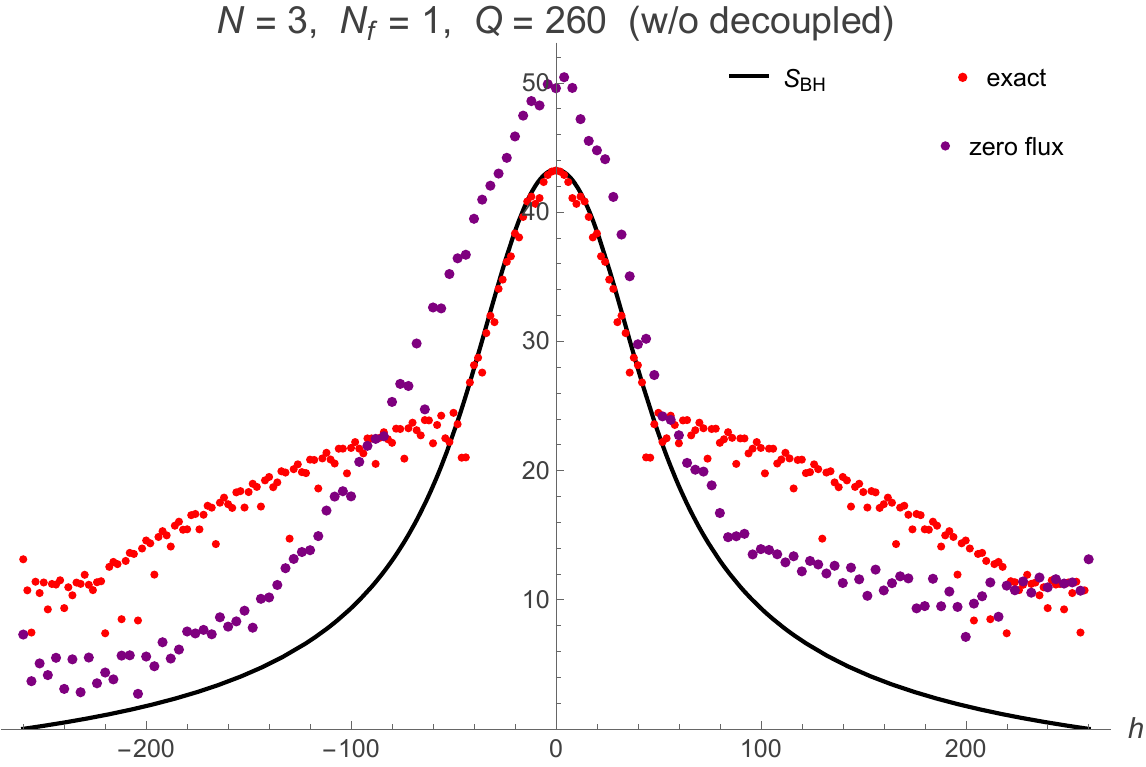}
    \end{minipage}

    \caption{The $N=3$, $N_f=1$ ADHM index restricted to the zero-flux sector (purple) compared with the exact index (red). The left plot includes the decoupled sector while the right plot excludes it.}
    \label{fig:ADHM_N3_y2_pnp}
\end{figure}

The matrix integral \eqref{eq:ADHM_index} for the ADHM index involves a sum over non-perturbative sectors labeled by monopole numbers $\mathfrak{m}\in\mathbb{Z}^N$. 
We use the zero-flux truncation $\mathcal I_m^{(0)}$ introduced in Sec.~\ref{sec:flux-vortex}.
Since this sector is neutral under $\mathrm{U}(1)_T$, the fugacity $\xi_T$ no longer appears. The left plot of Fig.~\ref{fig:ADHM_N3_y2_pnp} shows the zero-flux index (purple) at $N=3$ compared against the corresponding raw index (red). The zero-flux index follows the raw index relatively closely for $h\gtrsim -100$, but no longer shows a plateau for $h\lesssim-100$. The asymmetric behavior at large $|h|$ matches the discussion in Sec.~\ref{subsec:Coulomb-Higgs} and the bound \eqref{eq:ADHM_bare_monopole_contributions}, which increasingly restricts the contributing flux sectors as $h$ approaches $Q$. 
We may also consider the zero-flux index with the decoupled sector removed. After restricting to $\mathfrak{m}=0$, only the $\mathrm{U}(1)$ part of the adjoint hypermultiplet remains as decoupled sector. Therefore,
\begin{equation}
    \mathcal{I}^{(0)}_{m\text{ (no dec.)}}\left(x, \xi_{\ell}, \xi_F\right)=\frac{\mathcal{I}^{(0)}_m\left(x, \xi_{\ell}, \xi_F\right)}{\mathcal{I}^\mathrm{(N=1)}_\mathrm{hyper}\left(x, \xi_{\ell}, \xi_F\right)},
\end{equation}
where
\begin{equation}\label{eq:ADHM_index_U(1) hyper}
\mathcal{I}^\mathrm{(N=1)}_\mathrm{hyper}\left(x, \xi_{\ell}, \xi_F\right)=\text{PE}\left[\frac{x \left( \sqrt{\xi_\ell \xi_F} + \sqrt{\frac{\xi_F}{\xi_\ell}} \right) - x^3 \left( \frac{1}{\sqrt{\xi_\ell \xi_F}} + \sqrt{\frac{\xi_\ell}{\xi_F}} \right)}{1 - x^4}\right].
\end{equation}
The resulting index is shown in purple in the right plot of Fig.~\ref{fig:ADHM_N3_y2_pnp}. Although this sector demonstrates roughly BH-like degeneracy of states, it does not appear sufficient to capture DBH behavior.

\paragraph{Vorticity cutoff in the factorized index}\label{app:vort_cutoff}
We now consider truncations of the factorized ADHM index \eqref{eq:factorized_schematic}. We use the notation $\mathcal I_v^{(p)}$ introduced in Sec.~\ref{sec:flux-vortex}, writing $v_\text{cut}=p$. As discussed there, $v_\text{cut}=Q_\text{max}$ is sufficient to obtain exact coefficients through order $x^{Q_\text{max}}$. The effect of applying sharper vorticity cutoffs was considered in \cite{Hwang:2025hfs}. Although the computed coefficients are no longer exact, the resulting decrease in computation time makes reaching higher $Q_\text{max}$ feasible. For the single fugacity index, it was found that cutoffs as low as $v_\text{cut}\approx Q_\text{max}/2$ introduce only minor errors in the coefficients.

\begin{figure}[h]
    \centering
    \begin{minipage}{0.48\linewidth}
        \centering
        \includegraphics[width=\linewidth]{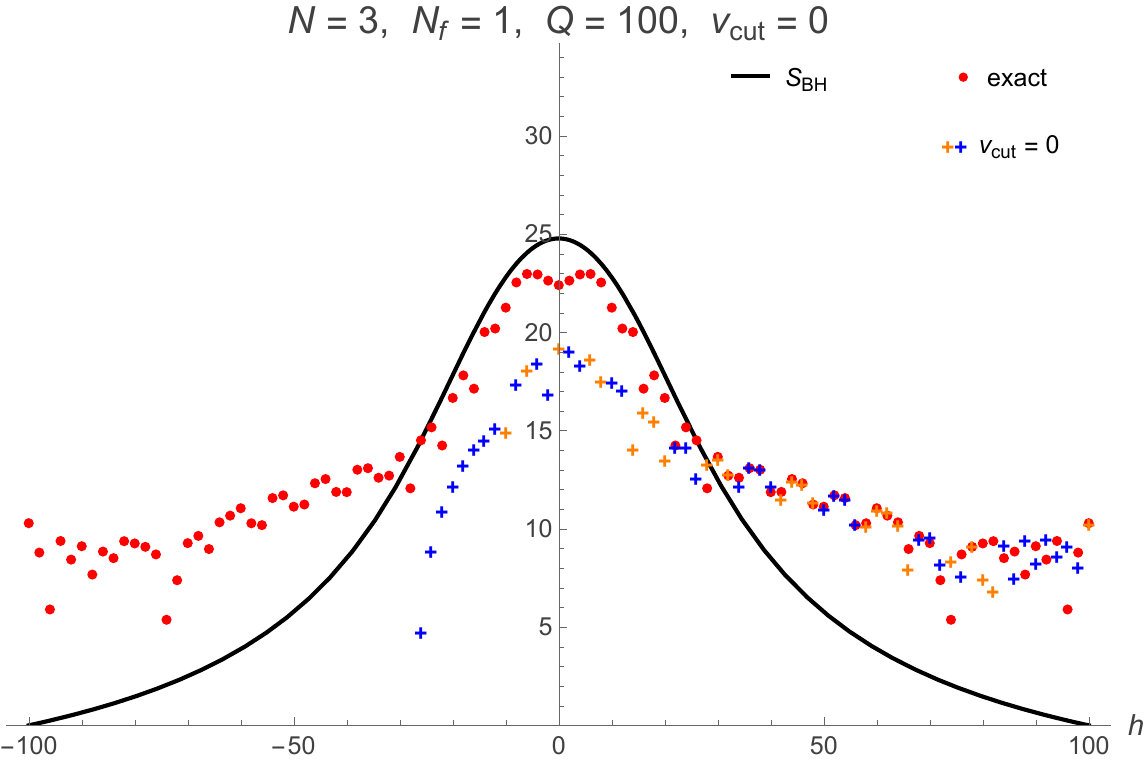}
    \end{minipage}
    \hfill
    \begin{minipage}{0.48\linewidth}
        \centering
        \includegraphics[width=\linewidth]{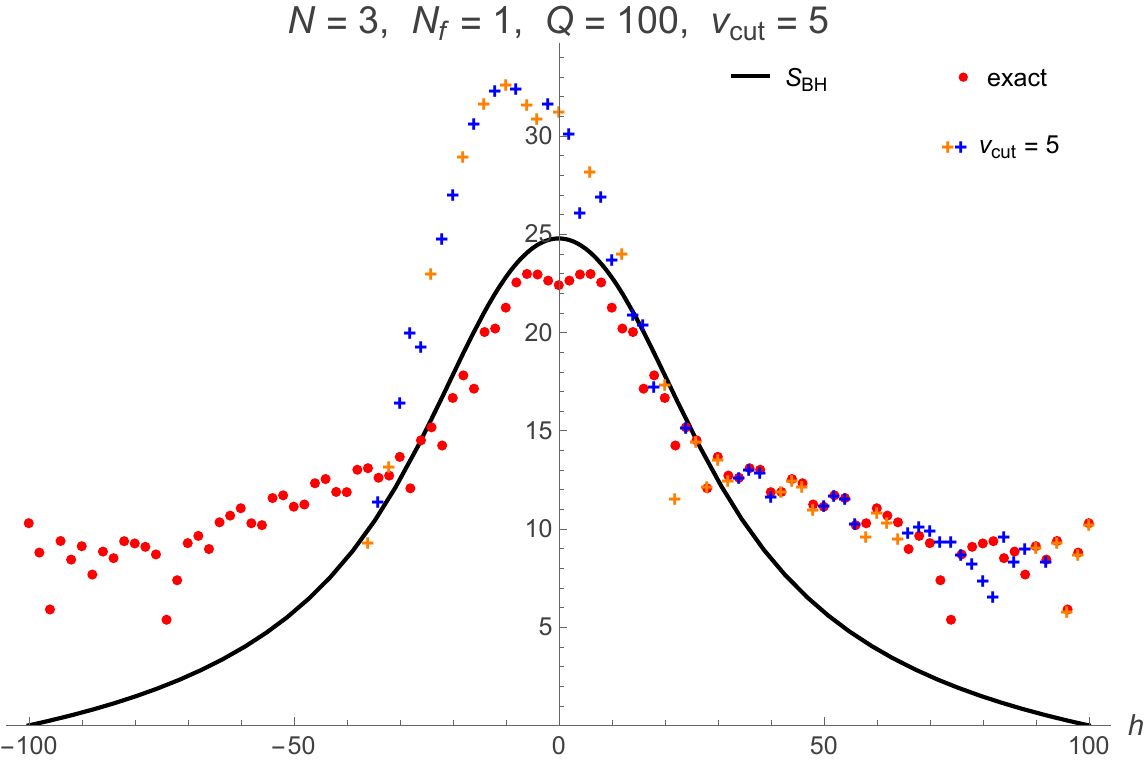}
    \end{minipage}
    \hfill
    \begin{minipage}{0.48\linewidth}
        \centering
        \includegraphics[width=\linewidth]{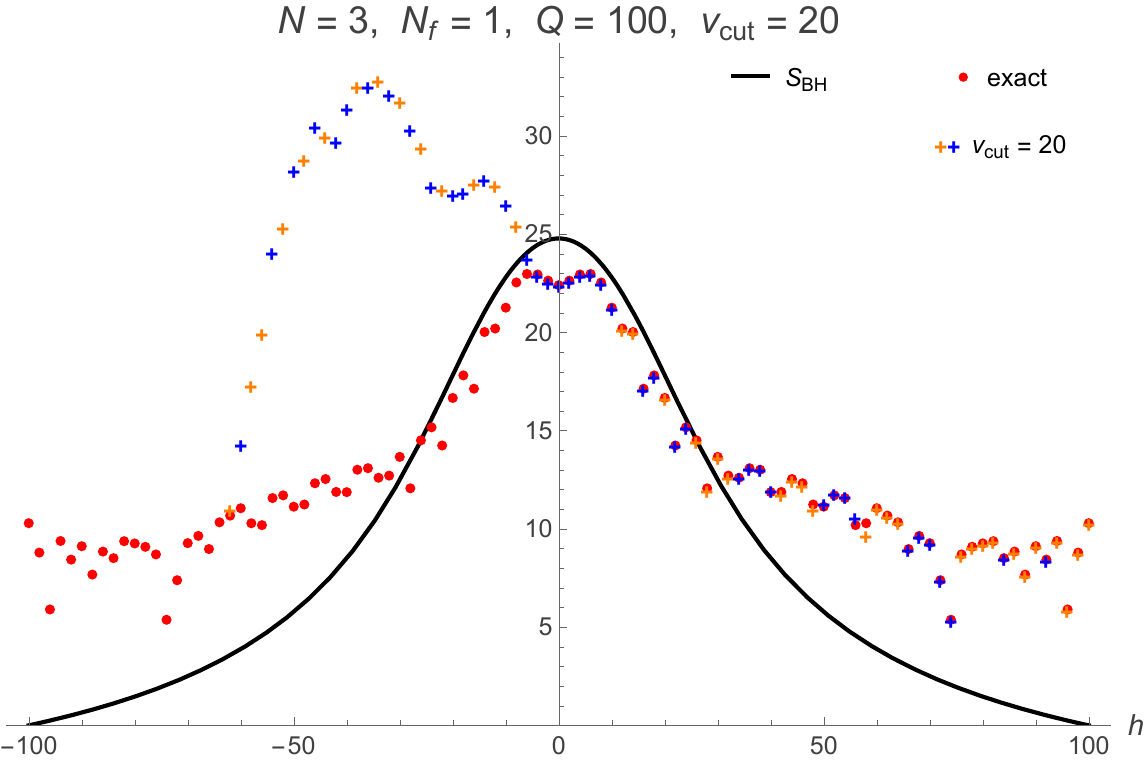}
    \end{minipage}
    \hfill
    \begin{minipage}{0.48\linewidth}
        \centering
        \includegraphics[width=\linewidth]{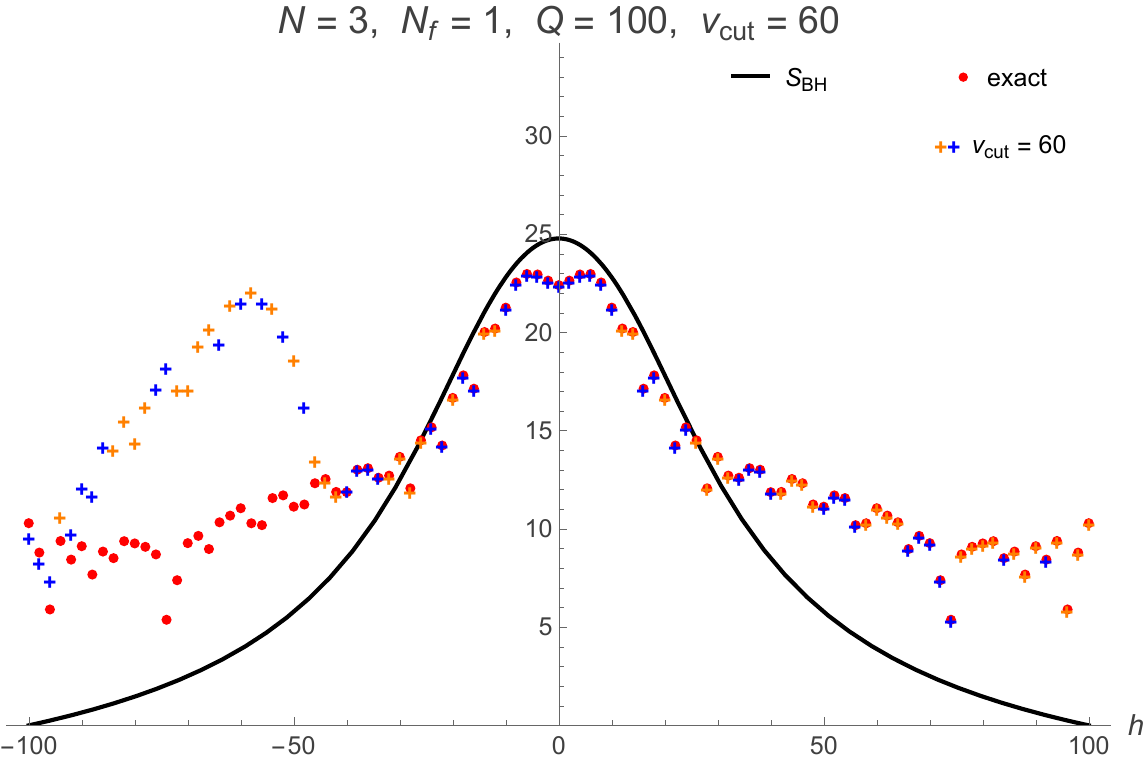}
    \end{minipage}

    \caption{The $N=3$ index with $Q=100$ for a range of vorticity cutoffs. The cutoff index is denoted by the '+' markers, with the orange and blue colors denoting positive and negative coefficients, respectively. $v_\text{cut}\geq 100$ yields the exact index, which is shown in red. }
    \label{fig:ADHM_N3_y2_cutoff}
\end{figure}

Here we test the cutoff approximation for the two-fugacity index. 
Fig.~\ref{fig:ADHM_N3_y2_cutoff} shows the effect of various $v_\text{cut}$ for $N=3$, $Q=100$. Lowering the cutoff leads quickly to inaccurate behavior on the $h<0$ side of the plots, and by $v_\text{cut}<20$ the index also deviates substantially in the pure BH region. The coefficients of the cut-off index are actually higher than the exact values in many cases, meaning that cancellations between different vortex sectors occur within the exact index. However, the signs of the coefficients (orange and blue) are highly oscillatory in the regions of excess. This oscillation explains why no excess was detected in the single fugacity results of \cite{Hwang:2025hfs}, since the total sum over all coefficients at fixed $Q$ is instead dominated by the more stable peak near $h=0$.

\begin{figure}[h]
    \centering
    \includegraphics[width=0.48\linewidth]{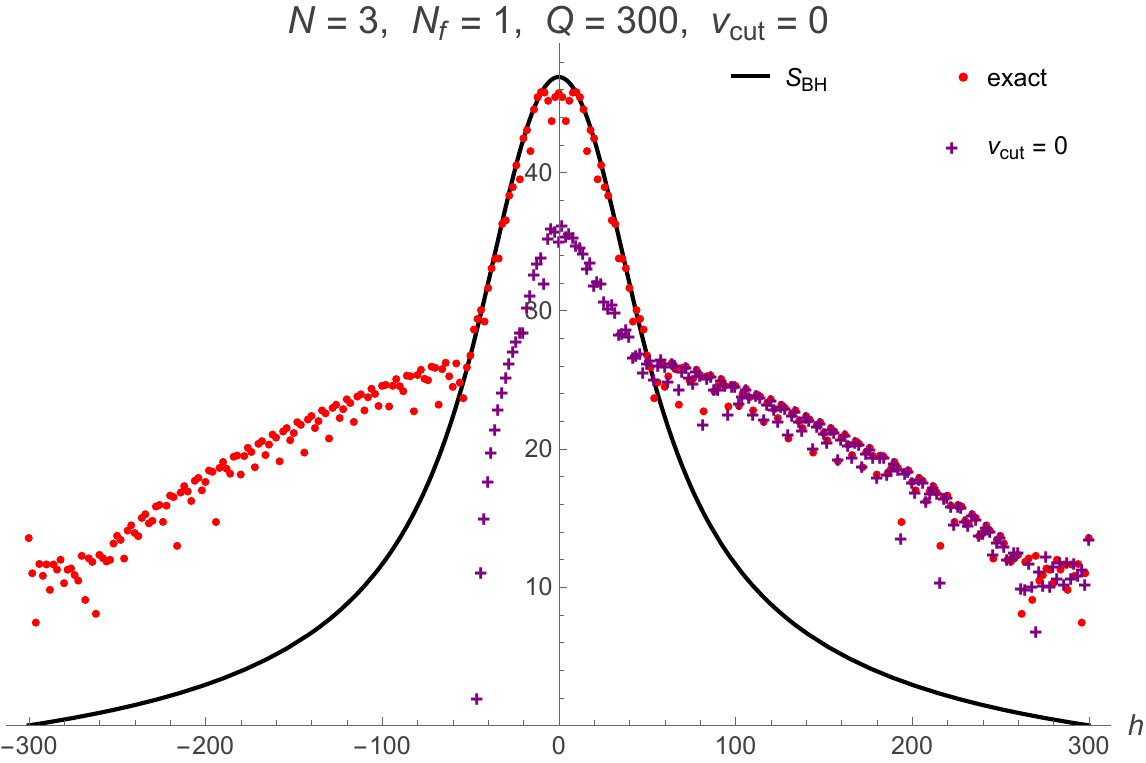}
    \caption{The  $v_\text{cut}=0$ limit for the $N=3$, $Q=300$ index.}
    \label{fig:ADHM_N3_y2_Q300cut}
\end{figure}

\begin{figure}[h]
    \centering
    \includegraphics[width=0.48\linewidth]{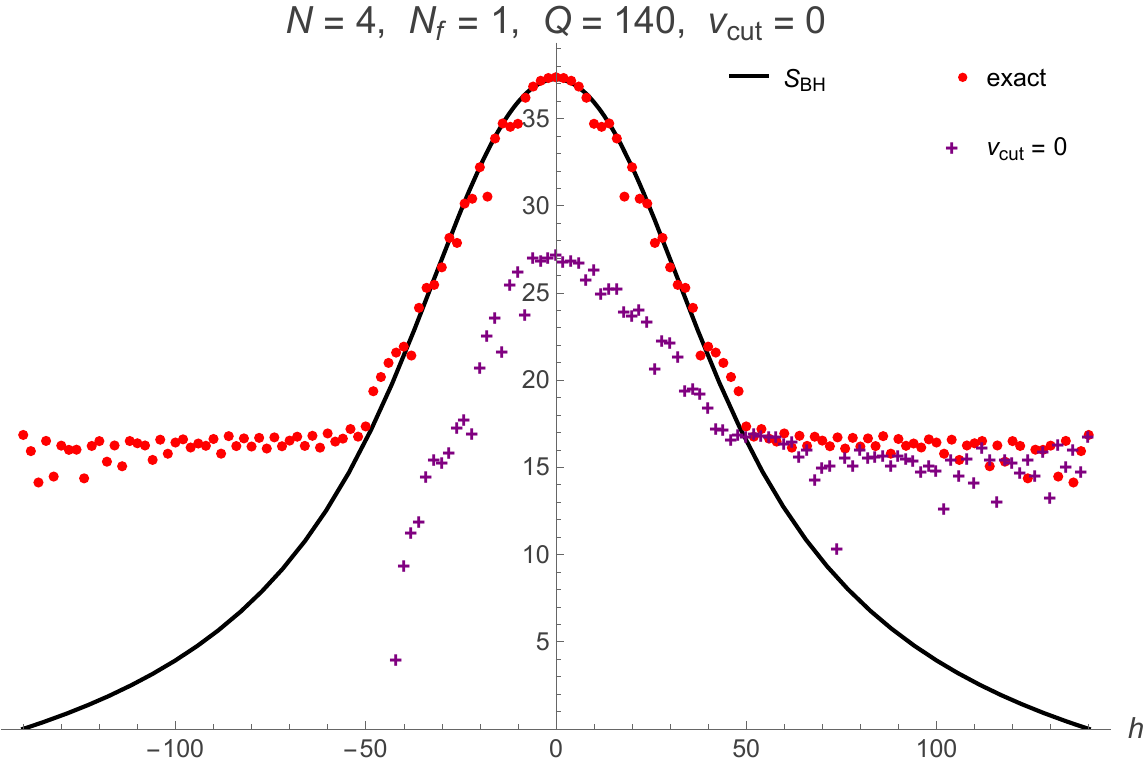}
    \quad
    \includegraphics[width=0.48\linewidth]{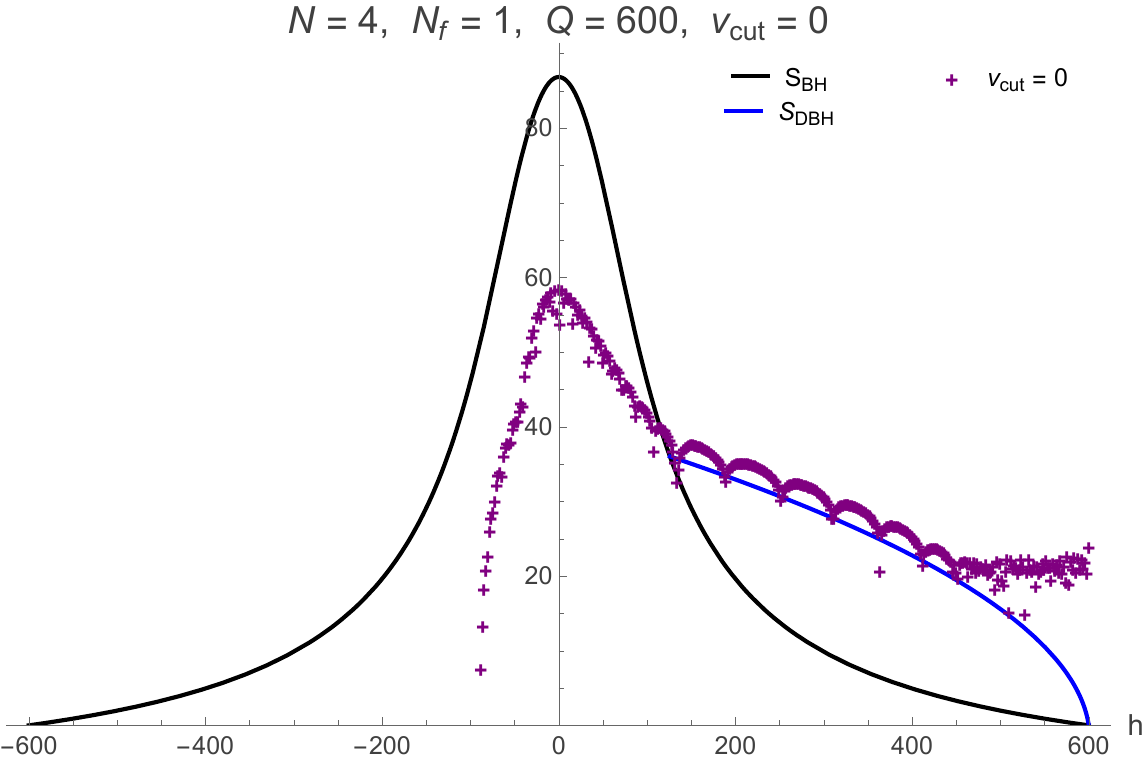}

    \caption{The $N=4$ index for $Q=140$ and $Q=600$. For $Q=140$, the exact computation (red) can be compared to the zero-vortex result (purple). The region $h>50$ is qualitatively similar, though not identical, as in  Fig.~\ref{fig:ADHM_N3_y2_cutoff}. For $Q=600$ the exact computation was not tractable, but the index without vortices matches a DBH curve.}
    \label{fig:ADHM_N4_y2_cutoff}
\end{figure}

Despite the failure of the cut-off index to capture the overall shape of the full index even qualitatively, it captures the dressed phase remarkably well. The shoulders appear even more persistent than the graviton region $h\gtrsim70$. Fig.~\ref{fig:ADHM_N3_y2_Q300cut} shows that this convenient agreement extends to larger $Q$ even when the vortex contributions are completely neglected, $v_{\rm cut}=0$.
If this extends to higher $N$ and $Q$, one could conceivably capture the DBH regime even where the exact index is beyond reach. Fig.~\ref{fig:ADHM_N4_y2_cutoff} presents results for $N=4$, where only $Q$ up to $140$ was reachable with exact computations. In the right plot at $Q=600$, the $v_\text{cut}=0$ computation visibly shows agreement with a DBH curve in the region $150<h<600$.

\section{Large $k$ at fixed $N$}\label{sec:k-gg-N}

Given the connections between the index and black hole physics at remarkably small $N$ discussed in the main part, one may wonder whether other criteria which are typically required for holographic behavior to emerge may also be relaxed while maintaining a connection to black hole physics. With this motivation in mind, we briefly comment on the limit of large $k$ at fixed $N$, where the `t Hooft coupling $\lambda=N/k$ vanishes and semi-classical gravity is not expected to provide an accurate description. This limit was discussed for the single-fugacity index in \cite{Belin:2026glc}, where the degeneracies at large charge and fixed $N$ were found to saturate for $k\rightarrow\infty$.
\begin{figure}
	\includegraphics[width=0.48\linewidth]{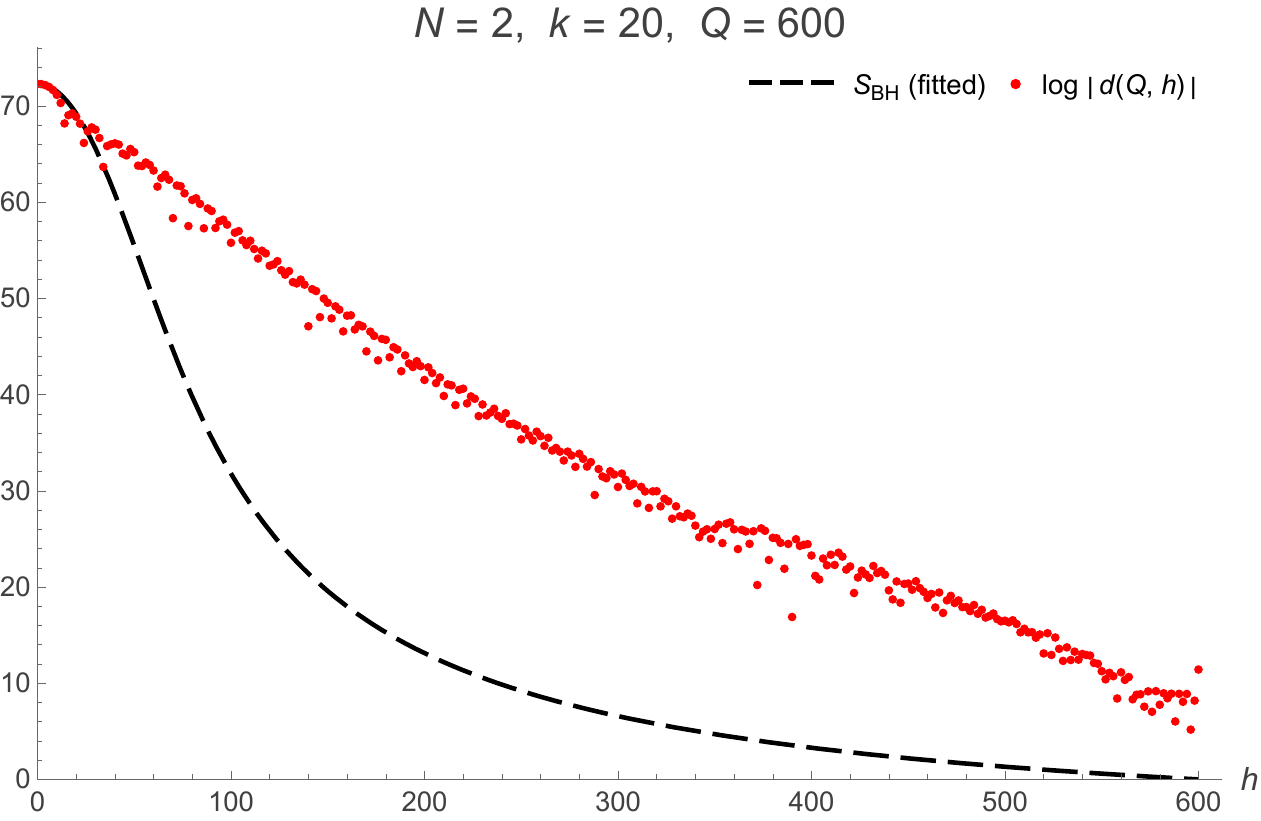}
	\quad
	\includegraphics[width=0.48\linewidth]{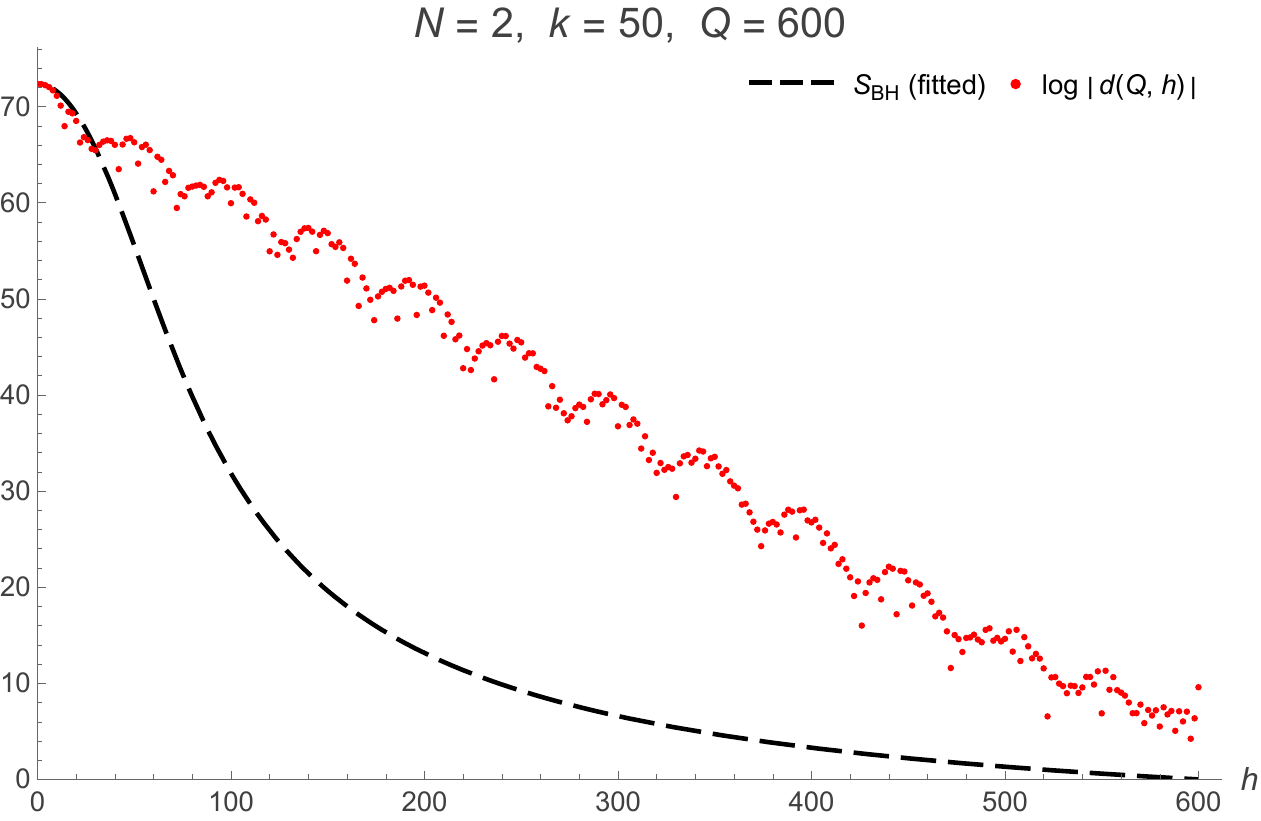}
	
	\caption{The ABJM index with $N=2$ and $k=20,50$. The maximum remains constant when increasing $k$, reflecting \cite{Belin:2026glc}. Attempting to fit the black hole entropy function (dashed curves) to these indices yields poor agreement.}\label{fig:large_k}
\end{figure}
We show the indicial entropies obtained from the two-fugacity index at $N=2$ for two large but finite $k$ in Fig.~\ref{fig:large_k}. While the curves are still peaked at $h=0$, their shape is triangular and distinct from black hole curves, in line with expectations. The height of the central peak varies slowly with $k$ in the regime of large $k$, in line with the discussion in \cite{Belin:2026glc}.

\bibliographystyle{JHEP}
\bibliography{step1ref}

\end{document}